\documentclass[
	english, 
	A4, oneside
]{article}

\def\documenttitle {Temporal Stamp Classifier: Classifying Short Sequence of Astronomical Alerts}
\def\documentsubtitle {}
\def\documentdate {\today}
\def\journalname {Revista, conferencia del artículo}

\usepackage[dvipsnames,table,usenames]{xcolor} 

\definecolor{cardinalred}{RGB}{140, 21, 21}
\definecolor{dkcyan}{RGB}{0, 123, 167}
\definecolor{dkgray}{RGB}{90, 90, 90}
\definecolor{dkgreen}{RGB}{0, 150, 0}
\definecolor{gray}{RGB}{127, 127, 127}
\definecolor{lbrown}{RGB}{255, 252, 249}
\definecolor{lgray}{RGB}{240, 240, 240}
\definecolor{mauve}{RGB}{150, 0, 210}
\definecolor{mitred}{RGB}{161, 0, 47}
\definecolor{ocre}{RGB}{243, 102, 25}

\def\documentfontsize {10}        
\def\documentinterline {1}         
\def\documentparindent {15}        
\def\documentparskip {0}           
\def\fontdocument {times}      
\def\fonttypewriter {tmodern}      
\def\fonturl {same}                
\def\graphicxdraft {false}         
\def\indentfirstpar {false}        
\def\pointdecimal {true}           
\def\showlayoutlines {false}       
\def\showlinenumbers {false}       
\def\twopagesclearformat {blank}   

\def\hfstyle {style1}              
\def\titleauthorspacing {0.35}     
\def\titleauthormarginbottom {0.3} 
\def\titleauthormaxwidth {0.85}    
\def\titlebold {true}              
\def\titlestyle {style1}           

\def\captionalignment {justified}  
\def\captionfontsize {small}       
\def\captionlabelformat {simple}   
\def\captionlabelsep {colon}       
\def\captionlessmarginimage {0.1}  
\def\captionlrmargin {0}           
\def\captionlrmarginmc {0}         
\def\captionmarginimage {0}        
\def\captionmarginimages {0}       
\def\captionmarginimagesmc {0}     
\def\captionmarginmultimg {0}      
\def\captionnumcode {arabic}       
\def\captionnumequation {arabic}   
\def\captionnumfigure {arabic}     
\def\captionnumsubfigure {alph}    
\def\captionnumsubtable {alph}     
\def\captionnumtable {arabic}      
\def\captionsubchar {.}            
\def\captiontbmarginfigure {9.35}  
\def\captiontbmargintable {7}      
\def\captiontextbold {true}        
\def\captiontextsubnumbold {false} 
\def\codecaptiontop {true}         
\def\equationcaptioncenter {true}  
\def\figurecaptiontop {false}      
\def\marginaligncaptbottom {0.1}   
\def\marginaligncapttop {-0.75}    
\def\marginalignedcaptbottom {0.1} 
\def\marginalignedcapttop {-0.75}  
\def\margineqncaptionbottom {0}    
\def\margineqncaptiontop {-0.7}    
\def\margingathercaptbottom {0.1}  
\def\margingathercapttop {-0.7}    
\def\margingatheredcaptbottom{0.1} 
\def\margingatheredcapttop {-0.7}  
\def\sectioncaptiondelimiter {.}   
\def\showsectioncaptioncode {none} 
\def\showsectioncaptioneqn {none}  
\def\showsectioncaptionfig {none}  
\def\showsectioncaptionmat {none}  
\def\showsectioncaptiontab {none}  
\def\subcaptionfsize{footnotesize} 
\def\subcaptionlabelformat{parens} 
\def\subcaptionlabelsep {space}    
\def\tablecaptiontop {true}        

\def\apacitebothers {et al.}       
\def\apaciterefcitecharclose {]}   
\def\apaciterefcitecharopen {[}    
\def\apaciterefnumber {false}      
\def\apaciterefsep {2}             
\def\apaciteshowurl {false}        
\def\apacitestyle {apacite}        
\def\appendixindepobjnum {true}    
\def\backrefpagecite {false}       
\def\bibtexenvrefsecnum {false}    
\def\bibtexrefsep {2}              
\def\bibtexstyle {ieeetr}          
\def\bibtextextalign {justify}     
\def\fontsizerefbibl {\small}      
\def\natbibrefcitecharclose {]}    
\def\natbibrefcitecharopen {[}     
\def\natbibrefcitecompress {true}  
\def\natbibrefcitesepcomma {true}  
\def\natbibrefcitetype {numbers}   
\def\natbibrefsep {2}              
\def\natbibrefstyle {ieeetr}    
\def\stylecitereferences {natbib}  
\def\twocolumnreferences {false}   

\def\abstractmarginbottom {0.0}    
\def\abstractmargintop {0}         
\def\animatedimageautoplay {true}  
\def\animatedimagecontrols {false} 
\def\animatedimageloop {true}      
\def\columnsepwidth {2.1}          
\def\defaultimagefolder {img/}     
\def\equationleftalign {false}     
\def\equationrestart {none}        
\def\footnotelmargin {10}          
\def\footnoterestart {none}        
\def\footnoterulefigure {false}    
\def\footnoterulepage {true}       
\def\footnoteruletable {false}     
\def\footnotetopmargin {0.5}       
\def\footnotetwocolumn {false}     
\def\fpremovetopbottomcenter{true} 
\def\imagedefaultplacement {H}     
\def\marginalignbottom {-0.4}      
\def\marginalignedbottom {-0.2}    
\def\marginalignedtop {-0.4}       
\def\marginaligntop {-0.4}         
\def\marginfloatimages {-13}       
\def\margingatheredbottom {-0.1}   
\def\margingatheredtop {-0.4}      
\def\marginimagebottom {-0.2}      
\def\marginimagemultright {0.5}    
\def\marginimagemulttop {-0.3}     
\def\marginimagetop {0}            
\def\marginlinenumbers {6}         
\def\numberedequation {true}       
\def\sitemsmargini {20}            
\def\sitemsmarginii {17}           
\def\sitemsmarginiii {0}           
\def\sitemsmarginiv {0}            
\def\sourcecodebgmarginbottom {0}  
\def\sourcecodebgmarginleft {0}    
\def\sourcecodebgmarginright {0}   
\def\sourcecodebgmargintop {0}     
\def\sourcecodefontf {\ttfamily}   
\def\sourcecodefonts {\small}      
\def\sourcecodeilfontf {\ttfamily} 
\def\sourcecodeilfonts {\small}    
\def\sourcecodenumbersep {6}       
\def\sourcecodenumbersize {\tiny}  
\def\sourcecodeskipabove {0.75}    
\def\sourcecodeskipbelow {0.95}    
\def\sourcecodetabsize {3}         
\def\tablenotesameline {true}      
\def\tablenotesfontsize{\footnotesize} 
\def\tablepaddingh {0.75}          
\def\tablepaddingv {1.15}          
\def\tikzdefaultplacement {H}      

\def\anumsecaddtocounter {false}   
\def\charaftersectionnum {}        
\def\charappendixsection {}        
\def\charbetwchaptersection {.}    
\def\charbetwssectionsssect {.}    
\def\charbetwsubsectionssect {.}   
\def\formatnumchapter {\Alph}    
\def\formatnumsection {\Roman}    
\def\formatnumssection {\Alph}   
\def\formatnumsssection {\Alph}  
\def\formatnumssssection {\Alph} 
\def\paragfontsize {\normalsize}   
\def\paragfontstyle {\bfseries}    
\def\paragspacingbottom {4}        
\def\paragspacingleft {0}          
\def\paragspacingtop {8}           
\def\paragsubfontsize{\normalsize} 
\def\paragsubfontstyle {\bfseries} 
\def\paragsubspacingbottom {4}     
\def\paragsubspacingleft {0}       
\def\paragsubspacingtop {8}        
\def\sectionfontsize {\Large}      
\def\sectionfontstyle {\bfseries}  
\def\sectionspacingbottom {10}     
\def\sectionspacingleft {0}        
\def\sectionspacingtop {15}        
\def\spacingaftersection {\quad}   
\def\ssectionfontsize {\large}     
\def\ssectionfontstyle {\bfseries} 
\def\ssectionspacingbottom {8}     
\def\ssectionspacingleft {0}       
\def\ssectionspacingtop {12}       
\def\sssectionfontsize{\normalsize}
\def\sssectionfontstyle{\bfseries} 
\def\sssectionspacingbottom {6}    
\def\sssectionspacingleft {0}      
\def\sssectionspacingtop {10}      
\def\ssssectionfontstyle{\bfseries}
\def\ssssectionfontsz{\normalsize} 
\def\ssssectionspacingbottom {4}   
\def\ssssectionspacingleft {0}     
\def\ssssectionspacingtop {8}      

\def\captioncolor {black}          
\def\captiontextcolor {black}      
\def\highlightcolor {yellow}       
\def\linkcolor {black}             
\def\maintextcolor {black}         
\def\numcitecolor {black}          
\def\pagescolor {white}            
\def\paragcolor {black}            
\def\paragsubcolor {black}         
\def\sectioncolor {black}          
\def\showborderonlinks {false}     
\def\sourcecodebgcolor {lgray}     
\def\ssectioncolor {black}         
\def\sssectioncolor {black}        
\def\ssssectioncolor {black}       
\def\tablelinecolor {black}        
\def\tablerowfirstcolor {none}     
\def\tablerowsecondcolor {gray!20} 
\def\urlcolor {magenta}            

\def\pagemarginbottom {1.91}       
\def\pagemarginleft {1.27}         
\def\pagemarginright {1.27}        
\def\pagemargintop {1.91}          

\def\cfgbookmarksopenlevel {1}     
\def\cfgpdfbookmarkopen {true}     
\def\cfgpdfcenterwindow {true}     
\def\cfgpdfcopyright {}            
\def\cfgpdfdisplaydoctitle {true}  
\def\cfgpdffitwindow {false}       
\def\cfgpdfkeywords {}             
\def\cfgpdflayout {OneColumn}      
\def\cfgpdfmenubar {true}          
\def\cfgpdfpageview {FitH}         
\def\cfgpdfsecnumbookmarks {true}  
\def\cfgpdftoolbar {true}          
\def\cfgshowbookmarkmenu {false}   
\def\indexdepth {4}                
\def\pdfcompilecompression {9}     
\def\pdfcompileobjcompression {2}  
\def\pdfcompileversion {7}         
\def\usepdfmetadata {true}         

\def\nameabstract {ABSTRACT}           
\def\nameappendixsection {Appendix}     
\def\namechapter {Chapter}           
\def\nameltappendixsection {Appendix}    
\def\nameltcont{Índice de Contenidos} 
\def\namelteqn {Índice de Ecuaciones} 
\def\nameltfigure {Índice de Figuras} 
\def\nameltsrc {Índice de Códigos}    
\def\namelttable {Índice de Tablas}   
\def\nameltwfigure {Figure}           
\def\nameltwsrc {Código}              
\def\nameltwtable {Table}             
\def\namemathcol {Corolario}          
\def\namemathdefn {Definición}        
\def\namemathej {Ejemplo}             
\def\namemathlem {Lema}               
\def\namemathobs {Observación}        
\def\namemathprp {Proposición}        
\def\namemaththeorem {Teorema}        
\def\namepageof { de }                
\def\namereferences {References}     

\let\counterwithin\relax
\let\underbar\relax
\let\underline\relax

\def\unaccentedoperators {}
\def\decimalpoint {}
\def\bibname {}

\makeatletter
\def\underline#1{\relax\ifmmode\@@underline{#1}\else $\@@underline{\hbox{#1}}\m@th$\relax\fi}
\def\underbar#1{\underline{\sbox\tw@{#1}\dp\tw@\z@\box\tw@}}
\makeatother

\usepackage{iftex}
\usepackage{ifthen}

\ifPDFTeX
	\def\compilertype {pdf2latex}
\else\ifXeTeX
	\def\compilertype {xelatex}
\else\ifLuaTeX
	\def\compilertype {lualatex}
\else
	\errmessage{Compilador no soportado}
	\stop
	\fi\fi
\fi

\usepackage{tracklang}
\ifthenelse{\equal{\stylecitereferences}{natbib}}{
	\usepackage[nottoc,notlof,notlot]{tocbibind}
	\ifthenelse{\equal{\natbibrefcitecompress}{true}}{
		\usepackage[sort&compress]{natbib}
	}{
		\usepackage{natbib}
	}
}{
\ifthenelse{\equal{\stylecitereferences}{apacite}}{
	\usepackage[nottoc,notlof,notlot]{tocbibind}
	\usepackage{apacite}
}{
\ifthenelse{\equal{\stylecitereferences}{bibtex}}{
}{
\ifthenelse{\equal{\stylecitereferences}{custom}}{
}{}}}
}
\IfTrackedLanguage{spanish}{
	\usepackage[es-nosectiondot,es-lcroman,es-noquoting]{babel}
}{
	\usepackage{babel}
}

\usepackage{sectsty}

\ifthenelse{\equal{\compilertype}{pdf2latex}}{
	\usepackage[utf8]{inputenc}}{
}

\newcommand{\throwbadconfig}[4][]{
	\ifthenelse{\equal{#1}{noheader}}{
		\errmessage{LaTeX Warning: #4}
	}{
		\ifthenelse{\equal{#1}{noheader-nostop}}{
			\errmessage{LaTeX Warning: #4}
		}{
			\errmessage{LaTeX Warning: #2 (\noexpand #3= #3). Valores esperados: #4}
		}
	}
	\ifthenelse{\equal{#1}{nostop}}{}{
		\ifthenelse{\equal{#1}{noheader-nostop}}{}{
			\stop
		}
	}
}

\ifthenelse{\equal{\equationleftalign}{true}}{
	\usepackage[fleqn]{amsmath}
}{
	\usepackage{amsmath}
}

\usepackage{scrextend}
\usepackage{anyfontsize}
\changefontsizes{\documentfontsize pt}

\usepackage{amsbsy}        
\usepackage{amssymb}       
\usepackage{amsthm}        
\usepackage{animate}       
\usepackage{array}         
\usepackage{bigstrut}      
\usepackage{bm}            
\usepackage{booktabs}      
\usepackage{caption}       
\usepackage{chngcntr}      
\usepackage{color}         
\usepackage{datetime}      
\usepackage{floatpag}      
\usepackage{floatrow}      
\usepackage{framed}        
\usepackage{gensymb}       
\usepackage{graphicx}      
\usepackage{lipsum}        
\usepackage{listings}      
\usepackage{longtable}     
\usepackage{mathrsfs}      
\usepackage{mathtools}     
\usepackage{multicol}      
\usepackage{needspace}     
\usepackage{pdflscape}     
\usepackage{pdfpages}      
\usepackage{physics}       
\usepackage{realboxes}     
\usepackage{rotating}      
\usepackage{selinput}      
\usepackage{setspace}      
\usepackage{soul}          
\usepackage{stfloats}      
\usepackage{subcaption}    
\usepackage{textcomp}      
\usepackage{wrapfig}       
\usepackage{xspace}        
\usepackage{xurl}          

\usepackage[export]{adjustbox} 
\usepackage[makeroom]{cancel} 
\usepackage[inline]{enumitem} 
\usepackage[titles]{tocloft} 
\usepackage[figure,table,lstlisting]{totalcount} 
\usepackage[normalem]{ulem} 
\usepackage[nointegrals]{wasysym} 

\ifthenelse{\equal{\graphicxdraft}{true}}{
	\usepackage[
		allfiguresdraft,
		filename,
		size={scriptsize},
		style={tt}
	]{draftfigure}}{
}

\ifthenelse{\equal{\compilertype}{pdf2latex}}{
	\usepackage{listingsutf8}}{
}

\ifthenelse{\equal{\footnotetwocolumn}{true}}{
	\usepackage{dblfnote}}{
}

\ifthenelse{\equal{\footnoterulepage}{true}}{
	\usepackage[bottom,hang]{footmisc} 
}{
	\usepackage[bottom,norule,hang]{footmisc}
}

\ifthenelse{\equal{\backrefpagecite}{true}}{
	\usepackage[pdfencoding=auto,psdextra,backref=page,bookmarks=false]{hyperref} 
}{
	\usepackage[pdfencoding=auto,psdextra,bookmarks=false]{hyperref} 
}

\def\showappendixsecindex {false}
\ifthenelse{\equal{\showappendixsecindex}{true}}{
}{
	\usepackage{appendix}
}

\ifthenelse{\equal{\compilertype}{lualatex}}{
	\usepackage[top=\pagemargintop cm,bottom=\pagemarginbottom cm,left=\pagemarginleft cm,right=\pagemarginright cm, footnotesep=\footnotetopmargin cm]{geometry}
}{
	\usepackage{geometry}
}

\ifthenelse{\equal{\tablenotesameline}{true}}{
	\usepackage[para]{threeparttable}
}{
	\usepackage{threeparttable}
}

\ifthenelse{\equal{\indentfirstpar}{true}}{
	\usepackage{indentfirst}}{
}

\usepackage{bookmark}      
\usepackage{fancyhdr}      
\usepackage{float}         
\usepackage{hyperxmp}      
\usepackage{multirow}      
\usepackage{notoccite}     
\usepackage{titlesec}      

\ifthenelse{\equal{\fontdocument}{lmodern}}{
	\usepackage{lmodern}
}{
\ifthenelse{\equal{\fontdocument}{arial}}{
	\usepackage{helvet}
	
}{
\ifthenelse{\equal{\fontdocument}{arial2}}{
	\usepackage{arial}
}{
\ifthenelse{\equal{\fontdocument}{times}}{
	\usepackage{mathptmx}
}{
\ifthenelse{\equal{\fontdocument}{mathptmx}}{
	\usepackage{mathptmx}
}{
\ifthenelse{\equal{\fontdocument}{helvet}}{
	
	\usepackage[scaled=0.95]{helvet}
	\usepackage[helvet]{sfmath}
}{
\ifthenelse{\equal{\fontdocument}{opensans}}{
	\usepackage[default,scale=0.95]{opensans}
}{
\ifthenelse{\equal{\fontdocument}{mathpazo}}{
	\usepackage{mathpazo}
}{
\ifthenelse{\equal{\fontdocument}{cambria}}{
	\usepackage{caladea}
}{
\ifthenelse{\equal{\fontdocument}{libertine}}{
	\usepackage[libertine]{newtxmath}
	\usepackage[tt=false]{libertine}
}{
\ifthenelse{\equal{\fontdocument}{custom}}{
}{

\ifthenelse{\equal{\fontdocument}{accanthis}}{
	\usepackage{accanthis}
}{
\ifthenelse{\equal{\fontdocument}{alegreya}}{
	\usepackage{Alegreya}
	
}{
\ifthenelse{\equal{\fontdocument}{alegreyasans}}{
	\usepackage[sfdefault]{AlegreyaSans}
	
}{
\ifthenelse{\equal{\fontdocument}{algolrevived}}{
	\usepackage{algolrevived}
}{
\ifthenelse{\equal{\fontdocument}{almendra}}{
	\usepackage{almendra}
}{
\ifthenelse{\equal{\fontdocument}{antpolt}}{
	\usepackage{antpolt}
}{
\ifthenelse{\equal{\fontdocument}{antpoltlight}}{
	\usepackage[light]{antpolt}
}{
\ifthenelse{\equal{\fontdocument}{anttor}}{
	\usepackage[math]{anttor}
}{
\ifthenelse{\equal{\fontdocument}{anttorcondensed}}{
	\usepackage[condensed,math]{anttor}
}{
\ifthenelse{\equal{\fontdocument}{anttorlight}}{
	\usepackage[light,math]{anttor}
}{
\ifthenelse{\equal{\fontdocument}{anttorlightcondensed}}{
	\usepackage[light,condensed,math]{anttor}
}{
\ifthenelse{\equal{\fontdocument}{arev}}{

	\usepackage{arev}
}{
\ifthenelse{\equal{\fontdocument}{arimo}}{
	\usepackage[sfdefault]{arimo}
	
}{
\ifthenelse{\equal{\fontdocument}{arvo}}{
	\usepackage{Arvo}
}{
\ifthenelse{\equal{\fontdocument}{baskervald}}{
	\usepackage{baskervald}
}{
\ifthenelse{\equal{\fontdocument}{baskervaldx}}{
	\usepackage[lf]{Baskervaldx}
	\usepackage[bigdelims,vvarbb]{newtxmath}
	\usepackage[cal=boondoxo]{mathalfa}
	
}{
\ifthenelse{\equal{\fontdocument}{berasans}}{
	\usepackage[scaled]{berasans}
	
}{
\ifthenelse{\equal{\fontdocument}{beraserif}}{
	\usepackage{bera}
}{
\ifthenelse{\equal{\fontdocument}{biolinum}}{
	\usepackage{libertine}
	
}{
\ifthenelse{\equal{\fontdocument}{bitter}}{
	\usepackage{bitter}
}{
\ifthenelse{\equal{\fontdocument}{boisik}}{
	
	\usepackage{boisik}
}{
\ifthenelse{\equal{\fontdocument}{bookman}}{
	\usepackage{bookman}
}{
\ifthenelse{\equal{\fontdocument}{cabin}}{
	\usepackage[sfdefault]{cabin}
	
}{
\ifthenelse{\equal{\fontdocument}{cabincondensed}}{
	\usepackage[sfdefault,condensed]{cabin}
	
}{
\ifthenelse{\equal{\fontdocument}{caladea}}{
	\usepackage{caladea}
}{
\ifthenelse{\equal{\fontdocument}{cantarell}}{
	\usepackage[default]{cantarell}
}{
\ifthenelse{\equal{\fontdocument}{carlito}}{
	\usepackage[sfdefault]{carlito}
	
}{
\ifthenelse{\equal{\fontdocument}{charterbt}}{
	\usepackage[bitstream-charter]{mathdesign}
}{
\ifthenelse{\equal{\fontdocument}{chivolight}}{
	\usepackage[familydefault,light]{Chivo}
}{
\ifthenelse{\equal{\fontdocument}{chivoregular}}{
	\usepackage[familydefault,regular]{Chivo}
}{
\ifthenelse{\equal{\fontdocument}{clara}}{
	\usepackage{clara}
}{
\ifthenelse{\equal{\fontdocument}{clearsans}}{
	\usepackage[sfdefault]{ClearSans}
	
}{
\ifthenelse{\equal{\fontdocument}{cochineal}}{
	\usepackage{cochineal}
}{
\ifthenelse{\equal{\fontdocument}{coelacanth}}{
	\usepackage[nf]{coelacanth}
	\let\oldnormalfont\normalfont
	\def\normalfont {\oldnormalfont\mdseries}
}{
\ifthenelse{\equal{\fontdocument}{coelacanthextralight}}{
	\usepackage[el,nf]{coelacanth}
	\let\oldnormalfont\normalfont
	\def\normalfont {\oldnormalfont\mdseries}
}{
\ifthenelse{\equal{\fontdocument}{coelacanthlight}}{
	\usepackage[l,nf]{coelacanth}
	\let\oldnormalfont\normalfont
	\def\normalfont {\oldnormalfont\mdseries}
}{
\ifthenelse{\equal{\fontdocument}{comfortaa}}{
	\usepackage[default]{comfortaa}
}{
\ifthenelse{\equal{\fontdocument}{comicneue}}{
	\usepackage[default]{comicneue}
}{
\ifthenelse{\equal{\fontdocument}{comicneueangular}}{
	\usepackage[default,angular]{comicneue}
}{
\ifthenelse{\equal{\fontdocument}{computerconcrete}}{
	\usepackage{concmath}
}{
\ifthenelse{\equal{\fontdocument}{computerconcreteeuler}}{

	\usepackage{beton}
	\usepackage{euler}
}{
\ifthenelse{\equal{\fontdocument}{computermodern}}{
}{
\ifthenelse{\equal{\fontdocument}{computermodernbright}}{
	\usepackage{cmbright}
}{
\ifthenelse{\equal{\fontdocument}{crimson}}{
	\usepackage{crimson}
}{
\ifthenelse{\equal{\fontdocument}{crimsonpro}}{
	\usepackage{CrimsonPro}
	\let\oldnormalfont\normalfont
	\def\normalfont {\oldnormalfont\mdseries}
}{
\ifthenelse{\equal{\fontdocument}{crimsonproextralight}}{
	\usepackage[extralight]{CrimsonPro}
	\let\oldnormalfont\normalfont
	\def\normalfont {\oldnormalfont\mdseries}
}{
\ifthenelse{\equal{\fontdocument}{crimsonprolight}}{
	\usepackage[light]{CrimsonPro}
	\let\oldnormalfont\normalfont
	\def\normalfont {\oldnormalfont\mdseries}
}{
\ifthenelse{\equal{\fontdocument}{crimsonpromedium}}{
	\usepackage[medium]{CrimsonPro}
	\let\oldnormalfont\normalfont
	\def\normalfont {\oldnormalfont\mdseries}
}{
\ifthenelse{\equal{\fontdocument}{cyklop}}{
	\usepackage{cyklop}
}{
\ifthenelse{\equal{\fontdocument}{dejavusans}}{
	\usepackage{DejaVuSans}
	
}{
\ifthenelse{\equal{\fontdocument}{dejavusanscondensed}}{
	\usepackage{DejaVuSansCondensed}
	
}{
\ifthenelse{\equal{\fontdocument}{domitian}}{
	\usepackage{mathpazo}
	\usepackage{domitian}
	
}{
\ifthenelse{\equal{\fontdocument}{droidsans}}{
	\usepackage[defaultsans]{droidsans}
	
}{
\ifthenelse{\equal{\fontdocument}{electrum}}{
	\usepackage[lf]{electrum}
}{	
\ifthenelse{\equal{\fontdocument}{erewhon}}{
	\usepackage[proportional,scaled=1.064]{erewhon}
	\usepackage[erewhon,vvarbb,bigdelims]{newtxmath}
	
}{
\ifthenelse{\equal{\fontdocument}{fbb}}{
	\usepackage{fbb}
}{
\ifthenelse{\equal{\fontdocument}{fetamont}}{
	\usepackage{fetamont}
	
}{
\ifthenelse{\equal{\fontdocument}{firasans}}{
	\usepackage[sfdefault]{FiraSans}
	
}{
\ifthenelse{\equal{\fontdocument}{firasansnewtxsf}}{
	\usepackage[sfdefault]{FiraSans}
	\usepackage{newtxsf}
}{
\ifthenelse{\equal{\fontdocument}{fourier}}{
	\usepackage{fourier}
}{
\ifthenelse{\equal{\fontdocument}{fouriernc}}{
	\usepackage{fouriernc}
}{
\ifthenelse{\equal{\fontdocument}{gfsartemisia}}{
	
	\usepackage{gfsartemisia}
}{
\ifthenelse{\equal{\fontdocument}{gfsartemisiaeuler}}{

	\usepackage{gfsartemisia-euler}
}{
\ifthenelse{\equal{\fontdocument}{heuristica}}{
	\usepackage{heuristica}
	\usepackage[heuristica,vvarbb,bigdelims]{newtxmath}
	
}{
\ifthenelse{\equal{\fontdocument}{iwona}}{
	\usepackage[math]{iwona}
}{
\ifthenelse{\equal{\fontdocument}{iwonacondensed}}{
	\usepackage[condensed,math]{iwona}
}{
\ifthenelse{\equal{\fontdocument}{iwonalight}}{
	\usepackage[light,math]{iwona}
}{
\ifthenelse{\equal{\fontdocument}{iwonalightcondensed}}{
	\usepackage[light,condensed,math]{iwona}
}{
\ifthenelse{\equal{\fontdocument}{kerkis}}{
	\usepackage{kmath,kerkis}
}{
\ifthenelse{\equal{\fontdocument}{kurier}}{
	\usepackage[math]{kurier}
}{
\ifthenelse{\equal{\fontdocument}{kuriercondensed}}{
	\usepackage[condensed,math]{kurier}
}{
\ifthenelse{\equal{\fontdocument}{kurierlight}}{
	\usepackage[light,math]{kurier}
}{
\ifthenelse{\equal{\fontdocument}{kurierlightcondensed}}{
	\usepackage[light,condensed,math]{kurier}
}{
\ifthenelse{\equal{\fontdocument}{lato}}{
	\usepackage[default]{lato}
}{
\ifthenelse{\equal{\fontdocument}{libertinus}}{
	\usepackage{libertinus}
}{
\ifthenelse{\equal{\fontdocument}{librebaskerville}}{
	\usepackage{librebaskerville}
}{
\ifthenelse{\equal{\fontdocument}{librebodoni}}{
	\usepackage{LibreBodoni}
}{
\ifthenelse{\equal{\fontdocument}{librecaslon}}{
	\usepackage{librecaslon}
}{
\ifthenelse{\equal{\fontdocument}{libris}}{
	\usepackage{libris}
	
}{
\ifthenelse{\equal{\fontdocument}{lxfonts}}{
	\usepackage{lxfonts}
}{
\ifthenelse{\equal{\fontdocument}{merriweather}}{
	\usepackage[sfdefault]{merriweather}
}{
\ifthenelse{\equal{\fontdocument}{merriweatherlight}}{
	\usepackage[sfdefault,light]{merriweather}
}{
\ifthenelse{\equal{\fontdocument}{mintspirit}}{
	\usepackage[default]{mintspirit}
}{
\ifthenelse{\equal{\fontdocument}{mlmodern}}{
	\usepackage{mlmodern}
}{
\ifthenelse{\equal{\fontdocument}{montserratalternatesextralight}}{
	\usepackage[defaultfam,extralight,tabular,lining,alternates]{montserrat}
	
}{
\ifthenelse{\equal{\fontdocument}{montserratalternatesregular}}{
	\usepackage[defaultfam,tabular,lining,alternates]{montserrat}
	
}{
\ifthenelse{\equal{\fontdocument}{montserratalternatesthin}}{
	\usepackage[defaultfam,thin,tabular,lining,alternates]{montserrat}
	
}{
\ifthenelse{\equal{\fontdocument}{montserratextralight}}{
	\usepackage[defaultfam,extralight,tabular,lining]{montserrat}
	
}{
\ifthenelse{\equal{\fontdocument}{montserratlight}}{
	\usepackage[defaultfam,light,tabular,lining]{montserrat}
	
}{
\ifthenelse{\equal{\fontdocument}{montserratregular}}{
	\usepackage[defaultfam,tabular,lining]{montserrat}
	
}{
\ifthenelse{\equal{\fontdocument}{montserratthin}}{
	\usepackage[defaultfam,thin,tabular,lining]{montserrat}
	
}{
\ifthenelse{\equal{\fontdocument}{newpx}}{
	\usepackage{newpxtext,newpxmath}
}{
\ifthenelse{\equal{\fontdocument}{nimbussans}}{
	\usepackage{nimbussans}
	
}{
\ifthenelse{\equal{\fontdocument}{noto}}{
	\usepackage[sfdefault]{noto}
	
}{
\ifthenelse{\equal{\fontdocument}{notoserif}}{
	\usepackage{notomath}
}{
\ifthenelse{\equal{\fontdocument}{opensansserif}}{
	\usepackage[default,oldstyle,scale=0.95]{opensans}
}{
\ifthenelse{\equal{\fontdocument}{overlock}}{
	\usepackage[sfdefault]{overlock}
	
}{
\ifthenelse{\equal{\fontdocument}{paratype}}{
	\usepackage{paratype}
	
}{
\ifthenelse{\equal{\fontdocument}{paratypesanscaption}}{
	\usepackage{PTSansCaption}
	
}{
\ifthenelse{\equal{\fontdocument}{paratypesansnarrow}}{
	\usepackage{PTSansNarrow}
	
}{
\ifthenelse{\equal{\fontdocument}{pxfonts}}{
	\usepackage{pxfonts}
}{
\ifthenelse{\equal{\fontdocument}{quattrocento}}{
	\usepackage[sfdefault]{quattrocento}
}{
\ifthenelse{\equal{\fontdocument}{raleway}}{
	\usepackage[default]{raleway}
}{
\ifthenelse{\equal{\fontdocument}{ralewayblack}}{
	\usepackage[black]{raleway}
}{
\ifthenelse{\equal{\fontdocument}{ralewayextralight}}{
	\usepackage[extralight]{raleway}
}{
\ifthenelse{\equal{\fontdocument}{ralewaymedium}}{
	\usepackage[medium]{raleway}
}{
\ifthenelse{\equal{\fontdocument}{ralewaylight}}{
	\usepackage[light]{raleway}
}{
\ifthenelse{\equal{\fontdocument}{ralewaythin}}{
	\usepackage[thin]{raleway}
}{
\ifthenelse{\equal{\fontdocument}{roboto}}{
	\usepackage[sfdefault]{roboto}
}{
\ifthenelse{\equal{\fontdocument}{robotocondensed}}{
	\usepackage[sfdefault,condensed]{roboto}
}{
\ifthenelse{\equal{\fontdocument}{robotolight}}{
	\usepackage[sfdefault,light]{roboto}
}{
\ifthenelse{\equal{\fontdocument}{robotolightcondensed}}{
	\usepackage[sfdefault,light,condensed]{roboto}
}{
\ifthenelse{\equal{\fontdocument}{robotothin}}{
	\usepackage[sfdefault,thin]{roboto}
}{
\ifthenelse{\equal{\fontdocument}{rosario}}{
	\usepackage[familydefault]{Rosario}
}{
\ifthenelse{\equal{\fontdocument}{sourcesanspro}}{
	\usepackage[default]{sourcesanspro}
}{
\ifthenelse{\equal{\fontdocument}{step}}{
	\usepackage[notext]{stix}
	\usepackage{step}
}{
\ifthenelse{\equal{\fontdocument}{stickstoo}}{
	\usepackage{stickstootext}
	\usepackage[stickstoo,vvarbb]{newtxmath}
}{
\ifthenelse{\equal{\fontdocument}{texgyrebonum}}{
	\usepackage{tgbonum}
}{
\ifthenelse{\equal{\fontdocument}{txfonts}}{
	\usepackage{txfonts}
}{
\ifthenelse{\equal{\fontdocument}{uarial}}{
	\usepackage{uarial}
	
}{
\ifthenelse{\equal{\fontdocument}{ugq}}{

}{
\ifthenelse{\equal{\fontdocument}{universalis}}{
	\usepackage[sfdefault]{universalis}
}{
\ifthenelse{\equal{\fontdocument}{universaliscondensed}}{
	\usepackage[condensed,sfdefault]{universalis}
}{
\ifthenelse{\equal{\fontdocument}{venturis}}{
	\usepackage[lf]{venturis}
	
}{
	\throwbadconfig[nostop]{Fuente desconocida}{\fontdocument}{(Fuentes recomendadas) lmodern,carial,arial2,times,mathptmx,helvet,opensans,mathpazo,cambria,libertine,custom}
	\throwbadconfig[noheader-nostop]{Fuente desconocida}{\fontdocument}{(Fuentes adicionales) accanthis,alegreya,alegreyasans,algolrevived,almendra,antpolt,antpoltlight,anttor,anttorcondensed,anttorlight,anttorlightcondensed,arev,arimo,arvo,baskervald,baskervaldx,berasans,beraserif,biolinum,bitter,boisik,bookman,cabin,cabincondensed,cantarell,caladea,carlito,charterbt,chivolight,chivoregular,clara,clearsans,cochineal,coelacanth,coelacanthextralight,coelacanthlight,comfortaa,comicneue,comicneueangular,computerconcrete,computerconcreteeuler,computermodern,computermodernbright,crimson,crimsonpro,crimsonproextralight,crimsonprolight,crimsonpromedium,cyklop}
	\throwbadconfig[noheader-nostop]{Fuente desconocida}{\fontdocument}{dejavusans,dejavusanscondensed,domitian,droidsans,electrum,erewhon,fbb,fetamont,firasans,firasansnewtxsf,fourier,fouriernc,gfsartemisia,gfsartemisiaeuler,heuristica,iwona,iwonacondensed,iwonalight,iwonalightcondensed,kerkis,kurier,kuriercondensed,kurierlight,kurierlightcondensed,lato,libertinus,librebaskerville,librebodoni,librecaslon,libris,lxfonts}
	\throwbadconfig[noheader]{Fuente desconocida}{\fontdocument}{merriweather,merriweatherlight,mintspirit,mlmodern,montserratalternatesextralight,montserratalternatesregular,montserratalternatesthin,montserratextralight,montserratlight,montserratregular,montserratthin,newpx,nimbussans,noto,notoserif,opensansserif,overlock,paratype,paratypesanscaption,paratypesansnarrow,pxfonts,quattrocento,raleway,ralewayblack,ralewayextralight,ralewaymedium,ralewaylight,ralewaythin,roboto,robotolight,robotolightcondensed,robotothin,rosario,sourcesanspro,step,stickstoo,uarial,texgyrebonum,txfonts,ugq,universalis,universaliscondensed,venturis}
	}}}}}}}}}}}}}}}}}}}}}}}}}}}}}}}}}}}}}}}}}}}}}}}}}}}}}}}}}}}}}}}}}}}}}}}}}}}}}}}}}}}}}}}}}}}}}}}}}}}}}}}}}}}}}}}}}}}}}}}}}}}}}}}}}}}}}}
}

\ifthenelse{\equal{\fonttypewriter}{custom}}{
}{
\ifthenelse{\equal{\fonttypewriter}{tmodern}}{
	
}{
\ifthenelse{\equal{\fonttypewriter}{anonymouspro}}{
	\usepackage[ttdefault=true]{AnonymousPro}
}{
\ifthenelse{\equal{\fonttypewriter}{ascii}}{
	\usepackage{ascii}
	
}{
\ifthenelse{\equal{\fonttypewriter}{beramono}}{
	\usepackage[scaled]{beramono}
}{
\ifthenelse{\equal{\fonttypewriter}{cascadiacode}}{
	\usepackage{cascadia-code}
}{
\ifthenelse{\equal{\fonttypewriter}{cmpica}}{
	\usepackage{addfont}
	\addfont{OT1}{cmpica}{\pica}
	\addfont{OT1}{cmpicab}{\picab}
	\addfont{OT1}{cmpicati}{\picati}
	
}{
\ifthenelse{\equal{\fonttypewriter}{cmodern}}{
}{
\ifthenelse{\equal{\fonttypewriter}{courier}}{
	\usepackage{courier}
}{
\ifthenelse{\equal{\fonttypewriter}{courier10}}{
	\usepackage{courierten}
}{
\ifthenelse{\equal{\fonttypewriter}{cmvtt}}{
	
}{
\ifthenelse{\equal{\fonttypewriter}{dejavusansmono}}{
	\usepackage[scaled]{DejaVuSansMono}
}{
\ifthenelse{\equal{\fonttypewriter}{droidsansmono}}{
	\usepackage[defaultmono]{droidsansmono}
}{
\ifthenelse{\equal{\fonttypewriter}{firamono}}{
	\usepackage[scale=0.85]{FiraMono}
}{
\ifthenelse{\equal{\fonttypewriter}{gomono}}{
	\usepackage[scale=0.85]{GoMono}
}{
\ifthenelse{\equal{\fonttypewriter}{inconsolata}}{
	\usepackage{inconsolata}
}{
\ifthenelse{\equal{\fonttypewriter}{nimbusmono}}{
	\usepackage{nimbusmono}
}{
\ifthenelse{\equal{\fonttypewriter}{newtxtt}}{
	\usepackage[zerostyle=d]{newtxtt}
}{
\ifthenelse{\equal{\fonttypewriter}{nimbusmono}}{
	\usepackage{nimbusmono}
}{
\ifthenelse{\equal{\fonttypewriter}{nimbusmononarrow}}{
	\usepackage{nimbusmononarrow}
}{
\ifthenelse{\equal{\fonttypewriter}{lcmtt}}{
	
}{
\ifthenelse{\equal{\fonttypewriter}{sourcecodepro}}{
	\usepackage[ttdefault=true,scale=0.85]{sourcecodepro}
}{
\ifthenelse{\equal{\fonttypewriter}{texgyrecursor}}{
	\usepackage{tgcursor}
}{
\ifthenelse{\equal{\fonttypewriter}{txtt}}{
	
}{
	\throwbadconfig{Fuente desconocida}{\fonttypewriter}{custom,anonymouspro,ascii,beramono,cascadiacode,cmpica,cmodern,courier,courier10,cvmtt,dejavusansmono,droidsansmono,firamono,gomono,inconsolata,kpmonospaced,lcmtt,newtxtt,nimbusmono,nimbusmononarrow,texgyrecursor,tmodern,txtt}
	}}}}}}}}}}}}}}}}}}}}}}}
}

\usepackage[T1]{fontenc} 
\ifthenelse{\equal{\showlayoutlines}{true}}{
	\usepackage{showframe}}{
}
\def\marginequationbottom {0}
\def\marginequationtop {0}
\def\margingatherbottom {0}
\def\margingathertop {0}
\ifthenelse{\equal{\showlinenumbers}{true}}{ 
	\usepackage[switch,columnwise,running]{lineno}
	\newcommand*\linenomathpatch[1]{
		\cspreto{#1}{\linenomath}%
		\cspreto{#1*}{\linenomath}%
		\csappto{end#1}{\endlinenomath}%
		\csappto{end#1*}{\endlinenomath}}
	\linenomathpatch{equation}
	\linenomathpatch{gather}
	\linenomathpatch{multline}
	\linenomathpatch{align}
	\linenomathpatch{alignat}
	\linenomathpatch{flalign}}{
}
\usepackage{commonunicode} 
\usepackage{csquotes} 
\ifthenelse{\equal{\compilertype}{pdf2latex}}{
	\inputencoding{utf8}}{
}

\global\def\GLOBALemptyvar {template:empty:var}   

\global\def\GLOBALcaptiondefn {\GLOBALemptyvar}   
\global\def\GLOBALchapternumenabled {false}       
\global\def\GLOBALenvappendix {false}             
\global\def\GLOBALenvimageadded {false}           
\global\def\GLOBALenvimagecf {false}              
\global\def\GLOBALenvimageinitialized {false}     
\global\def\GLOBALenvmulticol {false}             
\global\def\GLOBALsectionanumenabled {false}      
\global\def\GLOBALsubsectionanumenabled {false}   
\global\def\GLOBALsubsubsectionanumenabled{false} 
\global\def\GLOBALtablerowcolorindex {2}          
\global\def\GLOBALtablerowcolorswitch {false}     
\global\def\GLOBALtwoside {false}                 

\global\def\GLOBALformatnumchapter {\formatnumchapter}
\global\def\GLOBALformatnumsection {\formatnumsection}
\global\def\GLOBALformatnumssection {\formatnumssection}
\global\def\GLOBALformatnumsssection {\formatnumsssection}
\global\def\GLOBALformatnumssssection {\formatnumssssection}

\makeatletter
\if@twoside
\global\def\GLOBALtwoside {true}
\else
\fi 
\makeatother

\def\LOCALpercentchar#1{}
\edef\LOCALpercentchar{\expandafter\LOCALpercentchar\string\%}

\newcounter{templateEquations}      
\newcounter{templateFigures}        
\newcounter{templateIndexEquations} 
\newcounter{templateListings}       
\newcounter{templatePageCounter}    
\newcounter{templateTables}         

\newcounter{templateBookmarksLevelPrev}
\addtocounter{templateBookmarksLevelPrev}{-1}

\stepcounter{templatePageCounter}
\AtBeginShipout{\stepcounter{templatePageCounter}}

\newcounter{chapter}

\newlength{\coregluevarcm}
\newlength{\corefontwidth}
\newcommand{\throwerror}[2]{%
	\errmessage{LaTeX Error: \noexpand#1 #2 (linea \the\inputlineno)}%
	\stop
}

\newcommand{\throwwarning}[1]{%
	\errmessage{LaTeX Warning: #1 (linea \the\inputlineno)}%
}

\newcommand{\throwbadconfigondoc}[3]{%
	\errmessage{#1 \noexpand #2=#2. Valores esperados: #3}%
	\stop%
}

\makeatletter%
\newcommand{\checkmodulenotloaded}[1]{%
	\@ifpackageloaded{#1}{%
		\throwwarning{Template Error: No se pueden cargar paquetes (#1) antes de importar template.tex}%
		\stop%
	}{}%
}
\makeatother%

\newcommand{\checkvardefined}[1]{%
	\ifthenelse{\isundefined{#1}}{%
		\errmessage{LaTeX Warning: Variable \noexpand#1 no definida}%
		\stop}{%
	}%
}

\newcommand{\coretemplatemessage}[1]{%
	\message{Template: #1}%
}

\newcommand{\checkextravarexist}[2]{%
	\ifthenelse{\isundefined{#1}}{%
		\errmessage{LaTeX Warning: Variable \noexpand#1 no definida}%
		\ifx\hfuzz#2\hfuzz%
			\errmessage{LaTeX Warning: Defina la variable en el bloque de INFORMACION DEL DOCUMENTO al comienzo del archivo principal del template}%
		\else%
			\errmessage{LaTeX Warning: #2}%
		\fi}{%
	}%
}

\newcommand{\emptyvarerr}[3]{%
	\ifx\hfuzz#2\hfuzz%
		\errmessage{LaTeX Warning: \noexpand#1 #3 (linea \the\inputlineno)}%
	\fi
}

\newcommand{\setcaptionmargincm}[1]{
	\captionsetup{margin=#1cm}
}

\newcommand{\setpagemargincm}[4]{%
	\ifthenelse{\equal{\compilertype}{lualatex}}{%
	}{%
		\newgeometry{left=#1cm, top=#2cm, right=#3cm, bottom=#4cm, footnotesep=\footnotetopmargin cm}
	}
}

\newcommand{\resetindexcaption}{%
	\global\def\GLOBALcaptiondefn {\GLOBALemptyvar}%
	\hbadness=10000%
}

\newcommand{\changemargin}[2]{%
	\emptyvarerr{\changemargin}{#1}{Margen izquierdo no definido}%
	\emptyvarerr{\changemargin}{#2}{Margen derecho no definido}%
	\list{}{\rightmargin#2\leftmargin#1}\item[]%
}

\newcommand{\checkonlyonenvimage}{%
	\ifthenelse{\equal{\GLOBALenvimageinitialized}{true}}{}{%
		\throwwarning{Funciones \noexpand\addimage o \noexpand\addimageboxed no pueden usarse fuera del entorno \noexpand\images}\stop%
	}%
}

\newcommand{\checkoutsideenvimage}{%
	\ifthenelse{\equal{\GLOBALenvimageinitialized}{true}}{%
		\throwwarning{Esta funcion solo puede usarse fuera del entorno \noexpand\images}%
		\stop}{%
	}%
}

\newcommand{\checkinsidemulticol}{%
	\ifthenelse{\equal{\GLOBALenvmulticol}{false}}{%
		\throwwarning{Esta funcion solo puede usarse dentro de multicols}%
		\stop}{%
	}%
}

\newcommand{\checkoutsideappendix}{%
	\ifthenelse{\equal{\GLOBALenvappendix}{true}}{%
		\throwwarning{Esta funcion solo puede usarse fuera de anexo}%
		\stop}{%
	}%
}

\newcommand{\corecheckbooleanvar}[1]{%
	\emptyvarerr{\corecheckbooleanvar}{#1}{Variable no definida}%
	\ifthenelse{\equal{#1}{true}}{}{%
	\ifthenelse{\equal{#1}{false}}{}{%
		\throwwarning{Variable debe ser true o false}\stop%
	}}%
}

\newcommand{\verticallycentertext}[1]{%
	\emptyvarerr{\verticallycentertext}{#1}{Texto no definido}%
	\topskip0pt%
	\vspace*{\fill}%
	#1%
	\vspace*{\fill}%
}

\newcommand{\corevspacevarcm}[1]{%
	\ifthenelse{\equal{#1}{0}}{}{%
	\ifthenelse{\equal{#1}{0.0}}{}{%
		\vspace{\dimexpr#1 cm plus #1\coregluevarcm minus #1\coregluevarcm}%
	}}%
}

\makeatletter
\newcommand\addpathimage[1]{%
	\gappto\Ginput@path{{#1}}%
}
\makeatother

\newcommand{\corecheckfontsize}[1]{%
	\ifthenelse{\equal{#1}{normalsize}}{}{%
	\ifthenelse{\equal{#1}{small}}{}{%
	\ifthenelse{\equal{#1}{large}}{}{%
	\ifthenelse{\equal{#1}{Large}}{}{%
	\ifthenelse{\equal{#1}{LARGE}}{}{%
	\ifthenelse{\equal{#1}{huge}}{}{%
	\ifthenelse{\equal{#1}{Huge}}{}{%
	\ifthenelse{\equal{#1}{HUGE}}{}{%
	\ifthenelse{\equal{#1}{footnotesize}}{}{%
	\ifthenelse{\equal{#1}{scriptsize}}{}{%
	\ifthenelse{\equal{#1}{tiny}}{}{%
		\errmessage{LaTeX Warning: Tamano de fuente incorrecto (\noexpand #1= #1). Valores esperados: tiny,scriptsize,footnotesize,small,normalisize,large,Large,LARGE,huge,Huge,HUGE}%
		\stop%
		}}}}}}}}}}%
	}%
}

\newcommand{\aasin}[1][]{%
	\ifx\hfuzz#1\hfuzz%
		\ensuremath{\sin^{-1}#1}
	\else%
		\ensuremath{{\sin}^{-1}}
	\fi%
}

\newcommand{\aacos}[1][]{%
	\ifx\hfuzz#1\hfuzz%
		\ensuremath{\cos^{-1}#1}
	\else%
		\ensuremath{\cos^{-1}}
	\fi%
}

\newcommand{\aatan}[1][]{%
	\ifx\hfuzz#1\hfuzz%
		\ensuremath{\tan^{-1}#1}
	\else%
		\ensuremath{\tan^{-1}}
	\fi%
}

\newcommand{\aacsc}[1][]{%
	\ifx\hfuzz#1\hfuzz%
		\ensuremath{\csc^{-1}#1}
	\else%
		\ensuremath{\csc^{-1}}
	\fi%
}

\newcommand{\aasec}[1][]{%
	\ifx\hfuzz#1\hfuzz%
		\ensuremath{\sec^{-1}#1}
	\else%
		\ensuremath{\sec^{-1}}
	\fi%
}

\newcommand{\aacot}[1][]{%
	\ifx\hfuzz#1\hfuzz%
		\ensuremath{\cot^{-1}#1}
	\else%
		\ensuremath{\cot^{-1}}
	\fi%
}

\makeatletter
\def\longtilde#1{%
	\mathop{\vbox{\m@th\ialign{##\crcr\noalign{\kern3\p@}%
	\sortoftildefill\crcr\noalign{\kern3\p@\nointerlineskip}%
	$\hfil\displaystyle{#1}\hfil$\crcr}}}\limits%
}
\def\sortoftildefill {%
	$\m@th \setbox\z@\hbox{$\braceld$}%
	\braceld\leaders\vrule \@height\ht\z@ \@depth\z@\hfill\braceru$%
}
\makeatother

\ifthenelse{\isundefined{\C}}{\newcommand{\C}{C}}{}
\renewcommand{\C}{\ensuremath{\mathbb{C}}}

\ifthenelse{\isundefined{\G}}{\newcommand{\G}{G}}{}
\renewcommand{\G}{\ensuremath{\mathcal{G}}}

\ifthenelse{\isundefined{\H}}{\newcommand{\H}{H}}{}
\renewcommand{\H}{\ensuremath{\mathcal{H}}}

\ifthenelse{\isundefined{\U}}{\newcommand{\U}{U}}{}
\renewcommand{\U}{\ensuremath{\mathcal{U}}}

\ifthenelse{\equal{\fontdocument}{step}}{}{

}


\newcommand{\coreafterequationfn}{%
	\hbadness=10000%
}

\newcommand{\equationresize}[2]{%
	\emptyvarerr{\equationresize}{#1}{Dimension no definida}%
	\emptyvarerr{\equationresize}{#2}{Ecuacion a redimensionar no definida}%
	\resizebox{#1\linewidth}{!}{$#2$}%
}

\newcommand{\coreinsertequationcaption}[1]{%
	\begin{changemargin}{\captionlrmargin cm}{\captionlrmargin cm}%
		\ifthenelse{\equal{\equationcaptioncenter}{true}}{%
			\centering%
		}{%
			\justifying%
		}%
		\textcolor{\captiontextcolor}{%
			\linespread{0.5}\selectfont{%
				\begin{\captionfontsize}#1\end{\captionfontsize}%
			}%
		}%
	\end{changemargin}
}

\newcommand{\insertequation}[2][]{%
	\emptyvarerr{\insertequation}{#2}{Ecuacion no definida}%
	\ifthenelse{\equal{\numberedequation}{true}}{%
		\corevspacevarcm{\marginequationtop}%
		\begin{samepage}%
		\begin{equation}%
			\text{#1} #2
		\end{equation}
		\corevspacevarcm{\marginequationbottom}%
		\end{samepage}
		\coreafterequationfn%
	}{%
		\ifx\hfuzz#1\hfuzz%
		\else%
			\throwwarning{Label invalido en ecuacion sin numero}%
		\fi%
		\insertequationanum{#2}%
	}%
}

\newcommand{\insertequationanum}[1]{%
	\emptyvarerr{\insertequationanum}{#1}{Ecuacion no definida}%
	\corevspacevarcm{\marginequationtop}%
	\begin{samepage}%
	\begin{equation*}%
		\ensuremath{#1}
	\end{equation*}
	\corevspacevarcm{\marginequationbottom}%
	\end{samepage}
	\coreafterequationfn%
}

\newcommand{\insertindexequation}[3][]{%
	\emptyvarerr{\insertindexequation}{#2}{Ecuacion no definida}%
	\emptyvarerr{\insertindexequation}{#3}{Leyenda no definida}%
	\begin{equationindex}[#1]{#3}%
		#2
	\end{equationindex}
}

\newcommand{\insertequationleft}[2][]{%
	\emptyvarerr{\insertequationleft}{#2}{Ecuacion no definida}%
	\ifthenelse{\equal{\numberedequation}{true}}{%
		\vspace{\dimexpr\marginequationtop cm - \baselineskip}%
		\begin{samepage}%
		\begin{equation}
			\hfilneg \text{#1} #2 \hspace{10000pt minus 1fil}
		\end{equation}
		\vspace{\dimexpr-0.2\baselineskip + \marginequationbottom cm}%
		\end{samepage}
		\coreafterequationfn%
	}{%
		\ifx\hfuzz#1\hfuzz%
		\else%
			\throwwarning{Label invalido en ecuacion sin numero}%
		\fi%
		\insertequationleftanum{#2}%
	}%
}

\newcommand{\insertequationleftanum}[1]{%
	\emptyvarerr{\insertequationleftanum}{#1}{Ecuacion no definida}%
	\vspace{\dimexpr\marginequationtop cm - \baselineskip}%
	\begin{samepage}%
	\begin{equation*}
		\hfilneg \ensuremath{#1} \hspace{10000pt minus 1fil}
	\end{equation*}
	\vspace{\dimexpr-0.2\baselineskip + \marginequationbottom cm}%
	\end{samepage}
	\coreafterequationfn%
}

\newcommand{\insertequationright}[2][]{%
	\emptyvarerr{\insertequationright}{#2}{Ecuacion no definida}%
	\ifthenelse{\equal{\numberedequation}{true}}{%
		\vspace{\dimexpr\marginequationtop cm - \baselineskip}%
		\begin{samepage}%
		\begin{equation}
			\hspace{10000pt minus 1fil} \text{#1} #2 \hfilneg
		\end{equation}
		\vspace{\dimexpr-0.2\baselineskip + \marginequationbottom cm}%
		\end{samepage}
		\coreafterequationfn%
	}{%
		\ifx\hfuzz#1\hfuzz%
		\else%
			\throwwarning{Label invalido en ecuacion sin numero}%
		\fi%
		\insertequationrightanum{#2}%
	}%
}

\newcommand{\insertequationrightanum}[1]{%
	\emptyvarerr{\insertequationrightanum}{#1}{Ecuacion no definida}%
	\vspace{\dimexpr\marginequationtop cm - \baselineskip}%
	\begin{samepage}%
	\begin{equation*}
		\hspace{10000pt minus 1fil} \ensuremath{#1} \hfilneg
	\end{equation*}
	\vspace{\dimexpr-0.2\baselineskip + \marginequationbottom cm}%
	\end{samepage}
	\coreafterequationfn%
}

\newcommand{\insertequationcaptioned}[3][]{%
	\emptyvarerr{\insertequationcaptioned}{#2}{Ecuacion no definida}%
	\ifx\hfuzz#3\hfuzz%
		\insertequation[#1]{#2}%
	\else%
		\ifthenelse{\equal{\numberedequation}{true}}{%
			\corevspacevarcm{\marginequationtop}%
			\begin{samepage}%
			\begin{equation}
				\text{#1} #2
			\end{equation}
			\corevspacevarcm{\margineqncaptiontop}%
			\coreinsertequationcaption{#3}%
			\corevspacevarcm{\margineqncaptionbottom}%
			\end{samepage}
			\coreafterequationfn%
		}{%
			\ifx\hfuzz#1\hfuzz%
			\else%
				\throwwarning{Label invalido en ecuacion sin numero}%
			\fi%
			\insertequationcaptionedanum{#2}{#3}%
		}%
	\fi%
}

\newcommand{\insertequationcaptionedanum}[2]{%
	\emptyvarerr{\insertequationcaptionedanum}{#1}{Ecuacion no definida}%
	\ifx\hfuzz#2\hfuzz%
		\insertequationanum{#1}%
	\else%
		\corevspacevarcm{\marginequationtop}%
		\begin{samepage}%
		\begin{equation*}
			\ensuremath{#1}%
		\end{equation*}
		\corevspacevarcm{\margineqncaptiontop}%
		\coreinsertequationcaption{#2}%
		\corevspacevarcm{\margineqncaptionbottom}%
		\end{samepage}
		\coreafterequationfn%
	\fi%
}

\newcommand{\insertgather}[1]{%
	\emptyvarerr{\insertgather}{#1}{Ecuacion no definida}%
	\ifthenelse{\equal{\numberedequation}{true}}{%
		\corevspacevarcm{\margingathertop}%
		\begin{samepage}%
		\begin{gather}%
			\ensuremath{#1}
		\end{gather}
		\corevspacevarcm{\margingatherbottom}%
		\end{samepage}
		\coreafterequationfn%
	}{%
		\insertgatheranum{#1}%
	}%
}

\newcommand{\insertgatheranum}[1]{%
	\emptyvarerr{\insertgatheranum}{#1}{Ecuacion no definida}%
	\corevspacevarcm{\margingathertop}%
	\begin{samepage}%
	\begin{gather*}%
		\ensuremath{#1}
	\end{gather*}
	\corevspacevarcm{\margingatherbottom}%
	\end{samepage}
	\coreafterequationfn%
}

\newcommand{\insertgathercaptioned}[2]{%
	\emptyvarerr{\insertgathercaptioned}{#1}{Ecuacion no definida}%
	\ifx\hfuzz#2\hfuzz%
		\insertgather{#1}%
	\else%
		\ifthenelse{\equal{\numberedequation}{true}}{%
			\corevspacevarcm{\margingathertop}%
			\begin{samepage}%
			\begin{gather}%
				\ensuremath{#1}
			\end{gather}
			\corevspacevarcm{\margingathercapttop}%
			\coreinsertequationcaption{#2}%
			\corevspacevarcm{\margingathercaptbottom}%
			\end{samepage}
			\coreafterequationfn%
		}{%
			\insertgathercaptionedanum{#1}{#2}%
		}%
	\fi%
}

\newcommand{\insertgathercaptionedanum}[2]{%
	\emptyvarerr{\insertgathercaptionedanum}{#1}{Ecuacion no definida}%
	\ifx\hfuzz#2\hfuzz%
		\insertgatheranum{#1}%
	\else%
		\corevspacevarcm{\margingathertop}%
		\begin{samepage}%
		\begin{gather*}%
			\ensuremath{#1}
		\end{gather*}
		\corevspacevarcm{\margingathercapttop}%
		\coreinsertequationcaption{#2}%
		\corevspacevarcm{\margingathercaptbottom}%
		\end{samepage}
		\coreafterequationfn%
	\fi%
}

\newcommand{\insertgathered}[2][]{%
	\emptyvarerr{\insertgathered}{#2}{Ecuacion no definida}%
	\ifthenelse{\equal{\numberedequation}{true}}{%
		\corevspacevarcm{\marginequationtop}%
		\begin{samepage}%
		\begin{equation}
			\begin{gathered}
				\text{#1} \ensuremath{#2}
			\end{gathered}
		\end{equation}
		\corevspacevarcm{\margingatheredbottom}%
		\end{samepage}
	}{%
		\ifx\hfuzz#1\hfuzz%
		\else%
			\throwwarning{Label invalido en ecuacion (gathered) sin numero}%
		\fi%
		\corevspacevarcm{\margingatheredtop}%
		\begin{samepage}%
		\begin{gather*}%
			\ensuremath{#2}
		\end{gather*}
		\corevspacevarcm{\margingatheredbottom}%
		\end{samepage}
	}%
	\coreafterequationfn%
}

\newcommand{\insertgatheredanum}[1]{%
	\emptyvarerr{\insertgatheredanum}{#1}{Ecuacion no definida}%
	\corevspacevarcm{\margingatheredtop}%
	\begin{samepage}%
	\begin{gather*}
		\ensuremath{#1}
	\end{gather*}
	\vspace{\dimexpr-0.15cm + \margingatheredbottom cm}%
	\end{samepage}
	\coreafterequationfn%
}

\newcommand{\insertgatheredcaptioned}[3][]{%
	\emptyvarerr{\insertgatheredcaptioned}{#2}{Ecuacion no definida}%
	\ifx\hfuzz#3\hfuzz%
		\insertgathered[#1]{#2}%
	\else%
		\ifthenelse{\equal{\numberedequation}{true}}{%
			\corevspacevarcm{\marginequationtop}%
			\begin{samepage}%
			\begin{equation}
				\begin{gathered}
					\text{#1} \ensuremath{#2}
				\end{gathered}
			\end{equation}
			\corevspacevarcm{\margingatheredcapttop}%
			\coreinsertequationcaption{#3}%
			\corevspacevarcm{\margingatheredcaptbottom}%
			\end{samepage}
			\coreafterequationfn%
		}{%
			\ifx\hfuzz#1\hfuzz%
			\else%
				\throwwarning{Label invalido en ecuacion (gathered) sin numero}
			\fi%
			\insertgatheredcaptionedanum{#2}{#3}%
		}%
	\fi%
}

\newcommand{\insertgatheredcaptionedanum}[2]{%
	\emptyvarerr{\insertgatheredcaptionedanum}{#1}{Ecuacion no definida}%
	\ifx\hfuzz#2\hfuzz%
		\insertgatheredanum{#1}%
	\else%
		\corevspacevarcm{\margingatheredtop}%
		\begin{samepage}%
		\begin{gather*}
			\ensuremath{#1}
		\end{gather*}
		\vspace{\dimexpr-0.2cm + \margingatheredcapttop cm}%
		\coreinsertequationcaption{#2}%
		\vspace{\dimexpr-0.05cm + \margingatheredcaptbottom cm}%
		\end{samepage}
		\coreafterequationfn%
	\fi%
}

\newcommand{\insertalign}[1]{%
	\emptyvarerr{\insertalign}{#1}{Ecuacion no definida}%
	\ifthenelse{\equal{\numberedequation}{true}}{%
		\corevspacevarcm{\marginaligntop}%
		\begin{samepage}%
		\begin{align}
			\ensuremath{#1}
		\end{align}
		\corevspacevarcm{\marginalignbottom}%
		\end{samepage}
		\coreafterequationfn%
	}{%
		\insertalignanum{#1}%
	}%
}

\newcommand{\insertalignanum}[1]{%
	\emptyvarerr{\insertalignanum}{#1}{Ecuacion no definida}%
	\corevspacevarcm{\marginaligntop}%
	\begin{samepage}%
	\begin{align*}
		\ensuremath{#1}
	\end{align*}
	\corevspacevarcm{\marginalignbottom}%
	\end{samepage}
	\coreafterequationfn%
}

\newcommand{\insertaligncaptioned}[2]{%
	\emptyvarerr{\insertaligncaptioned}{#1}{Ecuacion no definida}%
	\ifx\hfuzz#2\hfuzz%
		\insertalign{#1}%
	\else%
		\ifthenelse{\equal{\numberedequation}{true}}{%
			\corevspacevarcm{\marginaligntop}%
			\begin{samepage}%
			\begin{align}
				\ensuremath{#1}
			\end{align}
			\corevspacevarcm{\marginaligncapttop}%
			\coreinsertequationcaption{#2}%
			\corevspacevarcm{\marginaligncaptbottom}%
			\end{samepage}
			\coreafterequationfn%
		}{%
			\insertaligncaptionedanum{#1}{#2}%
		}%
	\fi%
}

\newcommand{\insertaligncaptionedanum}[2]{%
	\emptyvarerr{\insertaligncaptionedanum}{#1}{Ecuacion no definida}%
	\ifx\hfuzz#2\hfuzz%
		\insertalignanum{#1}%
	\else%
		\corevspacevarcm{\marginaligntop}%
		\begin{samepage}%
		\begin{align*}
			\ensuremath{#1}
		\end{align*}
		\corevspacevarcm{\marginaligncapttop}%
		\coreinsertequationcaption{#2}%
		\corevspacevarcm{\marginaligncaptbottom}%
		\end{samepage}
		\coreafterequationfn%
	\fi%
}

\newcommand{\insertaligned}[2][]{%
	\emptyvarerr{\insertaligned}{#2}{Ecuacion no definida}%
	\ifthenelse{\equal{\numberedequation}{true}}{%
		\corevspacevarcm{\marginequationtop}%
		\begin{samepage}%
		\begin{equation}
			\begin{aligned}
				\text{#1} \ensuremath{#2}
			\end{aligned}
		\end{equation}
		\corevspacevarcm{\marginalignedbottom}%
		\end{samepage}
		\coreafterequationfn%
	}{%
		\ifx\hfuzz#1\hfuzz%
		\else%
			\throwwarning{Label invalido en ecuacion (aligned) sin numero}%
		\fi%
		\insertalignedanum{#2}%
	}%
}

\newcommand{\insertalignedanum}[1]{%
	\emptyvarerr{\insertalignedanum}{#1}{Ecuacion no definida}%
	\corevspacevarcm{\marginalignedtop}%
	\begin{samepage}%
	\begin{align*}
		\ensuremath{#1}
	\end{align*}
	\vspace{\dimexpr-0.2cm + \marginalignedbottom cm}%
	\end{samepage}
	\coreafterequationfn%
}

\newcommand{\insertalignedcaptioned}[3][]{%
	\emptyvarerr{\insertalignedcaptioned}{#2}{Ecuacion no definida}%
	\ifx\hfuzz#3\hfuzz%
		\insertaligned[#1]{#2}%
	\else%
		\ifthenelse{\equal{\numberedequation}{true}}{%
			\corevspacevarcm{\marginequationtop}%
			\begin{samepage}%
			\begin{equation}
				\begin{aligned}
					\text{#1} \ensuremath{#2}
				\end{aligned}
			\end{equation}
			\corevspacevarcm{\marginalignedcapttop}%
			\coreinsertequationcaption{#3}%
			\corevspacevarcm{\marginalignedcaptbottom}%
			\end{samepage}
			\coreafterequationfn%
		}{%
			\ifx\hfuzz#1\hfuzz%
			\else%
				\throwwarning{Label invalido en ecuacion (aligned) sin numero}%
			\fi%
			\insertalignedcaptionedanum{#2}{#3}%
		}%
	\fi%
}

\newcommand{\insertalignedcaptionedanum}[2]{%
	\emptyvarerr{\insertalignedcaptionedanum}{#1}{Ecuacion no definida}%
	\ifx\hfuzz#2\hfuzz%
		\insertalignedanum{#1}%
	\else%
		\corevspacevarcm{\marginequationtop}%
		\begin{samepage}%
		\begin{equation}
			\begin{aligned}
				\ensuremath{#1}
			\end{aligned}
		\end{equation}
		\corevspacevarcm{\marginalignedcapttop}%
		\coreinsertequationcaption{#2}%
		\corevspacevarcm{\marginalignedcaptbottom}%
		\end{samepage}
		\coreafterequationfn%
	\fi%
}

\global\def\GLOBALimagelink {\GLOBALemptyvar} 
\global\def\GLOBALimagenextmarginv {0 cm} 

\newlength{\coreimageshspace}
\newcommand{\addimage}[4][]{%
	\addimageboxed[#1]{#2}{#3}{0}{#4}%
}

\newcommand{\addimageboxed}[5][]{%
	\checkonlyonenvimage%
	\begingroup%
	\setlength{\fboxsep}{0 pt}%
	\setlength{\fboxrule}{#4 pt}%
	\ifthenelse{\equal{\GLOBALenvimageadded}{true}}{%
		\hspace{\dimexpr \marginimagemultright cm -\coreimageshspace}%
	}{} 
	\ifthenelse{\equal{#5}{\GLOBALemptyvar}}{ 
		\ifthenelse{\equal{\GLOBALimagelink}{\GLOBALemptyvar}}{
			\raisebox{\GLOBALimagenextmarginv}{%
				\fbox{\includegraphics[#3]{#2}}%
			}%
		}{
			\raisebox{\GLOBALimagenextmarginv}{%
				\fbox{\href{\GLOBALimagelink}{\includegraphics[#3]{#2}}}%
			}%
		}%
	}{ 
		\ifthenelse{\equal{\GLOBALimagelink}{\GLOBALemptyvar}}{
			\subfloat[#5#1]{%
				\raisebox{\GLOBALimagenextmarginv}{%
					\fbox{\includegraphics[#3]{#2}}%
				}%
			}%
		}{
			\subfloat[#5#1]{%
				\raisebox{\GLOBALimagenextmarginv}{%
					\fbox{\href{\GLOBALimagelink}{\includegraphics[#3]{#2}}}%
				}%
			}%
		}%
	}%
	\endgroup%
	\global\def\GLOBALenvimageadded {true}%
	\global\def\GLOBALimagenextmarginv {0 cm}%
}

\newcommand{\addimageanum}[2]{%
	\addimageboxed{#1}{#2}{0}{\GLOBALemptyvar}%
}

\newcommand{\addimageanimatedboxed}[7][]{%
	\checkonlyonenvimage%
	\begingroup%
	\setlength{\fboxsep}{0 pt}%
	\setlength{\fboxrule}{#6 pt}%
	\ifthenelse{\equal{\GLOBALenvimageadded}{true}}{%
		\hspace{\dimexpr \marginimagemultright cm - \coreimageshspace}%
	}{}%
	\ifthenelse{\equal{#7}{\GLOBALemptyvar}}{
		\ifthenelse{\equal{\animatedimageloop}{true}}{
			\ifthenelse{\equal{\animatedimageautoplay}{true}}{
				\raisebox{\GLOBALimagenextmarginv}{%
					\fbox{\animategraphics[loop,autoplay,#3]{#4}{#2-}{0}{#5}}%
				}%
			}{
				\raisebox{\GLOBALimagenextmarginv}{%
					\fbox{\animategraphics[loop,#3]{#4}{#2-}{0}{#5}}%
				}%
			}%
		}{
			\ifthenelse{\equal{\animatedimageautoplay}{true}}{
				\raisebox{\GLOBALimagenextmarginv}{%
					\fbox{\animategraphics[autoplay,#3]{#4}{#2-}{0}{#5}}%
				}%
			}{
				\raisebox{\GLOBALimagenextmarginv}{%
					\fbox{\animategraphics[#3]{#4}{#2-}{0}{#5}}%
				}%
			}%
		}%
	}{
		\subfloat[#7#1]{%
			\ifthenelse{\equal{\animatedimageloop}{true}}{
				\ifthenelse{\equal{\animatedimageautoplay}{true}}{
					\raisebox{\GLOBALimagenextmarginv}{%
						\fbox{\animategraphics[loop,autoplay,#3]{#4}{#2-}{0}{#5}}%
					}%
				}{
					\raisebox{\GLOBALimagenextmarginv}{%
						\fbox{\animategraphics[loop,#3]{#4}{#2-}{0}{#5}}%
					}%
				}%
			}{
				\ifthenelse{\equal{\animatedimageautoplay}{true}}{
					\raisebox{\GLOBALimagenextmarginv}{%
						\fbox{\animategraphics[autoplay,#3]{#4}{#2-}{0}{#5}}%
					}%
				}{
					\raisebox{\GLOBALimagenextmarginv}{%
						\fbox{\animategraphics[#3]{#4}{#2-}{0}{#5}}%
					}%
				}%
			}%
		}%
	}%
	\endgroup%
	\global\def\GLOBALenvimageadded {true}%
	\global\def\GLOBALimagenextmarginv {0 cm}%
}

\newcommand{\insertimageboxed}[5][]{%
	\emptyvarerr{\insertimageboxed}{#2}{Direccion de la imagen no definida}%
	\emptyvarerr{\insertimageboxed}{#3}{Parametros de la imagen no definidos}%
	\emptyvarerr{\insertimageboxed}{#4}{Ancho de la linea no definido}%
	\checkoutsideenvimage%
	\corevspacevarcm{\marginimagetop}%
	\begin{samepage}%
	\begin{figure}[H]%
		\begingroup%
			\setlength{\fboxsep}{0 pt}%
			\setlength{\fboxrule}{#4 pt}%
			\centering%
			\ifthenelse{\equal{\GLOBALimagelink}{\GLOBALemptyvar}}{
				\fbox{\includegraphics[#3]{#2}}%
			}{
				\fbox{\href{\GLOBALimagelink}{\includegraphics[#3]{#2}}}%
			}%
		\endgroup%
		\ifx\hfuzz#5\hfuzz%
			\corevspacevarcm{\captionlessmarginimage}%
		\else%
			\hspace{0cm}%
			\corevspacevarcm{\captionmarginimage}%
			\ifthenelse{\equal{\GLOBALcaptiondefn}{\GLOBALemptyvar}}{\caption{#5 #1}}{\caption[\GLOBALcaptiondefn]{#5 #1}}%
		\fi%
	\end{figure}
	\corevspacevarcm{\marginimagebottom}%
	\end{samepage}
	\resetindexcaption%
}

\newcommand{\insertanimatedimage}[6][]{%
	\insertanimatedimageboxed[#1]{#2}{#3}{#4}{#5}{0}{#6}%
}

\newcommand{\insertanimatedimageboxed}[7][]{%
	\emptyvarerr{\insertanimatedimage}{#2}{Direccion de la imagen no definida}%
	\emptyvarerr{\insertanimatedimage}{#3}{Parametros de la imagen no definidos}%
	\emptyvarerr{\insertanimatedimage}{#4}{FPS no definido}%
	\emptyvarerr{\insertanimatedimage}{#5}{Total imagenes no definido}%
	\emptyvarerr{\insertanimatedimage}{#6}{Ancho de la línea no definido}%
	\checkoutsideenvimage%
	\corevspacevarcm{\marginimagetop}%
	\begin{samepage}%
	\begin{figure}[H]%
		\begingroup%
			\setlength{\fboxsep}{0 pt}%
			\setlength{\fboxrule}{#6 pt}%
			\centering%
			\ifthenelse{\equal{\animatedimagecontrols}{true}}{
				\ifthenelse{\equal{\animatedimageloop}{true}}{
					\ifthenelse{\equal{\animatedimageautoplay}{true}}{
						\fbox{\animategraphics[controls,loop,autoplay,#3]{#4}{#2-}{0}{#5}}%
					}{
						\fbox{\animategraphics[controls,loop,#3]{#4}{#2-}{0}{#5}}%
					}%
				}{
					\ifthenelse{\equal{\animatedimageautoplay}{true}}{
						\fbox{\animategraphics[controls,autoplay,#3]{#4}{#2-}{0}{#5}}%
					}{
						\fbox{\animategraphics[controls,#3]{#4}{#2-}{0}{#5}}%
					}%
				}%
			}{
				\ifthenelse{\equal{\animatedimageloop}{true}}{
					\ifthenelse{\equal{\animatedimageautoplay}{true}}{
						\fbox{\animategraphics[loop,autoplay,#3]{#4}{#2-}{0}{#5}}%
					}{
						\fbox{\animategraphics[loop,#3]{#4}{#2-}{0}{#5}}%
					}%
				}{
					\ifthenelse{\equal{\animatedimageautoplay}{true}}{
						\fbox{\animategraphics[autoplay,#3]{#4}{#2-}{0}{#5}}%
					}{
						\fbox{\animategraphics[#3]{#4}{#2-}{0}{#5}}%
					}%
				}%
			}%
		\endgroup%
		\ifx\hfuzz#7\hfuzz%
			\corevspacevarcm{\captionlessmarginimage}%
		\else%
			\hspace{0cm}%
			\corevspacevarcm{\captionmarginimage}%
			\ifthenelse{\equal{\GLOBALcaptiondefn}{\GLOBALemptyvar}}{\caption{#7 #1}}{\caption[\GLOBALcaptiondefn]{#7 #1}}%
		\fi%
	\end{figure}
	\corevspacevarcm{\marginimagebottom}%
	\end{samepage}
	\resetindexcaption%
}

\newcommand{\insertimageboxedmc}[6][]{%
	\emptyvarerr{\insertimageboxedmc}{#2}{Direccion de la imagen no definida}%
	\emptyvarerr{\insertimageboxedmc}{#3}{Parametros de la imagen no definidos}%
	\emptyvarerr{\insertimageboxedmc}{#4}{Ancho de la linea no definido}%
	\emptyvarerr{\insertimageboxedmc}{#5}{Posicion de la imagen no definida}%
	\checkoutsideenvimage%
	\checkinsidemulticol%
	\checkoutsideappendix%
	\setcaptionmargincm{\captionlrmarginmc}%
	\ifthenelse{\equal{#5}{bottom}}{%
		\begin{samepage}%
		\begin{figure*}[!b]
	}{%
	\ifthenelse{\equal{#5}{top}}{%
		\begin{samepage}%
		\begin{figure*}[!t]
	}{%
	\ifthenelse{\equal{#5}{fixed2}}{%
		\end{multicols}
		\begin{samepage}%
		\begin{figure*}[!h]
	}{%
	\ifthenelse{\equal{#5}{fixed2b}}{%
		\end{multicols}
		\begin{samepage}%
		\begin{figure*}[!b]
	}{%
	\ifthenelse{\equal{#5}{fixed2t}}{%
		\end{multicols}
		\begin{samepage}%
		\begin{figure*}[!t]
	}{%
	\ifthenelse{\equal{#5}{fixed3}}{%
		\end{multicols}
		\begin{samepage}%
		\begin{figure*}[!h]
	}{%
	\ifthenelse{\equal{#5}{fixed3b}}{%
		\end{multicols}
		\begin{samepage}%
		\begin{figure*}[!b]
	}{%
	\ifthenelse{\equal{#5}{fixed3t}}{%
		\end{multicols}
		\begin{samepage}%
		\begin{figure*}[!t]
	}{%
	\ifthenelse{\equal{#5}{fixed4}}{%
		\end{multicols}
		\begin{samepage}%
		\begin{figure*}[!h]
	}{%
	\ifthenelse{\equal{#5}{fixed4b}}{%
		\end{multicols}
		\begin{samepage}%
		\begin{figure*}[!h]
	}{%
	\ifthenelse{\equal{#5}{fixed4t}}{%
		\end{multicols}
		\begin{samepage}%
		\begin{figure*}[!h]
	}{%
		\errmessage{LaTeX Warning: Posicion de imagen invalida, valores esperados: bottom,top,fixed2,fixed2b,fixed2t,fixed3,fixed3b,fixed3t,fixed4,fixed4b,fixed4t}
		\stop}}}}}}}}}}
	}%
		\begingroup%
			\setlength{\fboxsep}{0 pt}%
			\setlength{\fboxrule}{#4 pt}%
			\centering%
			\fbox{\includegraphics[#3]{#2}}%
		\endgroup%
		\ifx\hfuzz#6\hfuzz%
			\corevspacevarcm{\captionlessmarginimage}%
		\else%
			\hspace{0cm}%
			\corevspacevarcm{\captionmarginimage}%
			\ifthenelse{\equal{\GLOBALcaptiondefn}{\GLOBALemptyvar}}{\caption{#6 #1}}{\caption[\GLOBALcaptiondefn]{#6 #1}}%
		\fi%
	\end{figure*}
	\end{samepage}
	\ifthenelse{\equal{#5}{fixed2}}{%
		\begin{multicols}{2}%
	}{%
	\ifthenelse{\equal{#5}{fixed2b}}{%
		\begin{multicols}{2}%
	}{%
	\ifthenelse{\equal{#5}{fixed2t}}{%
		\begin{multicols}{2}%
	}{%
	\ifthenelse{\equal{#5}{fixed3}}{%
		\begin{multicols}{3}%
	}{%
	\ifthenelse{\equal{#5}{fixed3b}}{%
		\begin{multicols}{3}%
	}{%
	\ifthenelse{\equal{#5}{fixed3t}}{%
		\begin{multicols}{3}%
	}{%
	\ifthenelse{\equal{#5}{fixed4}}{%
		\begin{multicols}{4}%
	}{%
	\ifthenelse{\equal{#5}{fixed4b}}{%
		\begin{multicols}{4}%
	}{%
	\ifthenelse{\equal{#5}{fixed4t}}{%
		\begin{multicols}{4}%
	}{%
	}}}}}}}}}%
	\setcaptionmargincm{\captionlrmargin}%
	\resetindexcaption%
}

\newcommand{\inserttableimageboxed}[3]{%
	\emptyvarerr{\inserttableimageboxed}{#1}{Direccion de la imagen no definida}%
	\emptyvarerr{\inserttableimageboxed}{#2}{Parametros de la imagen no definidos}%
	\emptyvarerr{\inserttableimageboxed}{#3}{Ancho de la linea no definido}%
	\checkoutsideenvimage%
	\begingroup%
	\setlength{\fboxsep}{0 pt}%
	\setlength{\fboxrule}{#3 pt}%
	\raisebox{-1\totalheight}{\fbox{\includegraphics[#2]{#1}}}%
	\endgroup%
	\resetindexcaption%
}

\newcommand{\insertimageleftboxed}[5][]{%
	\emptyvarerr{\insertimageleftboxed}{#2}{Direccion de la imagen no definida}%
	\emptyvarerr{\insertimageleftboxed}{#3}{Ancho de la imagen no definido}%
	\emptyvarerr{\insertimageleftboxed}{#4}{Ancho de la linea no definido}%
	\checkoutsideenvimage%
	~%
	\vspace{-\baselineskip}%
	\par%
	\begin{wrapfigure}{l}{#3\linewidth}%
		\setcaptionmargincm{0}%
		\ifthenelse{\equal{\figurecaptiontop}{true}}{}{%
			\vspace{\marginfloatimages pt}%
		}%
		\begingroup%
			\setlength{\fboxsep}{0 pt}%
			\setlength{\fboxrule}{#4 pt}%
			\centering%
			\fbox{\includegraphics[width=\linewidth]{#2}}%
		\endgroup%
		\ifx\hfuzz#5\hfuzz%
			\corevspacevarcm{\captionlessmarginimage}%
		\else%
			\corevspacevarcm{\captionmarginimage}%
			\ifthenelse{\equal{\GLOBALcaptiondefn}{\GLOBALemptyvar}}{\caption{#5 #1}}{\caption[\GLOBALcaptiondefn]{#5 #1}}%
		\fi%
	\end{wrapfigure}
	\setcaptionmargincm{\captionlrmargin}%
	\resetindexcaption%
}

\newcommand{\insertimageleftlineboxed}[6][]{%
	\emptyvarerr{\insertimageleftlineboxed}{#2}{Direccion de la imagen no definida}%
	\emptyvarerr{\insertimageleftlineboxed}{#3}{Ancho de la imagen no definido}%
	\emptyvarerr{\insertimageleftlineboxed}{#4}{Ancho de la linea no definido}%
	\emptyvarerr{\insertimageleftlineboxed}{#5}{Altura en lineas de la imagen flotante izquierda no definida}
	\checkoutsideenvimage%
	~%
	\vspace{-\baselineskip}%
	\par%
	\begin{wrapfigure}[#5]{l}{#3\linewidth}%
		\setcaptionmargincm{0}%
		\ifthenelse{\equal{\figurecaptiontop}{true}}{}{%
			\vspace{\marginfloatimages pt}}%
		\begingroup%
			\setlength{\fboxsep}{0 pt}%
			\setlength{\fboxrule}{#4 pt}%
			\centering%
			\fbox{\includegraphics[width=\linewidth]{#2}}%
		\endgroup%
		\ifx\hfuzz#6\hfuzz%
			\corevspacevarcm{\captionlessmarginimage}%
		\else%
			\corevspacevarcm{\captionmarginimage}%
			\ifthenelse{\equal{\GLOBALcaptiondefn}{\GLOBALemptyvar}}{\caption{#6 #1}}{\caption[\GLOBALcaptiondefn]{#6 #1}}%
		\fi%
	\end{wrapfigure}
	\setcaptionmargincm{\captionlrmargin}%
	\resetindexcaption%
}

\newcommand{\insertimagerightboxed}[5][]{%
	\emptyvarerr{\insertimagerightboxed}{#2}{Direccion de la imagen no definida}%
	\emptyvarerr{\insertimagerightboxed}{#3}{Ancho de la imagen no defindo}%
	\emptyvarerr{\insertimagerightboxed}{#4}{Ancho de la linea no definido}%
	\checkoutsideenvimage%
	~%
	\vspace{-\baselineskip}%
	\par%
	\begin{wrapfigure}{r}{#3\linewidth}%
		\setcaptionmargincm{0}%
		\ifthenelse{\equal{\figurecaptiontop}{true}}{}{%
			\vspace{\marginfloatimages pt}%
		}%
		\begingroup%
			\setlength{\fboxsep}{0 pt}%
			\setlength{\fboxrule}{#4 pt}%
			\centering%
			\fbox{\includegraphics[width=\linewidth]{#2}}%
		\endgroup%
		\ifx\hfuzz#5\hfuzz%
			\corevspacevarcm{\captionlessmarginimage}%
		\else%
			\corevspacevarcm{\captionmarginimage}%
			\ifthenelse{\equal{\GLOBALcaptiondefn}{\GLOBALemptyvar}}{\caption{#5 #1}}{\caption[\GLOBALcaptiondefn]{#5 #1}}%
		\fi%
	\end{wrapfigure}
	\setcaptionmargincm{\captionlrmargin}%
	\resetindexcaption%
}

\newcommand{\insertimagerightlineboxed}[6][]{%
	\emptyvarerr{\insertimagerightlineboxed}{#2}{Direccion de la imagen no definida}%
	\emptyvarerr{\insertimagerightlineboxed}{#3}{Ancho de la imagen no defindo}%
	\emptyvarerr{\insertimagerightlineboxed}{#4}{Ancho de la linea no definido}%
	\emptyvarerr{\insertimagerightlineboxed}{#5}{Altura en lineas de la imagen flotante derecha no definida}%
	\checkoutsideenvimage%
	~%
	\vspace{-\baselineskip}%
	\par%
	\begin{wrapfigure}[#5]{r}{#3\linewidth}%
		\setcaptionmargincm{0}%
		\ifthenelse{\equal{\figurecaptiontop}{true}}{}{%
			\vspace{\marginfloatimages pt}%
		}%
		\begingroup%
			\setlength{\fboxsep}{0 pt}%
			\setlength{\fboxrule}{#4 pt}%
			\centering%
			\fbox{\includegraphics[width=\linewidth]{#2}}%
		\endgroup%
		\ifx\hfuzz#6\hfuzz%
			\corevspacevarcm{\captionlessmarginimage}%
		\else%
			\corevspacevarcm{\captionmarginimage}%
			\ifthenelse{\equal{\GLOBALcaptiondefn}{\GLOBALemptyvar}}{\caption{#6 #1}}{\caption[\GLOBALcaptiondefn]{#6 #1}}%
		\fi%
	\end{wrapfigure}
	\setcaptionmargincm{\captionlrmargin}%
	\resetindexcaption%
}

\newcommand{\insertimageleftboxedp}[6][]{%
	\emptyvarerr{\insertimageleftboxedp}{#2}{Direccion de la imagen no definida}%
	\emptyvarerr{\insertimageleftboxedp}{#3}{Ancho del objeto no definido}%
	\emptyvarerr{\insertimageleftboxedp}{#4}{Propiedades de la imagen no defindos}%
	\emptyvarerr{\insertimageleftboxedp}{#5}{Ancho de la linea no definido}%
	\checkoutsideenvimage%
	~%
	\vspace{-\baselineskip}%
	\par%
	\begin{wrapfigure}{l}{#3}%
		\setcaptionmargincm{0}%
		\ifthenelse{\equal{\figurecaptiontop}{true}}{}{%
			\vspace{\marginfloatimages pt}%
		}%
		\begingroup%
			\setlength{\fboxsep}{0 pt}%
			\setlength{\fboxrule}{#5 pt}%
			\centering%
			\fbox{\includegraphics[#4]{#2}}%
		\endgroup%
		\ifx\hfuzz#6\hfuzz%
			\corevspacevarcm{\captionlessmarginimage}%
		\else%
			\corevspacevarcm{\captionmarginimage}%
			\ifthenelse{\equal{\GLOBALcaptiondefn}{\GLOBALemptyvar}}{\caption{#6 #1}}{\caption[\GLOBALcaptiondefn]{#6 #1}}%
		\fi%
	\end{wrapfigure}
	\setcaptionmargincm{\captionlrmargin}%
	\resetindexcaption%
}

\newcommand{\insertimageleftlineboxedp}[7][]{%
	\emptyvarerr{\insertimageleftlineboxedp}{#2}{Direccion de la imagen no definida}%
	\emptyvarerr{\insertimageleftlineboxedp}{#3}{Ancho del objeto no definido}%
	\emptyvarerr{\insertimageleftlineboxedp}{#4}{Propiedades de la imagen no definidos}%
	\emptyvarerr{\insertimageleftlineboxedp}{#5}{Ancho de la linea no definido}%
	\emptyvarerr{\insertimageleftlineboxedp}{#6}{Altura en lineas de la imagen flotante izquierda no definida}%
	\checkoutsideenvimage%
	~%
	\vspace{-\baselineskip}%
	\par%
	\begin{wrapfigure}[#6]{l}{#3}%
		\setcaptionmargincm{0}%
		\ifthenelse{\equal{\figurecaptiontop}{true}}{}{%
			\vspace{\marginfloatimages pt}%
		}%
		\begingroup%
			\setlength{\fboxsep}{0 pt}%
			\setlength{\fboxrule}{#5 pt}%
			\centering%
			\fbox{\includegraphics[#4]{#2}}%
		\endgroup%
		\ifx\hfuzz#7\hfuzz%
			\corevspacevarcm{\captionlessmarginimage}%
		\else%
			\corevspacevarcm{\captionmarginimage}%
			\ifthenelse{\equal{\GLOBALcaptiondefn}{\GLOBALemptyvar}}{\caption{#7 #1}}{\caption[\GLOBALcaptiondefn]{#7 #1}}%
		\fi%
	\end{wrapfigure}
	\setcaptionmargincm{\captionlrmargin}%
	\resetindexcaption%
}

\newcommand{\insertimagerightboxedp}[6][]{%
	\emptyvarerr{\insertimagerightboxedp}{#2}{Direccion de la imagen no definida}%
	\emptyvarerr{\insertimagerightboxedp}{#3}{Ancho del objeto no definido}%
	\emptyvarerr{\insertimagerightboxedp}{#4}{Propiedades de la imagen no definidos}%
	\emptyvarerr{\insertimagerightboxedp}{#5}{Ancho de la linea no definido}%
	\checkoutsideenvimage%
	~%
	\vspace{-\baselineskip}%
	\par%
	\begin{wrapfigure}{r}{#3}%
		\setcaptionmargincm{0}%
		\ifthenelse{\equal{\figurecaptiontop}{true}}{}{%
			\vspace{\marginfloatimages pt}%
		}%
		\begingroup%
			\setlength{\fboxsep}{0 pt}%
			\setlength{\fboxrule}{#5 pt}%
			\centering%
			\fbox{\includegraphics[#4]{#2}}%
		\endgroup%
		\ifx\hfuzz#6\hfuzz%
			\corevspacevarcm{\captionlessmarginimage}%
		\else%
			\corevspacevarcm{\captionmarginimage}%
			\ifthenelse{\equal{\GLOBALcaptiondefn}{\GLOBALemptyvar}}{\caption{#6 #1}}{\caption[\GLOBALcaptiondefn]{#6 #1}}%
		\fi%
	\end{wrapfigure}
	\setcaptionmargincm{\captionlrmargin}%
	\resetindexcaption%
}

\newcommand{\insertimagerightlineboxedp}[7][]{%
	\emptyvarerr{\insertimagerightlineboxedp}{#2}{Direccion de la imagen no definida}%
	\emptyvarerr{\insertimagerightlineboxedp}{#3}{Ancho del objeto no definido}%
	\emptyvarerr{\insertimagerightlineboxedp}{#4}{Propiedades de la imagen no definidos}%
	\emptyvarerr{\insertimagerightlineboxedp}{#5}{Ancho de la linea no definido}%
	\emptyvarerr{\insertimagerightlineboxedp}{#6}{Altura en lineas de la imagen flotante derecha no definida}%
	\checkoutsideenvimage%
	~%
	\vspace{-\baselineskip}%
	\par%
	\begin{wrapfigure}[#6]{r}{#3}%
		\setcaptionmargincm{0}%
		\ifthenelse{\equal{\figurecaptiontop}{true}}{}{%
			\vspace{\marginfloatimages pt}%
		}%
		\begingroup%
			\setlength{\fboxsep}{0 pt}%
			\setlength{\fboxrule}{#5 pt}%
			\centering%
			\fbox{\includegraphics[#4]{#2}}%
		\endgroup%
		\ifx\hfuzz#7\hfuzz%
			\corevspacevarcm{\captionlessmarginimage}%
		\else%
			\corevspacevarcm{\captionmarginimage}%
			\ifthenelse{\equal{\GLOBALcaptiondefn}{\GLOBALemptyvar}}{\caption{#7 #1}}{\caption[\GLOBALcaptiondefn]{#7 #1}}%
		\fi%
	\end{wrapfigure}
	\setcaptionmargincm{\captionlrmargin}%
	\resetindexcaption%
}

\newcommand{\coreinsertkeyimage}[2]{%
	\expandafter\includegraphics\expandafter[#1]{\expandafter#2}%
}

\makeatletter
\define@key{Gin}{resolution}{\pdfimageresolution=#1\relax}
\makeatother

\global\def\GLOBALtitlerequirechapter {false}
\global\def\GLOBALtitleinitchapter {false}
\global\def\GLOBALtitleinitsection {false}
\global\def\GLOBALtitleinitsubsection {false}
\global\def\GLOBALtitleinitsubsubsection {false}
\global\def\GLOBALtitleinitsubsubsubsection {false}

\global\def\GLOBALtitlepresectionstr {}
\global\def\GLOBALtitlepresubsectionstr {}
\global\def\GLOBALtitlepresubsubsectionstr {}
\global\def\GLOBALtitlepresubsubsubsectionstr {}

\global\def\GLOBALtitlechapterenabled {true}

\def\coreintializetitlenumbering {%
	\renewcommand{\thechapter}{\GLOBALformatnumchapter{chapter}}%
	\ifthenelse{\equal{\GLOBALchapternumenabled}{false}}{%
		\renewcommand{\thesection}{%
			\GLOBALformatnumsection{section}%
		}%
	}{%
		\renewcommand{\thesection}{%
			\thechapter\charbetwchaptersection\GLOBALformatnumsection{section}%
		}%
	}%
	\ifthenelse{\equal{\GLOBALsectionanumenabled}{true}}{%
		\renewcommand{\thesubsection}{%
			\GLOBALformatnumssection{subsection}%
		}%
	}{%
		\renewcommand{\thesubsection}{%
			\GLOBALformatnumssection{subsection}%
		}%
	}%
	\ifthenelse{\equal{\GLOBALsubsectionanumenabled}{true}}{%
		\renewcommand{\thesubsubsection}{%
			\GLOBALformatnumsssection{subsubsection}%
		}%
	}{%
		\renewcommand{\thesubsubsection}{%
			\thesubsection\charbetwsubsectionssect\GLOBALformatnumsssection{subsubsection}%
		}%
	}%
	\ifthenelse{\equal{\GLOBALsubsubsectionanumenabled}{true}}{%
		\renewcommand{\thesubsubsubsection}{%
			\GLOBALformatnumssssection{subsubsubsection}%
		}%
	}{%
		\renewcommand{\thesubsubsubsection}{%
			\thesubsubsection\charbetwssectionsssect\GLOBALformatnumssssection{subsubsubsection}%
		}%
	}%
	\hbadness=10000%
}

\def\corecheckchapterenabled {%
	\ifthenelse{\equal{\GLOBALtitlechapterenabled}{false}}{
		\throwwarning{La insercion de capitulos esta desactivada}%
	}{}%
}

\def\corecheckchapterinitialized {%
	\ifthenelse{\equal{\GLOBALtitlerequirechapter}{true}}{%
		\ifthenelse{\equal{\GLOBALtitleinitchapter}{false}}{%
			\throwwarning{Se requiere un nuevo capitulo}%
		}{}%
	}{}%
}

\def\corechecksectioninitialized {%
	\ifthenelse{\equal{\GLOBALtitleinitsection}{false}}{%
		\throwwarning{Se requiere una nueva seccion}%
	}{}%
}

\def\corechecksubsectioninitialized {%
	\ifthenelse{\equal{\GLOBALtitleinitsubsection}{false}}{%
		\throwwarning{Se requiere una nueva subseccion}%
	}{}%
}

\def\corechecksubsubsectioninitialized {%
	\ifthenelse{\equal{\GLOBALtitleinitsubsubsection}{false}}{%
		\throwwarning{Se requiere una nueva subsubseccion}%
	}{}%
}

\pretocmd{\chapter}{%
	\corecheckchapterenabled%
	\ifthenelse{\equal{\showsectioncaptioncode}{chap}}{
		\addtocounter{templateListings}{\value{lstlisting}}%
		\setcounter{lstlisting}{0}%
	}{}%
	\ifthenelse{\equal{\showsectioncaptioneqn}{chap}}{
		\addtocounter{templateEquations}{\value{equation}}%
		\setcounter{equation}{0}%
	}{}%
	\ifthenelse{\equal{\equationrestart}{chap}}{
		\addtocounter{templateEquations}{\value{equation}}%
		\setcounter{equation}{0}%
	}{}%
	\ifthenelse{\equal{\showsectioncaptionfig}{chap}}{
		\addtocounter{templateFigures}{\value{figure}}%
		\setcounter{figure}{0}%
	}{}%
	\ifthenelse{\equal{\showsectioncaptiontab}{chap}}{
		\addtocounter{templateTables}{\value{table}}%
		\setcounter{table}{0}%
	}{}%
	\global\def\GLOBALchapternumenabled {true}%
	\global\def\GLOBALsectionanumenabled {false}%
	\global\def\GLOBALsubsectionanumenabled {false}%
	\global\def\GLOBALsubsubsectionanumenabled {false}%
	\global\def\GLOBALtitleinitchapter {true}%
	\global\def\GLOBALtitleinitsection {false}%
	\global\def\GLOBALtitleinitsubsection {false}%
	\global\def\GLOBALtitleinitsubsubsection {false}%
	\global\def\GLOBALtitleinitsubsubsubsection {false}%
	\coreintializetitlenumbering%
}{}{}

\pretocmd{\section}{%
	\ifthenelse{\equal{\showsectioncaptioncode}{sec}}{
		\addtocounter{templateListings}{\value{lstlisting}}%
		\setcounter{lstlisting}{0}%
	}{}%
	\ifthenelse{\equal{\showsectioncaptioneqn}{sec}}{
		\addtocounter{templateEquations}{\value{equation}}%
		\setcounter{equation}{0}%
	}{}%
	\ifthenelse{\equal{\equationrestart}{sec}}{
		\addtocounter{templateEquations}{\value{equation}}%
		\setcounter{equation}{0}%
	}{}%
	\ifthenelse{\equal{\showsectioncaptionfig}{sec}}{
		\addtocounter{templateFigures}{\value{figure}}%
		\setcounter{figure}{0}%
	}{}%
	\ifthenelse{\equal{\showsectioncaptiontab}{sec}}{
		\addtocounter{templateTables}{\value{table}}%
		\setcounter{table}{0}%
	}{}%
	\global\def\GLOBALsectionanumenabled {false}%
	\global\def\GLOBALsubsectionanumenabled {false}%
	\global\def\GLOBALsubsubsectionanumenabled {false}%
	\global\def\GLOBALtitleinitsection {true}%
	\global\def\GLOBALtitleinitsubsection {false}%
	\global\def\GLOBALtitleinitsubsubsection {false}%
	\global\def\GLOBALtitleinitsubsubsubsection {false}%
	\corecheckchapterinitialized%
	\coreintializetitlenumbering%
}{}{}

\pretocmd{\subsection}{%
	\ifthenelse{\equal{\showsectioncaptioncode}{ssec}}{
		\addtocounter{templateListings}{\value{lstlisting}}%
		\setcounter{lstlisting}{0}%
	}{}%
	\ifthenelse{\equal{\showsectioncaptioneqn}{ssec}}{
		\addtocounter{templateEquations}{\value{equation}}%
		\setcounter{equation}{0}%
	}{}%
	\ifthenelse{\equal{\equationrestart}{ssec}}{
		\addtocounter{templateEquations}{\value{equation}}%
		\setcounter{equation}{0}%
	}{}%
	\ifthenelse{\equal{\showsectioncaptionfig}{ssec}}{
		\addtocounter{templateFigures}{\value{figure}}%
		\setcounter{figure}{0}%
	}{}%
	\ifthenelse{\equal{\showsectioncaptiontab}{ssec}}{
		\addtocounter{templateTables}{\value{table}}%
		\setcounter{table}{0}%
	}{}%
	\global\def\GLOBALsubsectionanumenabled {false}%
	\global\def\GLOBALsubsubsectionanumenabled {false}%
	\global\def\GLOBALtitleinitsubsection {true}%
	\global\def\GLOBALtitleinitsubsubsection {false}%
	\global\def\GLOBALtitleinitsubsubsubsection {false}%
	\corecheckchapterinitialized%
	\corechecksectioninitialized%
	\coreintializetitlenumbering%
}{}{}

\pretocmd{\subsubsection}{%
	\ifthenelse{\equal{\showsectioncaptioncode}{sssec}}{
		\addtocounter{templateListings}{\value{lstlisting}}%
		\setcounter{lstlisting}{0}%
	}{}%
	\ifthenelse{\equal{\showsectioncaptioneqn}{sssec}}{
		\addtocounter{templateEquations}{\value{equation}}%
		\setcounter{equation}{0}%
	}{}%
	\ifthenelse{\equal{\equationrestart}{sssec}}{
		\addtocounter{templateEquations}{\value{equation}}%
		\setcounter{equation}{0}%
	}{}%
	\ifthenelse{\equal{\showsectioncaptionfig}{sssec}}{
		\addtocounter{templateFigures}{\value{figure}}%
		\setcounter{figure}{0}%
	}{}%
	\ifthenelse{\equal{\showsectioncaptiontab}{sssec}}{
		\addtocounter{templateTables}{\value{table}}%
		\setcounter{table}{0}%
	}{}%
	\global\def\GLOBALsubsubsectionanumenabled {false}%
	\global\def\GLOBALtitleinitsubsubsection {true}%
	\global\def\GLOBALtitleinitsubsubsubsection {false}%
	\corecheckchapterinitialized%
	\corechecksectioninitialized%
	\corechecksubsectioninitialized%
	\coreintializetitlenumbering%
}{}{}

\def\corepatchaftersubsubsubsection {%
	\ifthenelse{\equal{\showsectioncaptioncode}{ssssec}}{
		\addtocounter{templateListings}{\value{lstlisting}}%
		\setcounter{lstlisting}{0}%
	}{}%
	\ifthenelse{\equal{\showsectioncaptioneqn}{ssssec}}{
		\addtocounter{templateEquations}{\value{equation}}%
		\setcounter{equation}{0}%
	}{}%
	\ifthenelse{\equal{\equationrestart}{ssssec}}{
		\addtocounter{templateEquations}{\value{equation}}%
		\setcounter{equation}{0}%
	}{}%
	\ifthenelse{\equal{\showsectioncaptionfig}{ssssec}}{
		\addtocounter{templateFigures}{\value{figure}}%
		\setcounter{figure}{0}%
	}{}%
	\ifthenelse{\equal{\showsectioncaptiontab}{ssssec}}{
		\addtocounter{templateTables}{\value{table}}%
		\setcounter{table}{0}%
	}{}%
	\global\def\GLOBALtitleinitsubsubsubsection {true}%
	\corecheckchapterinitialized%
	\corechecksectioninitialized%
	\corechecksubsectioninitialized%
	\corechecksubsubsectioninitialized%
}

\makeatletter
\newcommand*\coredisabledchapter{%
	\@ifstar{\coredisabledchapterstar}{\@dblarg\coredisabledchapternostar}}
\newcommand*\coredisabledchapterstar[1]{%
	\noindent\textcolor{red}{Error (chapter):} \newline#1%
	\throwwarning{La insercion de capitulos esta desactivada}%
}
\def\coredisabledchapternostar[#1]#2{%
	\noindent\textcolor{red}{Error (chapter):} #1%
	\throwwarning{La insercion de capitulos esta desactivada}%
}
\makeatother
\let\oldchapter\chapter

\newcommand{\disablechapter}{%
	\let\chapter\coredisabledchapter%
	\global\def\GLOBALtitlechapterenabled {false}%
}

\newcommand{\enablechapter}{%
	\let\chapter\oldchapter%
	\global\def\GLOBALtitlechapterenabled {true}%
}

\newcommand{\chapteranum}[1]{%
	\corecheckchapterenabled%
	\emptyvarerr{\chapteranum}{#1}{Titulo no definido}%
	\phantomsection%
	\needspace{3\baselineskip}%
	\chapter*{#1}%
	\addcontentsline{toc}{chapter}{#1}%
	\ifthenelse{\equal{\anumsecaddtocounter}{true}}{\stepcounter{chapter}}{}%
	\changeheadertitle{#1}%
	\setcounter{section}{0}%
	\global\def\GLOBALchapternumenabled {false}%
	\coreintializetitlenumbering%
}

\newcommand{\sectionanum}[1]{%
	\emptyvarerr{\sectionanum}{#1}{Titulo no definido}%
	\phantomsection%
	\needspace{3\baselineskip}%
	\section*{#1}%
	\addcontentsline{toc}{section}{#1}%
	\ifthenelse{\equal{\anumsecaddtocounter}{true}}{\stepcounter{section}}{}%
	\changeheadertitle{#1}%
	\setcounter{subsection}{0}%
	\global\def\GLOBALsectionanumenabled {true}%
	\coreintializetitlenumbering%
}

\newcommand{\sectionanumnoi}[1]{%
	\emptyvarerr{\sectionanumnoi}{#1}{Titulo no definido}%
	\phantomsection%
	\needspace{3\baselineskip}%
	\section*{#1}%
	\ifthenelse{\equal{\anumsecaddtocounter}{true}}{\stepcounter{section}}{}%
	\changeheadertitle{#1}%
	\setcounter{subsection}{0}%
	\global\def\GLOBALsectionanumenabled {true}%
	\coreintializetitlenumbering%
}

\newcommand{\sectionanumheadless}[1]{%
	\emptyvarerr{\sectionanumnoheadless}{#1}{Titulo no definido}%
	\section*{#1}%
	\addcontentsline{toc}{section}{#1}%
	\ifthenelse{\equal{\anumsecaddtocounter}{true}}{\stepcounter{section}}{}%
	\setcounter{subsection}{0}%
	\global\def\GLOBALsectionanumenabled {true}%
	\coreintializetitlenumbering%
}

\newcommand{\sectionanumnoiheadless}[1]{%
	\emptyvarerr{\sectionanumnoiheadless}{#1}{Titulo no definido}%
	\section*{#1}%
	\ifthenelse{\equal{\anumsecaddtocounter}{true}}{\stepcounter{section}}{}%
	\setcounter{subsection}{0}%
	\global\def\GLOBALsectionanumenabled {true}%
	\coreintializetitlenumbering%
}

\newcommand{\subsectionanum}[1]{%
	\emptyvarerr{\subsectionanum}{#1}{Subtitulo no definido}%
	\subsection*{#1}%
	\addcontentsline{toc}{subsection}{#1}
	\ifthenelse{\equal{\anumsecaddtocounter}{true}}{\stepcounter{subsection}}{}%
	\setcounter{subsubsection}{0}%
	\global\def\GLOBALsubsectionanumenabled {true}%
	\coreintializetitlenumbering%
}

\newcommand{\subsectionanumnoi}[1]{%
	\emptyvarerr{\subsectionanumnoi}{#1}{Subtitulo no definido}%
	\subsection*{#1}%
	\ifthenelse{\equal{\anumsecaddtocounter}{true}}{\stepcounter{subsection}}{}%
	\setcounter{subsubsection}{0}%
	\global\def\GLOBALsubsectionanumenabled {true}%
	\coreintializetitlenumbering%
}

\newcommand{\subsubsectionanum}[1]{%
	\emptyvarerr{\subsubsectionanum}{#1}{Sub-subtitulo no definido}%
	\subsubsection*{#1}%
	\addcontentsline{toc}{subsubsection}{#1}%
	\ifthenelse{\equal{\anumsecaddtocounter}{true}}{\stepcounter{subsubsection}}{}%
	\setcounter{subsubsubsection}{0}%
	\global\def\GLOBALsubsubsectionanumenabled {true}%
	\coreintializetitlenumbering%
}

\newcommand{\subsubsectionanumnoi}[1]{%
	\emptyvarerr{\subsubsectionanumnoi}{#1}{Sub-subtitulo no definido}%
	\subsubsection*{#1}%
	\ifthenelse{\equal{\anumsecaddtocounter}{true}}{\stepcounter{subsubsection}}{}%
	\setcounter{subsubsubsection}{0}%
	\global\def\GLOBALsubsubsectionanumenabled {true}%
	\coreintializetitlenumbering%
}

\newcommand{\subsubsubsectionanum}[1]{%
	\emptyvarerr{\subsubsubsectionanum}{#1}{Sub-sub-subtitulo no definido}%
	\subsubsubsection*{#1}%
	\addcontentsline{toc}{subsubsubsection}{#1}%
	\ifthenelse{\equal{\anumsecaddtocounter}{true}}{\stepcounter{subsubsubsection}}{}%
}

\newcommand{\subsubsubsectionanumnoi}[1]{%
	\emptyvarerr{\subsubsubsectionanumnoi}{#1}{Sub-sub-subtitulo no definido}%
	\subsubsection*{#1}%
	\ifthenelse{\equal{\anumsecaddtocounter}{true}}{\stepcounter{subsubsubsection}}{}%
}

\newcommand{\changeheadertitle}[1]{%
	\emptyvarerr{\changeheadertitle}{#1}{Titulo no definido}%
	\markboth{#1}{}%
}

\newcommand{\newchapter}[1]{%
	\emptyvarerr{\newchapter}{#1}{Titulo no definido}%
	\clearpage%
	\stepcounter{section}%
	\phantomsection%
	\needspace{3\baselineskip}%
	\vspace* {3cm}%
	\noindent {\huge{\textbf{\namechapter\ \thesection}}} \\%
	\vspace* {0.5cm} \\%
	\noindent {\Huge{\textbf{#1}}} \\%
	\vspace {0.5cm} \\%
	\addcontentsline{toc}{section}{\protect\numberline{\thesection}#1}%
	\markboth{#1}{}%
}

\ifthenelse{\isundefined{\newcolumn}}{%
	\newcommand{\newcolumn}{%
		\checkinsidemulticol\vfill\null\columnbreak%
	}
}{%
	\renewcommand{\newcolumn}{%
		\checkinsidemulticol\vfill\null\columnbreak%
	}
}

\newcommand{\itemresize}[2]{%
	\emptyvarerr{\itemresize}{#1}{Tamano del nuevo objeto no definido}%
	\emptyvarerr{\itemresize}{#2}{Objeto a redimensionar no definido}%
	\resizebox{#1\linewidth}{!}{#2}%
}

\def\corecleardoublepage {%
	\clearpage %
	\ifthenelse{\equal{\GLOBALtwoside}{true}}{%
		\ifodd\thepage %
		\else%
			\emptypagespredocformat%
		\fi%
	}{}%
}

\newcommand{\coretriggeronpage}[2]{%
	\ifthenelse{\isodd{\value{templatePageCounter}}}{%
		#2%
	}{%
		#1%
	}%
}

\newcommand{\quotes}[1]{%
	\enquote*{#1}%
}

\newcommand{\scite}[1]{%
	\textsuperscript{\cite{#1}}%
}

\newcommand{\hreftext}[1]{%
	\ifthenelse{\equal{\fonturl}{same}}{%
		#1%
	}{%
	\ifthenelse{\equal{\fonturl}{tt}}{%
		\texttt{#1}%
	}{%
	\ifthenelse{\equal{\fonturl}{rm}}{%
		\textrm{#1}%
	}{%
	\ifthenelse{\equal{\fonturl}{sf}}{%
		\textsf{#1}%
	}{}}}}%
}

\newcommand{\insertemail}[1]{%
	\href{mailto:#1}{\hreftext{#1}}%
}

\newcommand{\settablerowcolors}[1]{%
	\emptyvarerr{\settablerowcolors}{#1}{Posicion de fila no definida}%
	\ifthenelse{\equal{\GLOBALtablerowcolorswitch}{false}}{
		\ifthenelse{\equal{\tablerowfirstcolor}{none}}{%
			\ifthenelse{\equal{\tablerowsecondcolor}{none}}{%
				\rowcolors{#1}{}{}%
			}{%
				\rowcolors{#1}{\tablerowsecondcolor}{}%
			}%
		}{%
			\ifthenelse{\equal{\tablerowsecondcolor}{none}}{%
				\rowcolors{#1}{}{\tablerowfirstcolor}%
			}{%
				\rowcolors{#1}{\tablerowsecondcolor}{\tablerowfirstcolor}%
			}%
		}%
	}{
		\ifthenelse{\equal{\tablerowfirstcolor}{none}}{%
			\ifthenelse{\equal{\tablerowsecondcolor}{none}}{%
				\rowcolors{#1}{}{}%
			}{%
				\rowcolors{#1}{}{\tablerowsecondcolor}%
			}%
		}{%
			\ifthenelse{\equal{\tablerowsecondcolor}{none}}{%
				\rowcolors{#1}{\tablerowfirstcolor}{}%
			}{%
				\rowcolors{#1}{\tablerowfirstcolor}{\tablerowsecondcolor}%
			}%
		}%
	}%
	
	\global\def\GLOBALtablerowcolorindex {#1}%
}

\newcommand{\settablerowcolorslast}{%
	\ifthenelse{\equal{\GLOBALtablerowcolorswitch}{false}}{
		\ifthenelse{\equal{\tablerowfirstcolor}{none}}{%
			\ifthenelse{\equal{\tablerowsecondcolor}{none}}{%
				\rowcolors{\GLOBALtablerowcolorindex}{}{}%
			}{%
				\rowcolors{\GLOBALtablerowcolorindex}{\tablerowsecondcolor}{}%
			}%
		}{%
			\ifthenelse{\equal{\tablerowsecondcolor}{none}}{%
				\rowcolors{\GLOBALtablerowcolorindex}{}{\tablerowfirstcolor}%
			}{%
				\rowcolors{\GLOBALtablerowcolorindex}{\tablerowsecondcolor}{\tablerowfirstcolor}%
			}%
		}%
	}{
		\ifthenelse{\equal{\tablerowfirstcolor}{none}}{%
			\ifthenelse{\equal{\tablerowsecondcolor}{none}}{%
				\rowcolors{\GLOBALtablerowcolorindex}{}{}%
			}{%
				\rowcolors{\GLOBALtablerowcolorindex}{}{\tablerowsecondcolor}%
			}%
		}{%
			\ifthenelse{\equal{\tablerowsecondcolor}{none}}{%
				\rowcolors{\GLOBALtablerowcolorindex}{\tablerowfirstcolor}{}%
			}{%
				\rowcolors{\GLOBALtablerowcolorindex}{\tablerowfirstcolor}{\tablerowsecondcolor}%
			}%
		}%
	}%
}

\newcommand{\enabletablerowcolor}[1][]{%
	\ifx\hfuzz#1\hfuzz%
		\settablerowcolors{2}%
	\else%
		\settablerowcolors{#1}%
	\fi%
}

\newcommand{\switchtablerowcolors}{%
	\ifthenelse{\equal{\GLOBALtablerowcolorswitch}{false}}{%
		\global\def\GLOBALtablerowcolorswitch {true}%
	}{%
		\global\def\GLOBALtablerowcolorswitch {false}%
	}%
	\settablerowcolorslast%
}

\newcommand{\settablecellpadding}[2]{%
	\emptyvarerr{\settablecellpadding}{#1}{Padding horizontal no definido}%
	\emptyvarerr{\settablecellpadding}{#2}{Padding vertical no definido}%
	\setlength{\tabcolsep}{#1 em} 
	\def\arraystretch {#2} 
}

\newcommand{\resettablecellpadding}{%
	\settablecellpadding{\tablepaddingh}{\tablepaddingv}%
}

\newcommand{\changepagesize}[3][]{%
	\emptyvarerr{\changepagesize}{#2}{Ancho de la pagina no definida}%
	\emptyvarerr{\changepagesize}{#3}{Altura de la pagina no definida}%
	\ifthenelse{\equal{\compilertype}{lualatex}}{%
		\throwwarning{Funcion no valida en compilador lualatex}%
	}{%
		\clearpage%
		\ifthenelse{\equal{#1}{}}{%
			\newgeometry{left=\pagemarginleft cm, top=\pagemargintop cm, right=\pagemarginright cm, bottom=\pagemarginbottom cm, paperwidth=#2 cm, paperheight=#3 cm}%
		}{%
		\ifthenelse{\equal{#1}{0}}{%
			\newgeometry{left=\pagemarginleft cm, top=\pagemargintop cm, right=\pagemarginright cm, bottom=\pagemarginbottom cm, paperwidth=#2 cm, paperheight=#3 cm}%
		}{%
		\ifthenelse{\equal{#1}{90}}{%
			\newgeometry{left=\pagemarginleft cm, top=\pagemargintop cm, right=\pagemarginright cm, bottom=\pagemarginbottom cm, paperwidth=#3 cm, paperheight=#2 cm}%
		}{%
			\throwbadconfig{Orientacion de pagina no valido}{\changepagesize}{0,90}}}%
		}%
	}%
}

\newcommand{\changepagesizeformat}[2][]{%
	\emptyvarerr{\changepagesizeformat}{#2}{Formato de pagina no definido}%
	\ifthenelse{\equal{#2}{4A0}}{%
		\changepagesize[#1]{168.2}{237.8}%
	}{%
	\ifthenelse{\equal{#2}{2A0}}{%
		\changepagesize[#1]{118.9}{168.2}%
	}{%
	\ifthenelse{\equal{#2}{A0}}{%
		\changepagesize[#1]{84.1}{118.9}%
	}{%
	\ifthenelse{\equal{#2}{A1}}{%
		\changepagesize[#1]{59.4}{84.1}%
	}{%
	\ifthenelse{\equal{#2}{A2}}{%
		\changepagesize[#1]{42.0}{84.1}%
	}{%
	\ifthenelse{\equal{#2}{A3}}{%
		\changepagesize[#1]{29.7}{42.0}%
	}{%
	\ifthenelse{\equal{#2}{A4}}{%
		\changepagesize[#1]{21.0}{29.7}%
	}{%
	\ifthenelse{\equal{#2}{A5}}{%
		\changepagesize[#1]{14.8}{21.0}%
	}{%
	\ifthenelse{\equal{#2}{A6}}{%
		\changepagesize[#1]{10.5}{14.8}%
	}{%
	\ifthenelse{\equal{#2}{letter}}{%
		\changepagesize[#1]{21.59}{27.94}%
	}{%
	\ifthenelse{\equal{#2}{legal}}{%
		\changepagesize[#1]{21.59}{35.6}%
	}{%
	\ifthenelse{\equal{#2}{foolscap}}{%
		\changepagesize[#1]{20.3}{33.0}%
	}{%
	\ifthenelse{\equal{#2}{executive}}{%
		\changepagesize[#1]{18.41}{26.67}%
	}{%
	\ifthenelse{\equal{#2}{ledger}}{%
		\changepagesize[#1]{27.94}{43.18}%
	}{%
	\ifthenelse{\equal{#2}{tabloid}}{%
		\changepagesize[#1]{43.18}{27.94}%
	}{%
	\ifthenelse{\equal{#2}{ANSIC}}{%
		\changepagesize[#1]{55.9}{43.2}%
	}{%
	\ifthenelse{\equal{#2}{ANSID}}{%
		\changepagesize[#1]{86.4}{55.9}%
	}{%
	\ifthenelse{\equal{#2}{ANSIE}}{%
		\changepagesize[#1]{111.8}{86.4}%
	}{%
	\ifthenelse{\equal{#2}{B0}}{%
		\changepagesize[#1]{100}{141.4}%
	}{%
	\ifthenelse{\equal{#2}{B1}}{%
		\changepagesize[#1]{70.7}{100}%
	}{%
	\ifthenelse{\equal{#2}{B2}}{%
		\changepagesize[#1]{50}{70.7}%
	}{%
	\ifthenelse{\equal{#2}{B3}}{%
		\changepagesize[#1]{35.3}{50}%
	}{%
	\ifthenelse{\equal{#2}{B4}}{%
		\changepagesize[#1]{25}{35.3}%
	}{%
	\ifthenelse{\equal{#2}{B5}}{%
		\changepagesize[#1]{17.6}{25}%
	}{%
	\ifthenelse{\equal{#2}{B6}}{%
		\changepagesize[#1]{12.5}{17.6}%
	}{%
		\throwbadconfig{Estilo de pagina no valido}{\changepagesizeformat}{4A0,2A0,A0,A1,A2,A3,A4,A5,A6,letter,legal,foolscap,executive,ledger,tabloid,ANSIC,ANSID,ANSIE,B0,B1,B2,B3,B4,B5,B6}}}}}}}}}}}}}}}}}}}}}}}}}%
	}%
}

\global\def\GLOBALtwocolumnap {l}
\global\def\GLOBALtwocolumnav {t}
\global\def\GLOBALtwocolumnbp {l}
\global\def\GLOBALtwocolumnbv {t}
\global\def\GLOBALthreecolumnap {l}
\global\def\GLOBALthreecolumnav {t}
\global\def\GLOBALthreecolumnbp {l}
\global\def\GLOBALthreecolumnbv {t}
\global\def\GLOBALthreecolumncp {l}
\global\def\GLOBALthreecolumncv {t}

\newcommand{\corecheckcolumnvvalue}[1]{%
	\ifthenelse{\equal{#1}{c}}{}{%
	\ifthenelse{\equal{#1}{t}}{}{%
	\ifthenelse{\equal{#1}{b}}{}{%
		\errmessage{LaTeX Warning: Posicion vertical columna invalido, valores esperados: c,t,b}%
	}}}%
}

\newcommand{\corecheckcolumnpvalue}[1]{%
	\ifthenelse{\equal{#1}{c}}{}{%
	\ifthenelse{\equal{#1}{l}}{}{%
	\ifthenelse{\equal{#1}{r}}{}{%
		\errmessage{LaTeX Warning: Alineacion columna invalida, valores esperados: c,l,r}%
	}}}%
}

\newcommand{\createtwocolumn}[6][]{%
	\setcaptionmargincm{0}%
	\begin{samepage}%
	\begin{flushleft}%
		\vspace{-0.5\baselineskip}%
		\begin{minipage}{1\linewidth}%
			\begin{minipage}[t][#1][\GLOBALtwocolumnav]{#2\linewidth}%
				\ifthenelse{\equal{\GLOBALtwocolumnap}{c}}{%
					\begin{center}#5\end{center}%
				}{%
				\ifthenelse{\equal{\GLOBALtwocolumnap}{l}}{%
					\begin{raggedright}#5\end{raggedright}%
				}{%
				\ifthenelse{\equal{\GLOBALtwocolumnap}{r}}{%
					\hfill\begin{raggedleft}#5\end{raggedleft}%
				}{%
					\errmessage{LaTeX Warning: Alineacion columna izquierda incorrecta, valores esperados: c,l,r}%
				}}}%
			\end{minipage}%
			\hspace{#4 cm}%
			\begin{minipage}[t][#1][\GLOBALtwocolumnbv]{#3\linewidth}%
				\ifthenelse{\equal{\GLOBALtwocolumnbp}{c}}{%
					\begin{center}#6\end{center}%
				}{%
				\ifthenelse{\equal{\GLOBALtwocolumnbp}{l}}{%
					\begin{raggedright}#6\end{raggedright}%
				}{%
				\ifthenelse{\equal{\GLOBALtwocolumnbp}{r}}{%
					\hfill\begin{raggedleft}#6\end{raggedleft}%
				}{%
					\errmessage{LaTeX Warning: Alineacion columna derecha incorrecta, valores esperados: c,l,r}%
				}}}%
			\end{minipage}
		\end{minipage}
	\end{flushleft}
	~ \vspace{-0.5\baselineskip}%
	\end{samepage}
	\setcaptionmargincm{\captionlrmargin}%
}

\newcommand{\createthreecolumn}[9][]{%
	\setcaptionmargincm{0}%
	\begin{samepage}%
	\begin{flushleft}%
		\vspace{-0.5\baselineskip}%
		\begin{minipage}{1\linewidth}%
			\begin{minipage}[t][#1][\GLOBALthreecolumnav]{#2\linewidth}%
				\ifthenelse{\equal{\GLOBALthreecolumnap}{c}}{%
					\begin{center}#7\end{center}%
				}{%
				\ifthenelse{\equal{\GLOBALthreecolumnap}{l}}{%
					\begin{raggedright}#7\end{raggedright}%
				}{%
				\ifthenelse{\equal{\GLOBALthreecolumnap}{r}}{%
					\hfill\begin{raggedleft}#7\end{raggedleft}%
				}{%
					\errmessage{LaTeX Warning: Alineacion columna izquierda incorrecta, valores esperados: c,l,r}%
				}}}%
			\end{minipage}
			\hspace{#5 cm}%
			\begin{minipage}[t][#1][\GLOBALthreecolumnbv]{#3\linewidth}%
				\ifthenelse{\equal{\GLOBALthreecolumnbp}{c}}{%
					\begin{center}#8\end{center}%
				}{%
				\ifthenelse{\equal{\GLOBALthreecolumnbp}{l}}{%
					\begin{raggedright}#8\end{raggedright}%
				}{%
				\ifthenelse{\equal{\GLOBALthreecolumnbp}{r}}{%
					\hfill\begin{raggedleft}#8\end{raggedleft}%
				}{%
					\errmessage{LaTeX Warning: Alineacion columna central incorrecta, valores esperados: c,l,r}%
				}}}%
			\end{minipage}
			\hspace{#6 cm}%
			\begin{minipage}[t][#1][\GLOBALthreecolumncv]{#4\linewidth}%
				\ifthenelse{\equal{\GLOBALthreecolumncp}{c}}{%
					\begin{center}#9\end{center}%
				}{%
				\ifthenelse{\equal{\GLOBALthreecolumncp}{l}}{%
					\begin{raggedright}#9\end{raggedright}%
				}{%
				\ifthenelse{\equal{\GLOBALthreecolumncp}{r}}{%
					\hfill\begin{raggedleft}#9\end{raggedleft}%
				}{%
					\errmessage{LaTeX Warning: Alineacion columna derecha incorrecta, valores esperados: c,l,r}%
				}}}%
			\end{minipage}
		\end{minipage}
	\end{flushleft}
	~ \vspace{-0.5\baselineskip}%
	\end{samepage}
	\setcaptionmargincm{\captionlrmargin}%
}

\def\predocpageromannumber {false}
\def\predocresetpagenumber {false}

\newwrite\fileauthornames
\newwrite\fileauthordata
\newwrite\fileauthornamesdata

\immediate\openout\fileauthornames=\jobname.authn
\immediate\openout\fileauthordata=\jobname.authd
\immediate\openout\fileauthornamesdata=\jobname.authnd
\AtBeginDocument{%
	\immediate\closeout\fileauthornames
	\immediate\closeout\fileauthordata
	\immediate\closeout\fileauthornamesdata
}

\newcounter{authornumber}
\ifthenelse{\equal{\titlestyle}{style1}}{%
	\def\LOCALhasname {false}
	\newcommand{\titleauthorname}[2][]{%
		\ifthenelse{\equal{\LOCALhasname}{true}}{\hspace{\titleauthorspacing cm}}{}%
		\normalsize{#2} \ifthenelse{\equal{#1}{}}{}{\textsuperscript{(#1)}}%
		\def\LOCALhasname {true}%
	}%
	\newcommand{\titleauthordata}[4][]{%
		\ifthenelse{\equal{\LOCALhasname}{true}}{\vspace{0.025cm}\\}{}%
		\indent \footnotesize{\textsuperscript{(#1)} #2%
			\ifthenelse{\equal{#3}{}}{}{, \insertemail{#3}}%
			\ifthenelse{\equal{#4}{}}{}{, ORCID \href{https://orcid.org/#4}{#4}}}%
		\def\LOCALhasname {true}%
	}%
	\newcommand{\inserttitle}{%
		\begin{center}%
			\ifthenelse{\equal{\titlebold}{true}}{%
				\textbf{\Large{\documenttitle}} \\%
			}{%
				\Large{\documenttitle} \\%
			}%
			\vspace{0.6cm}%
			\begin{minipage}[t]{\titleauthormaxwidth\textwidth}
				\centering
				\input{\jobname.authn}%
			\end{minipage}
			\vspace{\titleauthormarginbottom cm}%
			\def\LOCALhasname {false}%
			~ \\%
			\begin{minipage}[t]{\titleauthormaxwidth\textwidth}%
				\centering%
				\input{\jobname.authd}%
			\end{minipage}
		\end{center}
		\vspace{0.05cm}%
		\normalsize%
	}%
}{%
\ifthenelse{\equal{\titlestyle}{style2}}{%
	\def\LOCALhasname {false}
	\newcommand{\titleauthorname}[2][]{%
		\ifthenelse{\equal{\LOCALhasname}{true}}{\hspace{\titleauthorspacing cm}}{}%
		\normalsize{#2} \ifthenelse{\equal{#1}{}}{}{\textsuperscript{(#1)}}%
		\def\LOCALhasname {true}%
	}%
	\newcommand{\titleauthordata}[4][]{%
		\ifthenelse{\equal{\LOCALhasname}{true}}{\vspace{0.025cm}\\}{}%
		\indent \footnotesize{\textsuperscript{(#1)} #2%
		\ifthenelse{\equal{#3}{}}{}{, \insertemail{#3}}%
		\ifthenelse{\equal{#4}{}}{}{, ORCID \href{https://orcid.org/#4}{#4}}}%
		\def\LOCALhasname {true}%
		\vspace{-0.05cm}%
	}%
	\newcommand{\inserttitle}{%
		\begin{center}%
			\ifthenelse{\equal{\titlebold}{true}}{%
				\textbf{\Large{\documenttitle}} \\%
			}{%
				\Large{\documenttitle} \\%
			}%
			\vspace{0.7cm}%
			\input{\jobname.authn}%
			\vspace{\titleauthormarginbottom cm}%
		\end{center}
		\par%
		\def\LOCALhasname {false}%
		\begin{minipage}[t]{\titleauthormaxwidth\textwidth}%
			\input{\jobname.authd}%
		\end{minipage}
		\vspace{0.3cm}%
		\normalsize%
	}%
}{%
\ifthenelse{\equal{\titlestyle}{style3}}{%
	\def\LOCALhasname {false}
	\newcommand{\titleauthorname}[2][]{%
		\ifthenelse{\equal{\LOCALhasname}{true}}{,\ }{}%
		\normalsize{#2}%
		\def\LOCALhasname {true}%
	}%
	\newcommand{\inserttitle}{%
		\begin{center}%
			\ifthenelse{\equal{\titlebold}{true}}{%
				\textbf{\Large{\documenttitle}} \\%
			}{%
				\Large{\documenttitle} \\%
			}%
			\vspace{0.7cm}%
			\begin{minipage}[t]{\titleauthormaxwidth\textwidth}%
				\centering%
				\input{\jobname.authn}%
			\end{minipage}
			\vspace{\titleauthormarginbottom cm}%
			\def\LOCALhasname {false}%
		\end{center}
		\normalsize%
	}%
}{%
\ifthenelse{\equal{\titlestyle}{style4}}{%
	\def\LOCALhasname {false}
	\newcommand{\titleauthorname}[2][]{%
		\ifthenelse{\equal{\LOCALhasname}{true}}{,\ }{}%
		\normalsize{#2}%
		\def\LOCALhasname {true}%
	}%
	\newcommand{\inserttitle}{%
		\begin{center}%
			\ifthenelse{\equal{\titlebold}{true}}{%
				\textbf{\Large{\documenttitle}} \\%
			}{%
				\Large{\documenttitle} \\%
			}%
			\vspace{0.7cm}%
			\begin{minipage}[t]{\titleauthormaxwidth\textwidth}%
				\centering%
				\input{\jobname.authn}%
			\end{minipage}
			\vspace{\titleauthormarginbottom cm}%
			\def\LOCALhasname {false}%
			~ \\%
			\small{\documentdate} \\
			\vspace{0.3cm}%
		\end{center}%
		\normalsize%
	}%
}{%
\ifthenelse{\equal{\titlestyle}{style5}}{%
	\def\LOCALhasname {false}
	\newcommand{\titleauthornamedata}[5][]{%
		\ifthenelse{\equal{\LOCALhasname}{true}}{\hspace{\titleauthorspacing cm}}{}%
		\begin{tabular}[t]{C{4cm}}%
			\normalsize{#2} \\%
			\small{#3} \\%
			\ifthenelse{\equal{#4}{}}{}{\small{\insertemail{#4}} \\}%
			\ifthenelse{\equal{#5}{}}{}{\small{ORCID \href{https://orcid.org/#5}{#5}}}%
		\end{tabular}
		\def\LOCALhasname {true}%
	}%
	\newcommand{\inserttitle}{%
		\begin{center}%
			\ifthenelse{\equal{\titlebold}{true}}{%
				\textbf{\Large{\documenttitle}} \\%
			}{%
				\Large{\documenttitle} \\%
			}%
			\vspace{0.9cm}%
			\noindent%
			\settablecellpadding{0}{1.5}%
			\linespread{0.6}\selectfont{%
				\begin{minipage}[t]{\titleauthormaxwidth\textwidth}%
					\centering%
					\input{\jobname.authnd}%
				\end{minipage}
			}%
			\resettablecellpadding%
			\def\LOCALhasname {false}%
		\end{center}
		\normalsize%
	}%
}{%
	\throwbadconfig{Valor configuracion incorrecto}{\titlestyle}{style1 .. style5}}}}}
}

\BeforeBeginEnvironment{abstract}{%
	\vspace{\abstractmargintop cm}%
}
\AfterEndEnvironment{abstract}{%
	\vspace{\abstractmarginbottom cm}%
}

\newenvironment{keywords}{%
	\begin{small}%
		\textbf{Keywords:} %
		\begin{em}%
		}{%
		\end{em}%
	\end{small}%
	\normalfont\selectfont%
	\vspace{0.3cm}%
}

	{%
	\end{thebibliography}
	\endgroup 
}

\newcommand{\coreinitsourcecodep}[4]{%
	\emptyvarerr{\coreinitsourcecodep}{#2}{Estilo de codigo no definido}%
	\checkvalidsourcecodestyle{#2}%
	\ifthenelse{\equal{\showlinenumbers}{true}}{%
		\rightlinenumbers}{%
	}%
	\lstset{%
		backgroundcolor=\color{\sourcecodebgcolor}
	}%
	\ifthenelse{\equal{\codecaptiontop}{true}}{%
		\ifx\hfuzz#4\hfuzz%
			\ifx\hfuzz#3\hfuzz%
				\lstset{%
					escapeinside={(*@}{@*)},
					style=#2
				}%
			\else%
				\lstset{%
					escapeinside={(*@}{@*)},
					style=#2,
					#3
				}%
			\fi%
		\else%
			\ifx\hfuzz#3\hfuzz%
				\lstset{%
					caption={#4 #1},
					captionpos=t,
					escapeinside={(*@}{@*)},
					style=#2
				}%
			\else%
				\lstset{%
					caption={#4 #1},
					captionpos=t,
					escapeinside={(*@}{@*)},
					style=#2,
					#3
				}%
			\fi%
		\fi%
	}{%
		\ifx\hfuzz#4\hfuzz%
			\ifx\hfuzz#3\hfuzz%
				\lstset{%
					escapeinside={(*@}{@*)},
					style=#2
				}%
			\else%
				\lstset{%
					escapeinside={(*@}{@*)},
					style=#2,
					#3
				}%
			\fi%
		\else%
			\ifx\hfuzz#3\hfuzz%
				\lstset{%
					caption={#4 #1},
					captionpos=b,
					style=#2
				}%
			\else%
				\lstset{%
					caption={#4 #1},
					captionpos=b,
					escapeinside={(*@}{@*)},
					style=#2,
					#3
				}%
			\fi%
		\fi%
	}%
}

\lstnewenvironment{sourcecodep}[4][]{%
	\coreinitsourcecodep{#1}{#2}{#3}{#4}%
}{%
	\ifthenelse{\equal{\showlinenumbers}{true}}{%
		\leftlinenumbers}{%
	}%
}

\newcommand{\importsourcecodep}[5][]{%
	\coreinitsourcecodep{#1}{#2}{#3}{#5}%
	\inputlisting{#4}%
	\ifthenelse{\equal{\showlinenumbers}{true}}{%
		\leftlinenumbers}{%
	}%
}

\newcommand{\coreinitsourcecode}[3]{%
	\emptyvarerr{\coreinitsourcecode}{#2}{Estilo de codigo no definido}%
	\checkvalidsourcecodestyle{#2}%
	\ifthenelse{\equal{\showlinenumbers}{true}}{%
		\rightlinenumbers}{%
	}%
	\lstset{%
		backgroundcolor=\color{\sourcecodebgcolor}
	}%
	\ifthenelse{\equal{\codecaptiontop}{true}}{%
		\ifx\hfuzz#3\hfuzz%
			\lstset{%
				escapeinside={(*@}{@*)},
				style=#2
			}%
		\else%
			\lstset{%
				escapeinside={(*@}{@*)},
				caption={#3 #1},
				captionpos=t,
				style=#2
			}%
		\fi%
	}{%
		\ifx\hfuzz#3\hfuzz%
			\lstset{%
				escapeinside={(*@}{@*)},
				style=#2
			}%
		\else%
			\lstset{%
				escapeinside={(*@}{@*)},
				caption={#3 #1},
				captionpos=b,
				style=#2
			}%
		\fi%
	}%
}

\lstnewenvironment{sourcecode}[3][]{%
	\coreinitsourcecode{#1}{#2}{#3}%
}{%
	\ifthenelse{\equal{\showlinenumbers}{true}}{%
		\leftlinenumbers}{%
	}%
}

\newcommand{\importsourcecode}[4][]{%
	\coreinitsourcecode{#1}{#2}{#4}%
	\lstinputlisting{#3}%
	\ifthenelse{\equal{\showlinenumbers}{true}}{%
		\leftlinenumbers}{%
	}%
}

\newcommand\blfootnote[1]{%
  \begingroup
  \renewcommand\thefootnote{}\footnote{#1}%
  \addtocounter{footnote}{-1}%
  \endgroup
}

\newenvironment{images}[2][]{%
	\def\envimageslabelvar {#1}%
	\def\envimagescaptioncf {false}%
	\def\envimagescaptionvar {#2}%
	\global\def\GLOBALenvimageadded {false}%
	\global\def\GLOBALenvimageinitialized {true}%
	\corevspacevarcm{\marginimagetop}%
	\setcaptionmargincm{\captionmarginmultimg} 
	\begin{samepage}%
	\begin{figure}[H] \centering%
		\ifthenelse{\equal{\GLOBALenvimagecf}{true}}{%
			\ContinuedFloat%
			\global\def\GLOBALenvimagecf {false}%
			\def\envimagescaptioncf {true}%
		}{}%
		\corevspacevarcm{\marginimagemulttop}%
		}{%
		\setcaptionmargincm{\captionlrmargin}%
		\ifthenelse{\equal{\envimagescaptionvar}{}}{%
			\corevspacevarcm{\captionlessmarginimage}%
		}{%
			\corevspacevarcm{\captionmarginimages}%
			\ifthenelse{\equal{\envimagescaptioncf}{true}}{%
				\caption[]{\envimagescaptionvar\envimageslabelvar}%
			}{%
				\caption{\envimagescaptionvar\envimageslabelvar}%
			}%
		}%
	\end{figure}%
	\setcaptionmargincm{\captionlrmargin}%
	\corevspacevarcm{\marginimagebottom}%
	\end{samepage}
	\global\def\GLOBALenvimageinitialized {false}%
}

\newenvironment{imagesmc}[3][]{%
	\def\envimageslabelvar {#1}%
	\def\envimagesmcpos {#2}%
	\def\envimagescaptioncf {false}%
	\def\envimagescaptionvar {#3}%
	\global\def\GLOBALenvimageadded {false}%
	\global\def\GLOBALenvimageinitialized {true}%
	\checkinsidemulticol%
	\checkoutsideappendix%
	\setcaptionmargincm{\captionmarginmultimg} 
	\ifthenelse{\equal{#2}{bottom}}{%
		\begin{figure*}[!b] \centering%
	}{%
	\ifthenelse{\equal{#2}{top}}{%
		\begin{figure*}[!t] \centering%
	}{%
		\errmessage{LaTeX Warning: Posicion de imagen invalida, valores esperados: bottom,top}
		\stop
	}}%
		\ifthenelse{\equal{\GLOBALenvimagecf}{true}}{%
			\ContinuedFloat%
			\global\def\GLOBALenvimagecf {false}%
			\def\envimagescaptioncf {true}%
		}{}%
		\corevspacevarcm{\marginimagemulttop}%
	}{%
		\setcaptionmargincm{\captionlrmargin}%
		\ifthenelse{\equal{\envimagescaptionvar}{}}{%
			\corevspacevarcm{\captionlessmarginimage}%
		}{%
			\corevspacevarcm{\captionmarginimagesmc}%
			\ifthenelse{\equal{\envimagescaptioncf}{true}}{%
				\caption[]{\envimagescaptionvar\envimageslabelvar}%
			}{%
				\caption{\envimagescaptionvar\envimageslabelvar}%
			}%
		}%
	\end{figure*}%
	\setcaptionmargincm{\captionlrmarginmc}%
	\global\def\GLOBALenvimageinitialized {false}%
}

\newenvironment{equationindex}[2][]{%
	\def\coreequationindexcaption {#2}%
	\emptyvarerr{\coreequationindexcaption}{#2}{Leyenda no definida}%
	\corevspacevarcm{\margineqnindextop}%
	\begin{samepage}%
		\begin{equation}%
			\text{#1}%
		}{%
		\end{equation}
		\myindexequations{\coreequationindexcaption}%
		\corevspacevarcm{\margineqnindexbottom}%
	\end{samepage}
	\coreinsertequationcaption{\textit{\coreequationindexcaption}}%
	\addtocounter{templateIndexEquations}{1}%
	\coreafterequationfn%
}

\lstdefinestyle{abap}{
	language=ABAP
}

\lstdefinestyle{ada}{
	language=[2005]Ada
}

\lstdefinelanguage[x64]{Assembler}[x86masm]{Assembler}{
	morekeywords={
		CDQE,CMPSQ,CMPXCHG16B,CQO,IRETQ,JRCXZ,LODSQ,MOVSXD,POPFQ,PUSHFQ,r8,r8b,r8d,r8w,r9,r9b,r9d,r9w,r10,r10b,r10d,r10w,r11,r11b,r11d,r11w,r12,r12b,r12d,r12w,r13,r13b,r13d,r13w,r14,r14b,r14d,r14w,r15,r15b,r15d,r15w,rax,rbp,rbx,rcx,rdi,RDTSCP,rdx,rsi,rsp,SCASQ,STOSQ,SWAPGS
	}
}
\lstdefinestyle{assemblerx64}{
	language=[x64]Assembler
}
\lstdefinestyle{assemblerx86}{
	language=[x86masm]Assembler
}

\lstdefinestyle{awk}{
	language=[gnu]Awk
}

\lstdefinestyle{bash}{
	language=bash,
	breakatwhitespace=false,
	morecomment=[l]{rem},
	morecomment=[s]{::}{::},
	morekeywords={
		call,cp,dig,gcc,git,grep,ls,mv,python,rm,sudo,vim
	},
	sensitive=false
}

\lstdefinestyle{basic}{
	language=[Visual]Basic
}

\lstdefinestyle{c}{
	language=C,
	breakatwhitespace=false,
	keepspaces=true
}

\lstdefinestyle{caml}{
	language=[light]Caml
}

\lstdefinestyle{cmake}{
	language=[gnu] make,
	keywordstyle=[2]\color{dkcyan},
	morekeywords=[1]{
		add_custom_command,add_custom_target,add_definitions,add_executable,add_library,add_subdirectory,cmake_minimum_required,cmake_policy,configure_file,cuda_add_library,cuda_include_directories,else,elseif,endforeach,endfunction,endif,endmacro,execute_process,file,find_library,find_package,find_path,find_program,foreach,function,get_directory_property,get_filename_component,get_filename_component,get_source_file_property,get_target_property,if,include,include_directories,install,link_directories,list,macro,mark_as_advanced,message,option,PKG_CHECK_MODULES,project,set,SET_CHECK_CXX_FLAGS,set_property,set_source_files_properties,set_target_properties,string,target_compile_options,target_include_directories,target_link_libraries,unset
	},
	morekeywords=[2]{
		AND,APPEND,APPLE,ARCHIVE,CACHE,CMAKE_CURRENT_LIST_DIR,CMAKE_CXX_STANDARD,CMAKE_MODULE_PATH,CMAKE_SYSTEM_NAME,COMMAND,COMMENT,COMPILE_DEFINITIONS,CONFIG,DEFINED,DEPENDS,DESTINATION,DIRECTORY,ENDIF,ENV,EQUAL,ERROR_QUIET,EXISTS,FATAL_ERROR,FILES,FILES_MATCHING,FIND,FIND,FIND_LIBRARY,FORCE,GLOB,GREATER,IF,INCLUDE_DIRECTORIES,IS_ABSOLUTE,LESS,LIBRARY,LINK_PRIVATE,LIST,MAIN_DEPENDENCY,MAKE_DIRECTORY,MARK_AS_ADVANCED,MATCHALL,MATCHES,NOT,OBJECT,OFF,ON,OPTIONAL,OR,OUTPUT,OUTPUT_STRIP_TRAILING_WHITESPACE,OUTPUT_VARIABLE,PARENT_SCOPE,PATTERN,PRE_BUILD_COMMAND,PRE_LINK,PRIVATE,PROJECT_NAME,PROPERTIES,PROPERTY,PUBLIC,REGEX,RELEASE,RENAME,REQUIRED,RUNTIME,SET,STATIC,STREQUAL,SYSTEM,TARGET,TARGETS,TOUPPER,UNIX,VERSION,VERSION_EQUAL,VERSION_LESS,WIN32,WORKING_DIRECTORY
	}
}

\lstdefinestyle{cobol}{
	language=Cobol
}

\lstdefinestyle{cpp}{
	language=C++,
	breakatwhitespace=false,
	morekeywords={NULL}
}

\lstdefinestyle{csharp}{
	language=csh,
	morecomment=[l]{//},
	morecomment=[s]{/*}{*/},
	morekeywords={
		abstract,as,base,bool,break,byte,case,catch,char,checked,class,const,continue,decimal,default,delegate,do,double,else,enum,event,explicit,extern,false,finally,fixed,float,for,foreach,goto,if,implicit,in,int,interface,internal,is,lock,long,namespace,new,null,object,operator,out,override,params,private,protected,public,readonly,ref,return,sbyte,sealed,short,sizeof,stackalloc,static,string,struct,switch,this,throw,true,try,typeof,uint,ulong,unchecked,unsafe,ushort,using,virtual,void,volatile,while
	}
}

\lstdefinelanguage{CSS}{
	morecomment=[s]{/*}{*/},
	morekeywords={
		-moz-binding,-moz-border-bottom-colors,-moz-border-left-colors,-moz-border-radius,-moz-border-radius-bottomleft,-moz-border-radius-bottomright,-moz-border-radius-topleft,-moz-border-radius-topright,-moz-border-right-colors,-moz-border-top-colors,-moz-opacity,-moz-outline,-moz-outline-color,-moz-outline-style,-moz-outline-width,-moz-user-focus,-moz-user-input,-moz-user-modify,-moz-user-select,-replace,-set-link-source,-use-link-source,accelerator,azimuth,background,background-attachment,background-color,background-image,background-position,background-position-x,background-position-y,background-repeat,behavior,border,border-bottom,border-bottom-color,border-bottom-style,border-bottom-width,border-collapse,border-color,border-left,border-left-color,border-left-style,border-left-width,border-right,border-right-color,border-right-style,border-right-width,border-spacing,border-style,border-top,border-top-color,border-top-style,border-top-width,border-width,bottom,caption-side,clear,clip,color,content,counter-increment,counter-reset,cue,cue-after,cue-before,cursor,direction,display,elevation,empty-cells,filter,float,font,font-family,font-size,font-size-adjust,font-stretch,font-style,font-variant,font-weight,height,ime-mode,include-source,layer-background-color,layer-background-image,layout-flow,layout-grid,layout-grid-char,layout-grid-char-spacing,layout-grid-line,layout-grid-mode,layout-grid-type,left,letter-spacing,line-break,line-height,list-style,list-style-image,list-style-position,list-style-type,margin,margin-bottom,margin-left,margin-right,margin-top,marker-offset,marks,max-height,max-width,min-height,min-width,orphans,outline,outline-color,outline-style,outline-width,overflow,overflow-X,overflow-Y,padding,padding-bottom,padding-left,padding-right,padding-top,page,page-break-after,page-break-before,page-break-inside,pause,pause-after,pause-before,pitch,pitch-range,play-during,position,quotes,richness,right,ruby-align,ruby-overhang,ruby-position,scrollbar-3d-light-color,scrollbar-arrow-color,scrollbar-base-color,scrollbar-dark-shadow-color,scrollbar-face-color,scrollbar-highlight-color,scrollbar-shadow-color,scrollbar-track-color,size,speak,speak-header,speak-numeral,speak-punctuation,speech-rate,stress,table-layout,text-align,text-align-last,text-autospace,text-decoration,text-indent,text-justify,text-kashida-space,text-overflow,text-shadow,text-transform,text-underline-position,top,unicode-bidi,vertical-align,visibility,voice-family,volume,white-space,widows,width,word-break,word-spacing,word-wrap,writing-mode,z-index,zoom
	},
	morestring=[s]{:}{;},
	sensitive=true
}
\lstdefinestyle{css}{
	language=CSS,
	breakatwhitespace=true
}

\lstdefinestyle{csv}{
	language={}
}

\lstdefinestyle{cuda}{
	language=C++,
	breakatwhitespace=false,
	emph={
		cudaFree,cudaMalloc,__device__,__global__,__host__,__shared__,__syncthreads
	},
	emphstyle=\color{dkcyan}\ttfamily,
	morecomment=[l][\color{magenta}]{\#},
	moredelim=[s][\ttfamily]{<<<}{>>>}
}

\lstdefinestyle{dart}{
	language=Java,
	emph=[2]{
		findAllElements,findElements
	},
	morekeywords={
		*,get,library,List,num,set,String,var
	}
}

\lstdefinelanguage{docker}{
	comment=[l]{\#},
	keywords={
		ADD,CMD,COPY,ENTRYPOINT,ENV,EXPOSE,FROM,LABEL,MAINTAINER,ONBUILD,RUN,STOPSIGNAL,USER,VOLUME,WORKDIR
	},
	morestring=[b]',
	morestring=[b]"
}
\lstdefinestyle{docker}{
	language=docker,
	breakatwhitespace=true
}

\lstdefinestyle{elisp}{
	language=elisp
}

\lstdefinestyle{elixir}{
	morekeywords={
		case,catch,def,do,else,false,use,alias,receive,timeout,defmacro,defp,for,if,import,defmodule,defprotocol,nil,defmacrop,defoverridable,defimpl,super,fn,raise,true,try,end,with,unless
	},
	otherkeywords={
		<-,->, |>, \%\{, \}, \{, \, (, )
	},
	morecomment=[l]{\#},
	morecomment=[n]{/*}{*/},
	morecomment=[s][\color{purple}]{:}{\ },
	morestring=[s][\color{mauve}]"",
	sensitive=true
}

\lstdefinestyle{erlang}{
	language=erlang
}

\lstdefinestyle{fortran}{
	language=[95]Fortran,
	breakatwhitespace=false
}

\lstdefinestyle{fsharp}{
	morecomment=[l][\color{dkgreen}]{///},
	morecomment=[l][\color{dkgreen}]{//},
	morecomment=[s][\color{dkgreen}]{{(*}{*)}},
	morestring=[b]",
	morekeywords={
		abstract,and,Application,Array,Async,async,begin,cloud,do,else,end,false,finally,for,fun,function,if,in,inherit,interface,let,List,match,member,module,mutable,namespace,new,of,open,rec,return,Seq,static,System,then,true,try,type,use,while,with,yield
	},
	otherkeywords={
		by,do!,from,let!,order,return!,select,use!,var,where,yield!
	},
	sensitive=true
}

\lstdefinelanguage{GLSL}{
	alsoletter={\#},
	morekeywords=[1]{
		attribute,bool,break,bvec2,bvec3,bvec4,case,centroid,const,continue,default,discard,do,else,false,flat,float,for,highp,if,in,inout,int,invariant,isampler1D,isampler1DArray,isampler2D,isampler2DArray,isampler2DMS,isampler2DMSArray,isampler2DRect,isampler3D,isamplerBuffer,isamplerCube,ivec2,ivec3,ivec4,layout,lowp,mat2,mat2x2,mat2x3,mat2x4,mat3,mat3x2,mat3x3,mat3x4,mat4,mat4x2,mat4x3,mat4x4,mediump,noperspective,out,precision,return,sampler1D,sampler1DArray,sampler1DArrayShadow,sampler1DShadow,sampler2D,sampler2DArray,sampler2DArrayShadow,sampler2DMS,sampler2DMSArray,sampler2DRect,sampler2DRectShadow,sampler2DShadow,sampler3D,samplerBuffer,samplerCube,samplerCubeShadow,smooth,struct,switch,true,uint,uniform,usampler1D,usampler1DArray,usampler2D,usampler2DArray,usampler2DMS,usampler2DMSArray,usampler2DRect,usampler3D,usamplerBuffer,usamplerCube,uvec2,uvec3,uvec4,varying,vec2,vec3,vec4,void,while
	},
	morekeywords=[2]{
		abs,acos,acosh,all,any,asin,asinh,atan,atan,atanh,ceil,clamp,cos,cosh,cross,degrees,determinant,dFdx,dFdy,distance,dot,EmitVertex,EndPrimitive,equal,exp,exp2,faceforward,floatBitsToInt,floatBitsToUint,floor,fract,fwidth,greaterThan,greaterThanEqual,intBitsToFloat,inverse,inversesqrt,isinf,isnan,length,lessThan,lessThanEqual,log,log2,matrixCompMult,max,min,mix,mod,modf,noise1,noise2,noise3,noise4,normalize,not,notEqual,outerProduct,pow,radians,reflect,refract,round,roundEven,shadow1D,shadow1DLod,shadow1DProj,shadow1DProjLod,shadow2D,shadow2DLod,shadow2DProj,shadow2DProjLod,sign,sin,sinh,smoothstep,sqrt,step,tan,tanh,texelFetch,texelFetchOffset,texture,texture1D,texture1DProj,texture1DProjLod,texture2D,texture2DLod,texture2DProj,texture2DProjLod,texture3D,texture3DLod,texture3DProj,texture3DProjLod,textureCube,textureCubeLod,textureGrad,textureGradOffset,textureLod,textureLodOffset,textureOffset,textureProj,textureProjGrad,textureProjGradOffset,textureProjLod,textureProjLodOffset,textureProjOffset,textureSize,transpose,trunc,uintBitsToFloat
	},
	morekeywords=[3]{
		\#version,core,gl_ClipDistance,gl_ClipDistance,gl_ClipVertex,gl_DepthRange,gl_FragColor,gl_FragCoord,gl_FragData,gl_FragDepth,gl_FrontFacing,gl_InstanceID,gl_Layer,gl_MaxClipDistances,gl_MaxCombinedTextureImageUnits,gl_MaxDrawBuffers,gl_MaxDrawBuffers,gl_MaxFragmentInputComponents,gl_MaxFragmentUniformComponents,gl_MaxGeometryInputComponents,gl_MaxGeometryOutputComponents,gl_MaxGeometryOutputVertices,gl_MaxGeometryOutputVertices,gl_MaxGeometryTextureImageUnits,gl_MaxGeometryTotalOutputComponents,gl_MaxGeometryUniformComponents,gl_MaxGeometryVaryingComponents,gl_MaxTextureImageUnits,gl_MaxVaryingComponents,gl_MaxVaryingFloats,gl_MaxVertexAttribs,gl_MaxVertexOutputComponents,gl_MaxVertexTextureImageUnits,gl_MaxVertexUniformComponents,gl_PerVertex,gl_PointCoord,gl_PointSize,gl_Position,gl_PrimitiveID,gl_VertexID
	},
	morecomment=[l]{//},
	morecomment=[s]{/*}{*/}
}
\lstdefinestyle{glsl}{
	language=GLSL,
	keywordstyle=[3]\color{dkcyan}\ttfamily,
	prebreak=\raisebox{0ex}[0ex][0ex]{\ensuremath{\hookleftarrow}},
	sensitive=true,
	upquote=true
}

\lstdefinestyle{gnuplot}{
	language=Gnuplot
}

\lstdefinestyle{go}{
	language=Go
}

\lstdefinestyle{haskell}{
	language=haskell,
	morecomment=[l]\%
}

\lstdefinelanguage{HTML5}{
	language=html,
	alsoletter={<>=-},
	morecomment=[s]{<!--}{-->},
	ndkeywords={
		=,
		accept-charset=,accept=,accesskey=,action=,align=,alt=,async=,autocomplete=,autofocus=,autoplay=,autosave=,bgcolor=,border=,buffered=,challenge=,charset=,checked=,cite=,class=,code=,codebase=,color=,cols=,colspan=,content=,contenteditable=,contextmenu=,controls=,coords=,data=,datetime=,default=,defer=,dir=,dirname=,disabled=,download=,draggable=,dropzone=,enctype=,for=,form=,formaction=,headers=,height=,hidden=,high=,href=,hreflang=,http-equiv=,icon=,id=,ismap=,itemprop=,keytype=,kind=,label=,lang=,language=,list=,loop=,low=,manifest=,max=,maxlength=,media=,method=,min=,multiple=,name=,novalidate=,open=,optimum=,pattern=,ping=,placeholder=,poster=,preload=,pubdate=,radiogroup=,readonly=,rel=,required=,reversed=,rows=,rowspan=,sandbox=,scope=,scoped=,seamless=,selected=,shape=,size=,sizes=,span=,spellcheck=,src=,srcdoc=,srclang=,start=,step=,style=,summary=,tabindex=,target=,title=,type=,usemap=,value=,width=,wrap=,
		-moz-binding:,-moz-border-bottom-colors:,-moz-border-left-colors:,-moz-border-radius-bottomleft:,-moz-border-radius-bottomright:,-moz-border-radius-topleft:,-moz-border-radius-topright:,-moz-border-radius:,-moz-border-right-colors:,-moz-border-top-colors:,-moz-opacity:,-moz-outline-color:,-moz-outline-style:,-moz-outline-width:,-moz-outline:,-moz-transform:,-moz-user-focus:,-moz-user-input:,-moz-user-modify:,-moz-user-select:,-replace:,-set-link-source:,-use-link-source:,accelerator:,azimuth:,background-attachment:,background-color:,background-image:,background-position-x:,background-position-y:,background-position:,background-repeat:,background:,behavior:,border-bottom-color:,border-bottom-style:,border-bottom-width:,border-bottom:,border-collapse:,border-color:,border-left-color:,border-left-style:,border-left-width:,border-left:,border-right-color:,border-right-style:,border-right-width:,border-right:,border-spacing:,border-style:,border-top-color:,border-top-style:,border-top-width:,border-top:,border-width:,border:,bottom:,caption-side:,clear:,clip:,color:,content:,counter-increment:,counter-reset:,cue-after:,cue-before:,cue:,cursor:,direction:,display:,elevation:,empty-cells:,filter:,float:,font-family:,font-size-adjust:,font-size:,font-stretch:,font-style:,font-variant:,font-weight:,font:,height:,ime-mode:,include-source:,layer-background-color:,layer-background-image:,layout-flow:,layout-grid-char-spacing:,layout-grid-char:,layout-grid-line:,layout-grid-mode:,layout-grid-type:,layout-grid:,left:,letter-spacing:,line-break:,line-height:,list-style-image:,list-style-position:,list-style-type:,list-style:,margin-bottom:,margin-left:,margin-right:,margin-top:,margin:,marker-offset:,marks:,max-height:,max-width:,min-height:,min-width:,orphans:,outline-color:,outline-style:,outline-width:,outline:,overflow-X:,overflow-Y:,overflow:,padding-bottom:,padding-left:,padding-right:,padding-top:,padding:,page-break-after:,page-break-before:,page-break-inside:,page:,pause-after:,pause-before:,pause:,pitch-range:,pitch:,play-during:,position:,quotes:,richness:,right:,ruby-align:,ruby-overhang:,ruby-position:,scrollbar-3d-light-color:,scrollbar-arrow-color:,scrollbar-base-color:,scrollbar-dark-shadow-color:,scrollbar-face-color:,scrollbar-highlight-color:,scrollbar-shadow-color:,scrollbar-track-color:,size:,speak-header:,speak-numeral:,speak-punctuation:,speak:,speech-rate:,stress:,table-layout:,text-align-last:,text-align:,text-autospace:,text-decoration:,text-indent:,text-justify:,text-kashida-space:,text-overflow:,text-shadow:,text-transform:,text-underline-position:,top:,transform:,transition-duration:,transition-property:,transition-timing-function:,unicode-bidi:,vertical-align:,visibility:,voice-family:,volume:,white-space:,widows:,width:,word-break:,word-spacing:,word-wrap:,writing-mode:,z-index:,zoom:
	},
	otherkeywords={
		<,</,>,</a,<a,</a>,</abbr,<abbr,</abbr>,</address,<address,</address>,</area,<area,</area>,</area,<area,</area>,</article,<article,</article>,</aside,<aside,</aside>,</audio,<audio,</audio>,</audio,<audio,</audio>,</b,<b,</b>,</base,<base,</base>,</bdi,<bdi,</bdi>,</bdo,<bdo,</bdo>,</blockquote,<blockquote,</blockquote>,</body,<body,</body>,</br,<br,</br>,</button,<button,</button>,</canvas,<canvas,</canvas>,</caption,<caption,</caption>,</cite,<cite,</cite>,</code,<code,</code>,</col,<col,</col>,</colgroup,<colgroup,</colgroup>,</data,<data,</data>,</datalist,<datalist,</datalist>,</dd,<dd,</dd>,</del,<del,</del>,</details,<details,</details>,</dfn,<dfn,</dfn>,</div,<div,</div>,</dl,<dl,</dl>,</dt,<dt,</dt>,</em,<em,</em>,</embed,<embed,</embed>,</fieldset,<fieldset,</fieldset>,</figcaption,<figcaption,</figcaption>,</figure,<figure,</figure>,</footer,<footer,</footer>,</form,<form,</form>,</h1,<h1,</h1>,</h2,<h2,</h2>,</h3,<h3,</h3>,</h4,<h4,</h4>,</h5,<h5,</h5>,</h6,<h6,</h6>,</head,<head,</head>,</header,<header,</header>,</hr,<hr,</hr>,</html,<html,</html>,</i,<i,</i>,</iframe,<iframe,</iframe>,</img,<img,</img>,</input,<input,</input>,</ins,<ins,</ins>,</kbd,<kbd,</kbd>,</keygen,<keygen,</keygen>,</label,<label,</label>,</legend,<legend,</legend>,</li,<li,</li>,</link,<link,</link>,</main,<main,</main>,</map,<map,</map>,</mark,<mark,</mark>,</math,<math,</math>,</menu,<menu,</menu>,</menuitem,<menuitem,</menuitem>,</meta,<meta,</meta>,</meter,<meter,</meter>,</nav,<nav,</nav>,</noscript,<noscript,</noscript>,</object,<object,</object>,</ol,<ol,</ol>,</optgroup,<optgroup,</optgroup>,</option,<option,</option>,</output,<output,</output>,</p,<p,</p>,</param,<param,</param>,</pre,<pre,</pre>,</progress,<progress,</progress>,</q,<q,</q>,</rp,<rp,</rp>,</rt,<rt,</rt>,</ruby,<ruby,</ruby>,</s,<s,</s>,</samp,<samp,</samp>,</script,<script,</script>,</section,<section,</section>,</select,<select,</select>,</small,<small,</small>,</source,<source,</source>,</span,<span,</span>,</strong,<strong,</strong>,</style,<style,</style>,</summary,<summary,</summary>,</sup,<sup,</sup>,</svg,<svg,</svg>,</table,<table,</table>,</tbody,<tbody,</tbody>,</td,<td,</td>,</template,<template,</template>,</textarea,<textarea,</textarea>,</tfoot,<tfoot,</tfoot>,</th,<th,</th>,</thead,<thead,</thead>,</time,<time,</time>,</title,<title,</title>,</tr,<tr,</tr>,</track,<track,</track>,</u,<u,</u>,</ul,<ul,</ul>,</var,<var,</var>,</video,<video,</video>,</wbr,<wbr,</wbr>,/>,<!
	},
	sensitive=true,
	tag=[s]
}
\lstdefinestyle{html}{
	language=HTML5,
	alsodigit={.:;},
	alsolanguage=JavaScript,
	firstnumber=1,
	ndkeywordstyle=\color{dkgreen}\bfseries,
	numberfirstline=true
}

\lstdefinestyle{ini}{
	language={},
	commentstyle=\color{gray}\ttfamily,
	keywordstyle={\color{black}\bfseries},
	morecomment=[l]{;},
	morecomment=[l]{\#},
	morecomment=[s][\color{dkgreen}\bfseries]{[}{]},
	morekeywords={},
	otherkeywords={=,:}
}

\lstdefinestyle{java}{
	language=Java,
	breakatwhitespace=true,
	keepspaces=true
}

\lstdefinelanguage{JavaScript}{
	comment=[l]{//},
	keepspaces=true,
	keywords={
		break,else,false,for,function,if,in,new,null,return,true,typeof,var,while
	},
	morecomment=[s]{/*}{*/},
	morestring=[b]',
	morestring=[b]",
	morestring=[b]`,
	ndkeywords={
		await,async,case,catch,class,const,default,do,enum,export,extends,finally,from,implements,import,instanceof,let,static,super,switch,then,this,throw,try
	},
	ndkeywordstyle=\color{blue}\bfseries,
	sensitive=false
}
\lstdefinestyle{javascript}{
	language=JavaScript
}

\lstdefinestyle{json}{
	literate=*{0}{{{\color{cardinalred}0}}}{1}{1}{{{\color{cardinalred}1}}}{1}{2}
	{{{\color{cardinalred}2}}}{1}{3}{{{\color{cardinalred}3}}}{1}{4}{{{\color{cardinalred}4}}}
	{1}{5}{{{\color{cardinalred}5}}}{1}{6}{{{\color{cardinalred}6}}}{1}{7}{{{\color{cardinalred}7}}}
	{1}{8}{{{\color{cardinalred}8}}}{1}{9}{{{\color{cardinalred}9}}}{1}{:}
	{{{\color{dkcyan}{:}}}}{1}{,}{{{\color{dkcyan}{,}}}}{1}{\{}
	{{{\color{MidnightBlue}{\{}}}}{1}{\}}{{{\color{MidnightBlue}{\}}}}}
	{1}{[}{{{\color{MidnightBlue}{[}}}}{1}{]}{{{\color{MidnightBlue}{]}}}}{1},
	tabsize=2
}

\lstdefinestyle{julia}{
	keywordsprefix=\@,
	morecomment=[l]{\#},
	morekeywords={
		abstract,Any,applicable,assert,baremodule,begin,bitstype,Bool,break,catch,ccall,Complex64,Complex128,const,continue,convert,dlopen,dlsym,do,edit,else,elseif,end,eps,error,exit,export,finalizer,Float32,Float64,for,function,global,hash,if,im,immutable,import,importall,in,Inf,Int,Int8,Int16,Int32,Int64,invoke,is,isa,isequal,let,load,local,macro,method_exists,module,Nan,new,None,Nothing,ntuple,pi,promote,promote_type,quote,realmax,realmin,return,sizeof,subtype,system,throw,try,tuple,type,typealias,typemax,typemin,typeof,uid,Uint,Uint8,Uint16,Uint32,Uint64,using,while,whos
	},
	morestring=[b]',
	morestring=[b]",
	sensitive=true
}

\lstdefinestyle{kotlin}{
	comment=[l]{//},
	emph={delegate,filter,first,firstOrNull,forEach,lazy,map,mapNotNull,println,
		return@},
	emphstyle={\color{blue}},
	keywords={
		abstract,actual,as,as?,break,by,class,companion,continue,data,do,dynamic,else,enum,expect,false,final,for,fun,get,if,import,in,interface,internal,is,null,object,override,package,private,public,return,set,super,suspend,this,throw,true,try,typealias,val,var,vararg,when,where,while
	},
	morecomment=[s]{/*}{*/},
	morestring=[b]",
	morestring=[s]{"""*}{*"""},
	ndkeywords={
		@Deprecated,@JvmField,@JvmName,@JvmOverloads,@JvmStatic,@JvmSynthetic,Array,Byte,Double,Float,Int,Integer,Iterable,Long,Runnable,Short,String
	},
	ndkeywordstyle=\color{BurntOrange}\bfseries,
	sensitive=true
}

\lstdefinestyle{latex}{
	language=TeX,
	morekeywords={
		aacos,aasin,aatan,acos,addimage,addimageanum,addimageboxed,align,asin,atan,begin,bibitem,bibliography,bigstrut,boldmath,bookmarksetup,boxed,cancelto,caption,changeheadertitle,checkmark,checkvardefined,cite,clearpage,dd,degree,eqref,equal,frac,fracnpartial,fullcite,hline,href,ifthenelse,imageshspace,imagesnewline,imagesvspace,includefullhfpdf,includehfpdf,insertalign,insertalignanum,insertaligncaptioned,insertaligncaptioned,insertaligncaptionedanum,insertaligned,insertalignedanum,insertalignedcaptioned,insertalignedcaptionedanum,insertemail,insertemptypage,inserteqimage,insertequation,insertequationanum,insertequationcaptioned,insertequationcaptionedanum,insertgather,insertgatheranum,insertgathercaptioned,insertgathercaptionedanum,insertgathered,insertgatheredanum,insertgatheredcaptioned,insertgatheredcaptionedanum,insertimage,insertimageleft,insertimageright,insertindextitle,insertindextitlepage,insertphone,isundefined,itemresize,label,LaTeX,lipsum,lpow,makeatletter,makeatother,newcommand,newcounter,newp,newpage,pow,quotes,ref,renewcommand,section,sectionanum,setcounter,setlength,shortcite,sourcecode,sourcecodep,subsection,subsectionanum,subsubsection,subsubsectionanum,subsubsubsection,subsubsubsection,subsubsubsectionanum,textbf,textit,textregistered,textsuperscript,texttt,throwbadconfig,unboldmath,url,xspace
	}
}

\lstdefinestyle{lisp}{
	language=Lisp,
	morekeywords={if}
}

\lstdefinestyle{llvm}{
	language=LLVM
}

\lstdefinestyle{lua}{
	language={[5.3]Lua}
}

\lstdefinestyle{make}{
	language=[gnu] make
}

\lstdefinelanguage{Maple}{
	morecomment=[l]\#,
	morekeywords={
		and,assuming,break,by,catch,description,do,done,elif,else,end,error,export,fi,finally,for,from,global,if,implies,in,intersect,local,minus,mod,module,next,not,o,option,options,or,proc,quit,read,restart,return,save,stop,subset,then,to,try,union,use,uses,with,while,xor
	},
	morestring=[b]",
	morestring=[d],
	sensitive=true
} 
\lstdefinestyle{maple}{
	language=Maple
}

\lstdefinestyle{mathematica}{
	language=Mathematica
}

\lstdefinestyle{matlab}{
	language=Matlab,
	deletekeywords={fft},
	keepspaces=true,
	morecomment=[l]\%,
	morecomment=[n]{\%\{\^^M}{\%\}\^^M},
	morekeywords={
		addOptional,box,break,catch,cell,classdef,continue,deal,double,end,factorial,for,gradient,hessian,if,isa,ltitr,matlab2tikz,methods,minor,movegui,normcdf,normpdf,on,ones,parse,persistent,poissrnd,properties,repmat,solve,strcat,subs,syms,try,var,warning,xlim,ylim
	}
}

\lstdefinestyle{mercury}{
	language=Mercury
}

\lstdefinestyle{modula2}{
	language=Modula-2
}

\lstdefinestyle{objectivec}{
	language=[Objective]C,
	breakatwhitespace=false,
	keepspaces=true,
	moredirectives={
		import
	},
	morekeywords={
		@catch,@class,@dynamic,@encode,@end,@finally,@implementation,@interface,@package,@private,@property,@protected,@protocol,@public,@selector,@synchronized,@synthesize,@throw,@try,assign,BOOL,bycopy,byref,Class,copy,id,IMP,in,inout,Nil,nil,NO,nonatomic,oneway,out,readonly,readwrite,retain,SEL,self,super,YES,_cmd
	}
}

\lstdefinestyle{octave}{
	language=Octave,
	keepspaces=true,
	morecomment=[l]\%,
	morecomment=[n]{\%\{\^^M}{\%\}\^^M}
}

\lstdefinestyle{opencl}{
	language=C++,
	breakatwhitespace=false,
	emph={
		bool2,bool3,bool4,bool8,bool16,char2,char3,char4,char8,char16,complex,constant,event_t,float2,float3,float4,float8,float16,global,half2,half3,half4,half8,half16,image2d_t,image3d_t,imaginary,int2,int3,int4,int8,int16,kernel,local,long2,long3,long4,long8,long16,private,quad,quad2,quad3,quad4,quad8,quad16,sampler_t,short2,short3,short4,short8,short16,uchar2,uchar3,uchar4,uchar8,uchar16,uint2,uint3,uint4,uint8,uint16,ulong2,ulong3,ulong4,ulong8,ulong16,ushort2,ushort3,ushort4,ushort8,ushort16,__constant,__global,__kernel,__local,__private
	},
	emphstyle=\color{dkcyan}\ttfamily,
	morecomment=[l][\color{magenta}]{\#}
}

\lstdefinestyle{opensees}{
	language=tcl,
	breakatwhitespace=false,
	emph=[1]{
		-accel,-beamUniform,-dir,-dof,-ele,-eleRange,-file,-height,-increment,-initial,-iNode,-integration,-iterate,-jNode,-kNode,-mass,-mat,-matConcrete,-matShear,-matSteel,-max,-maxDim,-maxEta,-maxIter,-min,-minEta,-ndf,-ndm,-node,-nodeRange,-numSublevels,-numSubSteps,-perpDirn,-region,-rho,-sections,-thick,-time,-tol,-type,-width
	},
	emphstyle=[1]\color{black}\bfseries\em,
	keepspaces=true,
	morecomment=[l]{\#},
	morekeywords={
		algorithm,analysis,analyze,constraints,deformation,disp,eleLoad,element,equalDOF,fix,fixX,fixY,fixZmodel,geomTransf,initialize,integrator,layer,loadConst,mass,model,node,numberer,patch,pattern,printA,PySimple1Gen,reaction,recorder,region,rigidDiaphragm,section,system,test,uniaxialMaterial,wipe,wipeAnalysis
	},
	ndkeywords={
		9_4_QuadUP,20_8_BrickUP,AC3D8,Aggregator,ArcLength,ASI3D8,AV3D4,AxialSp,AxialSpHD,BandGeneral,BARSLIP,BasicBuilder,bbarBrick,bbarBrickUP,bbarQuad,bbarQuadUP,BeamColumnJoint,BeamContact2D,BeamContact3D,BeamEndContact3D,BFGS,Bilin,BilinearOilDamper,Bond_SP01,BoucWen,Brick20N,brickUP,Broyden,BWBN,Cast,CatenaryCable,CentralDifference,CFSSSWP,CFSWSWP,Concrete01,Concrete01WithSITC,Concrete02,Concrete03,Concrete04,Concrete06,Concrete07,ConcreteCM,ConcreteD,ConfinedConcrete01,constraintsTypeGravity,Corotational,corotTruss,corotTrussSection,CoupledZeroLength,DeformedShape,dispBeamColumn,dispBeamColumnInt,DisplacementControl,Dodd_Restrepo,Drift,ECC01,Elastic,elasticBeamColumn,ElasticBilin,ElasticMultiLinear,ElasticPP,ElasticPPGap,ElasticTimoshenkoBeam,ElasticTubularJoint,elastomericBearingBoucWen,elastomericBearingPlasticity,ElastomericX,Element,EnergyIncr,enhancedQuad,ENT,Explicitdifference,Fatigue,flatSliderBearing,forceBeamColumn,forceBeamColumn,FourNodeTetrahedron,FPBearingPTV,FRPConfinedConcrete,GeneralizedAlpha,Hardening,HDR,HHT,HyperbolicGapMaterial,Hysteretic,ImpactMaterial,InitStrainMaterial,InitStressMaterial,Joint2D,KikuchiAikenHDR,KikuchiAikenLRB,KikuchiBearing,KrylovNewton,Lagrange,LeadRubberX,LimitState,Linear,LoadControl,LoadControl,MinMax,MinUnbalDispNorm,mkdir,ModElasticBeam2d,ModifiedNewton,ModIMKPeakOriented,ModIMKPinching,MultiLinear,multipleShearSpring,MVLEM,Newmark,Newton,NewtonLineSearch,Node,NodeNumbers,nonlinearBeamColumn,NormDispIncr,numberer,Parallel,PathIndependentMaterial,pattern,PDelta,Pinching4,PinchingLimitStateMaterial,Plain,PyLiq1,PySimple1,quad,quadr,quadUP,QzSimple1,RambergOsgoodSteel,rayleigh,RCM,rect,ReinforcingSteel,RJWatsonEqsBearing,SAWS,SecantNewton,SelfCentering,Series,SFI_MVLEM,ShallowFoundationGen,ShellDKGQ,ShellDKGT,ShellMITC4,ShellNL,ShellNLDKGQ,ShellNLDKGT,SimpleContact2D,SimpleContact3D,singleFPBearing,SparseGeneral,SSPbrick,SSPbrickUP,SSPquad,SSPquadUP,Static,stdBrick,Steel01,Steel01,Steel02,Steel4,SteelMPF,straight,SurfaceLoad,TFP,Transient,TRBDF2,tri31,TripleFrictionPendulum,truss,trussSection,twoNodeLink,TzLiq1,TzSimple1,UniformExcitation,ViewScale,Viscous,ViscousDamper,VS3D4,YamamotoBiaxialHDR,zeroLength,zeroLengthContact,zeroLengthContactNTS2D,zeroLengthImpact3D,zeroLengthImpact3D,zeroLengthInterface2D,zeroLengthND,zeroLengthSection
	},
	ndkeywordstyle=\color{dkcyan}\ttfamily
}

\lstdefinestyle{pascal}{
	language=Pascal,
	morecomment=[l]{//},
	sensitive=false
}

\lstdefinestyle{perl}{
	language=Perl,
	alsoletter={\%},
	breakatwhitespace=false,
	keepspaces=true
}

\lstdefinestyle{php}{
	language=php,
	emph=[1]{
		php
	},
	emph=[2]{
		if,and,or,else
	},
	emph=[3]{
		abstract,as,const,else,elseif,endfor,endforeach,endif,extends,final,for,foreach,global,if,implements,private,protected,public,static,var
	},
	emphstyle=[1]\color{black},
	emphstyle=[2]\color{blue},
	keywords={
		abstract,and,array,as,break,callable,case,catch,class,clone,const,continue,declare,default,die,do,echo,else,elseif,empty,enddeclare,endfor,endforeach,endif,endswitch,endwhile,eval,exit,extends,final,finally,for,foreach,function,global,goto,if,implements,include,include_once,instanceof,insteadof,interface,isset,list,namespace,new,or,print,private,protected,public,require,require_once,return,static,switch,throw,trait,try,unset,use,var,while,xor,yield,__halt_compiler
	},
	showlines=true,
	upquote=true
}

\lstdefinestyle{plaintext}{
	language={},
	keepspaces=true,
	postbreak={},
	tabsize=4
}

\lstdefinestyle{postscript}{
	language=PostScript,
	keepspaces=true
}

\lstdefinestyle{powershell}{
	alsodigit={-},
	morecomment=[l]{\#},
	morecomment=[n]{<\#}{\#>},
	morekeywords={
		Add-Content,Add-PSSnapin,Clear-Content,Clear-History,Clear-Host,Clear-Item,Clear-ItemProperty,Clear-Variable,Compare-Object,Connect-PSSession,Convert-Path,ConvertFrom-String,Copy-Item,Copy-ItemProperty,Disable-PSBreakpoint,Disconnect-PSSession,Enable-PSBreakpoint,Enter-PSSession,Exit-PSSession,Export-Alias,Export-Csv,Export-PSSession,ForEach-Object,Format-Custom,Format-Hex,Format-List,Format-Table,Format-Wide,Get-Alias,Get-ChildItem,Get-Clipboard,Get-Command,Get-ComputerInfo,Get-Content,Get-History,Get-Item,Get-ItemProperty,Get-ItemPropertyValue,Get-Job,Get-Location,Get-Member,Get-Module,Get-Process,Get-PSBreakpoint,Get-PSCallStack,Get-PSDrive,Get-PSSession,Get-PSSnapin,Get-Service,Get-TimeZone,Get-Unique,Get-Variable,Get-WmiObject,Group-Object,help,Import-Alias,Import-Csv,Import-Module,Import-PSSession,Invoke-Command,Invoke-Expression,Invoke-History,Invoke-Item,Invoke-RestMethod,Invoke-WebRequest,Invoke-WmiMethod,Measure-Object,mkdir,Move-Item,Move-ItemProperty,New-Alias,New-Item,New-Module,New-PSDrive,New-PSSession,New-PSSessionConfigurationFile,New-Variable,Out-GridView,Out-Host,Out-Printer,Pop-Location,powershell_ise.exe,Push-Location,Receive-Job,Receive-PSSession,Remove-Item,Remove-ItemProperty,Remove-Job,Remove-Module,Remove-PSBreakpoint,Remove-PSDrive,Remove-PSSession,Remove-PSSnapin,Remove-Variable,Remove-WmiObject,Rename-Item,Rename-ItemProperty,Resolve-Path,Resume-Job,Select-Object,Select-String,Set-Alias,Set-Clipboard,Set-Content,Set-Item,Set-ItemProperty,Set-Location,Set-PSBreakpoint,Set-TimeZone,Set-Variable,Set-WmiInstance,Show-Command,Sort-Object,Start-Job,Start-Process,Start-Service,Start-Sleep,Stop-Job,Stop-Process,Stop-Service,Suspend-Job,Tee-Object,Trace-Command,Wait-Job,Where-Object,Write-Output
	},
	morekeywords={
		Do,Else,For,ForEach,Function,If,In,Until,While
	},
	morestring=[b]{"},
	morestring=[b]{'},
	morestring=[s]{@'}{'@},
	morestring=[s]{@"}{"@},
	sensitive=false
}

\lstdefinestyle{prolog}{
	language=Prolog
}

\lstdefinestyle{promela}{
	language=Promela
}

\lstdefinelanguage{Pseudocode}{
	language={},
	breakatwhitespace=false,
	commentstyle=\color{gray}\upshape,
	keepspaces=true,
	keywords={
		and,be,begin,break,datatype,do,elif,else,end,for,foreach,fun,function,if,in,input,let,not,null,or,output,pop,procedure,push,repeat,return,swap,until,while,xor
	},
	keywordstyle=\color{black}\bfseries,
	mathescape=true,
	morecomment=[l]{//},
	morecomment=[l]{\#},
	morecomment=[s]{/*}{*/},
	morecomment=[s]{/**}{*/},
	sensitive=false,
	stringstyle=\color{dkgray}\bfseries\em
}
\lstdefinestyle{pseudocode}{
	language=Pseudocode,
	backgroundcolor=\color{white},
	frame=tb,
	numbers=none
}
\lstdefinestyle{pseudocodecolor}{
	language=Pseudocode
}

\lstdefinelanguage{pythonEXTENDED}{
	language=Python,
	breakatwhitespace=false,
	emph={
		AbstractSet,Any,AsyncContextManager,AsyncGenerator,AsyncIterable,AsyncIterator,Awaitable,AwaitableGenerator,BinaryIO,ByteString,Callable,Collection,Container,ContextManager,Coroutine,Dict,False,ForwardRef,Generator,GenericMeta,Hashable,IO,ItemsView,Iterable,Iterator,KeysView,List,Mapping,MappingView,Match,Meta,MutableMapping,MutableSequence,MutableSet,NamedTuple,None,Pattern,Reversible,Sequence,Sized,SupportInts,SupportsAbs,SupportsBytes,SupportsComplex,SupportsFloat,SupportsIndex,SupportsRound,TextIO,True,Tuple,TypeAlias,TYPE_CHECKING,Union,ValuesView,__add__,__and__,__eq__,__floordiv__,__ge__,__gt__,__init__,__le__,__lt__,__main__,__mod__,__mul__,__name__,__ne__,__or__,__pow__,__repr__,__str__,__sub__,__truediv__,__xor__
	},
	emphstyle=\color{dkcyan}\ttfamily,
	keepspaces=true,
	morecomment=[s][\color{BurntOrange}]{@}{\ },
	morekeywords={
		as,assert,close,listdir,self,sorted,split,strip,with
	}
}
\lstdefinestyle{python}{
	language=pythonEXTENDED
}

\lstdefinestyle{qsharp}{
	mathescape=true,
	morecomment=[l]{//},
	morecomment=[l][\color{dkgreen}]{///},
	morekeywords={
		Adj,Adjoint,adjoint,and,apply,as,auto,BigInt,body,Bool,borrowing,Controlled,controlled,Ctl,distribute,Double,elif,else,fail,false,fixup,for,function,if,in,Int,intrinsic,invert,is,let,mutable,namespace,new,newtype,not,One,open,operation,or,Pauli,PauliI,PauliX,PauliY,PauliZ,Qubit,Range,repeat,Result,return,self,set,String,true,Unit,until,using,while,within,Zero
	},
	morekeywords=[2]{
		Assert,AssertProb,CCNOT,CNOT,Exp,ExpFrac,H,I,M,Measure,Message,R,R1,R1Frac,Random,Reset,ResetAll,RFrac,Rx,Ry,Rz,S,SWAP,T,X,Y,Z
	},
	sensitive=true
}

\lstdefinestyle{r}{
	language=R,
	alsoletter={.<-},
	alsoother={._$},
	deletekeywords={
		df,data,frame,length,as,character
	},
	morecomment=[l]\#,
	morestring=[d]',
	morestring=[d]",
	otherkeywords={
		!,!=,~,$,*,\&,\%/\%,\%*\%,\%\%,<-,<<-,/
	}
}

\lstdefinestyle{racket}{
	alsoletter={',`,-,/,>,<,\#,\%},
	morekeywords=[1]{
		define,define-macro,define-stream,define-syntax,lambda,stream-lambda
	},
	morekeywords=[2]{
		->,always_publish,and,\#',\#\%module-begin,\#lang,\#`,begin,begin-for-syntax,Boolean,call-with-current-continuation,call-with-input-file,call-with-output-file,callback,call/cc,case,cond,define-context,define-controller,define-struct/contract,define/contract,delay,do,else,environment,eval,fold,for,for-each,force,get,if,implement,in-range,Integer,label,let,let*,let*-values,let-syntax,let-values,letrec,letrec-syntax,map,maybe_publish,message-box,module,new,not,or,or/c,parent,provide,quasiquote,query,quote,rename-out,require,send,submod,syntax,syntax-case,syntax-rules,unquote,unquote-splicing,when,when-provided,when-required,with-syntax
	},
	morekeywords=[3]{
		export,import
	},
	morecomment=[l]{;},
	moredelim=**[is][\color{lgray}]{<<@<<}{>>@>>},
	moredelim=**[is][\itshape\color{mauve}]{<<;<<}{>>;>>},
	morecomment=[s]{\#|}{|\#},
	morestring=[s]{"}{"},
	sensitive=true
}

\lstdefinestyle{reil}{
	comment=[l]{;},
	keywords=[1]{
		ADD,add,and,AND,BISZ,bisz,bsh,BSH,div,DIV,jcc,JCC,LDM,ldm,MOD,mod,mul,MUL,nop,NOP,or,OR,stm,STM,STR,str,sub,SUB,undef,UNDEF,unkn,UNKN,XOR,xor
	},
	keywords=[3]{
		ah,AH,al,AL,AX,ax,bh,BH,BL,bl,bp,BP,bpl,BPL,BX,bx,ch,CH,cl,CL,cx,CX,DH,dh,di,DI,dil,DIL,dl,DL,DX,dx,EAX,eax,EBP,ebp,ebx,EBX,ECX,ecx,EDI,edi,edx,EDX,esi,ESI,esp,ESP,r8,R8,r8b,R8B,r8d,R8D,r8w,R8W,r9,R9,R9B,r9b,R9D,r9d,r9w,R9W,r10,R10,R10B,r10b,R10D,r10d,r10w,R10W,r11,R11,r11b,R11B,r11d,R11D,R11W,r11w,R12,r12,R12B,r12b,r12d,R12D,r12w,R12W,R13,r13,r13b,R13B,R13D,r13d,R13W,r13w,r14,R14,R14B,r14b,r14d,R14D,R14W,r14w,r15,R15,r15b,R15B,r15d,R15D,R15W,r15w,RAX,rax,rbp,RBP,rbx,RBX,RCX,rcx,RDI,rdi,rdx,RDX,RSI,rsi,RSP,rsp,SI,si,SIL,sil,SP,sp,spl,SPL
	},
	sensitive=true
}

\lstdefinestyle{ruby}{
	language=Ruby,
	breakatwhitespace=true,
	morestring=[s][]{\#\{}{\}},
	morestring=*[d]{"},
	sensitive=true
}

\lstdefinelanguage{Rust}{
	sensitive,
	alsodigit={},
	alsoletter={!},
	alsoother={},
	morecomment=[l]{//},
	morecomment=[s]{/*}{*/},
	moredelim=[s][{\itshape\color[rgb]{0,0,0.75}}]{\#[}{]},
	morekeywords=[2]{
		Add,AddAssign,Any,AsciiExt,AsInner,AsInnerMut,AsMut,AsRawFd,AsRawHandle,AsRawSocket,AsRef,Binary,BitAnd,BitAndAssign,Bitor,BitOr,BitOrAssign,BitXor,BitXorAssign,Borrow,BorrowMut,Boxed,BoxPlace,BufRead,BuildHasher,CastInto,CharExt,Clone,CoerceUnsized,CommandExt,Copy,Debug,DecodableFloat,Default,Deref,DerefMut,DirBuilderExt,DirEntryExt,Display,Div,DivAssign,DoubleEndedIterator,DoubleEndedSearcher,Drop,EnvKey,Eq,Error,ExactSizeIterator,ExitStatusExt,Extend,FileExt,FileTypeExt,Float,Fn,FnBox,FnMut,FnOnce,Freeze,From,FromInner,FromIterator,FromRawFd,FromRawHandle,FromRawSocket,FromStr,FullOps,FusedIterator,Generator,Hash,Hasher,Index,IndexMut,InPlace,Int,Into,IntoCow,IntoInner,IntoIterator,IntoRawFd,IntoRawHandle,IntoRawSocket,IsMinusOne,IsZero,Iterator,JoinHandleExt,LargeInt,LowerExp,LowerHex,MetadataExt,Mul,MulAssign,Neg,Not,Octal,OpenOptionsExt,Ord,OsStrExt,OsStringExt,Packet,PartialEq,PartialOrd,Pattern,PermissionsExt,Place,Placer,Pointer,Product,Put,RangeArgument,RawFloat,Read,Rem,RemAssign,Seek,Shl,ShlAssign,Shr,ShrAssign,Sized,SliceConcatExt,SliceExt,SliceIndex,Stats,Step,StrExt,Sub,SubAssign,Sum,Sync,TDynBenchFn,Terminal,Termination,ToOwned,ToSocketAddrs,ToString,Try,TryFrom,TryInto,UnicodeStr,Unsize,UpperExp,UpperHex,WideInt,Write
	},
	morekeywords=[2]{
		Send
	},
	morekeywords=[3]{
		bool,char,f32,f64,i8,i16,i32,i64,isize,str,u8,u16,u32,u64,unit,usize,i128,u128
	},
	morekeywords=[4]{
		Err,false,None,Ok,Some,true
	},
	morekeywords=[5]{
		assert!,assert_eq!,assert_ne!,cfg!,column!,compile_error!,concat!,concat_idents!,debug_assert!,debug_assert_eq!,debug_assert_ne!,env!,eprint!,eprintln!,file!,format!,format_args!,include!,include_bytes!,include_str!,line!,module_path!,option_env!,panic!,print!,println!,select!,stringify!,thread_local!,try!,unimplemented!,unreachable!,vec!,write!,writeln!
	},
	morekeywords={
		abstract,alignof,become,box,do,final,macro,offsetof,override,priv, proc,pure,sizeof,typeof,unsized,virtual,yield
	},
	morekeywords={
		as,const,let,move,mut,ref,static
	},
	morekeywords={
		break,continue,else,for,if,in,loop,match,return,while
	},
	morekeywords={
		crate,extern,mod,pub,super
	},
	morekeywords={
		dyn,enum,fn,impl,Self,self,struct,trait,type,union,use,where
	},
	morekeywords={
		unsafe
	},
	morestring=[b]{"}
}
\lstdefinestyle{rust}{
	language=Rust,
	keywordstyle=[2]\color[rgb]{0.75,0,0}, 
	keywordstyle=[3]\color[rgb]{0,0.5,0}, 
	keywordstyle=[4]\color[rgb]{0,0.5,0}, 
	keywordstyle=[5]\color[rgb]{0,0,0.75} 
}

\lstdefinestyle{scala}{
	language=scala,
	breakatwhitespace=true,
	morecomment=[l]{//},
	morecomment=[n]{/*}{*/},
	morekeywords={
		abstract,case,catch,class,def,do,else,extends,false,final,finally,for,if,implicit,import,match,mixin,new,null,object,override,package,private,protected,requires,return,sealed,super,this,throw,trait,true,try,type,val,var,while,with,yield
	},
	morestring=[b]',
	morestring=[b]",
	morestring=[b]""",
	otherkeywords={
		=>,<-,<\%,<:,>:,\#,@
	}
}

\lstdefinestyle{scheme}{
	language=Lisp,
	morecomment=[l]{;},
	morekeywords={
		and,begin,case,case-lambda,cond,cond-expand,define,delay,delay-force,do,else,force,guard,if,lambda,let,let*,let*-values,let-syntax,let-values,letrec,letrec*,letrec-syntax,make-parameter,make-promise,map,or,parameterize,promise?,quasiquote,quote,set!,syntax-rules,unless,when
	},
	morestring=[b]"
}

\lstdefinestyle{scilab}{
	language=Scilab
}

\lstdefinestyle{simula}{
	language=Simula
}

\lstdefinestyle{sparql}{
	language=SPARQL
}

\lstdefinestyle{sql}{
	language=SQL,
	breakatwhitespace=true
}

\lstdefinestyle{swift}{
	language=Swift
}

\lstdefinestyle{tcl}{
	language=tcl,
	breakatwhitespace=false,
	keepspaces=true,
	morecomment=[l]{\#}
}

\lstdefinestyle{vbscript}{
	language=[Visual]Basic,
	extendedchars=true
}

\lstdefinestyle{verilog}{
	language=Verilog
}

\lstdefinelanguage{VHDL}{
	morekeywords=[1]{
		ALL,all,and,architecture,begin,downto,end,entity,in,is,library,Not,of,or,out,port,use
	},
	morekeywords=[2]{
		IEEE,NUMERIC_STD,STD_LOGIC,std_logic,STD_LOGIC_1164,STD_LOGIC_ARITH,STD_LOGIC_UNSIGNED,STD_LOGIC_VECTOR,std_logic_vector
	},
	morecomment=[l]--
}
\lstdefinestyle{vhdl}{
	language=VHDL
}

\lstdefinelanguage{XML}{
	morecomment=[s]{<?}{?>},
	morekeywords={
		encoding,type,version,xmlns
	},
	morestring=[b]",
	morestring=[s]{>}{<}
}
\lstdefinestyle{xml}{
	language=XML,
	tabsize=2
}

\newcommand{\checkvalidsourcecodestyle}[1]{%
	\ifthenelse{\equal{#1}{abap}}{}{%
	\ifthenelse{\equal{#1}{ada}}{}{%
	\ifthenelse{\equal{#1}{assemblerx64}}{}{%
	\ifthenelse{\equal{#1}{assemblerx86}}{}{%
	\ifthenelse{\equal{#1}{awk}}{}{%
	\ifthenelse{\equal{#1}{bash}}{}{%
	\ifthenelse{\equal{#1}{basic}}{}{%
	\ifthenelse{\equal{#1}{c}}{}{%
	\ifthenelse{\equal{#1}{caml}}{}{%
	\ifthenelse{\equal{#1}{cmake}}{}{%
	\ifthenelse{\equal{#1}{cobol}}{}{%
	\ifthenelse{\equal{#1}{cpp}}{}{%
	\ifthenelse{\equal{#1}{csharp}}{}{%
	\ifthenelse{\equal{#1}{css}}{}{%
	\ifthenelse{\equal{#1}{csv}}{}{%
	\ifthenelse{\equal{#1}{cuda}}{}{%
	\ifthenelse{\equal{#1}{dart}}{}{%
	\ifthenelse{\equal{#1}{docker}}{}{%
	\ifthenelse{\equal{#1}{elisp}}{}{%
	\ifthenelse{\equal{#1}{elixir}}{}{%
	\ifthenelse{\equal{#1}{erlang}}{}{%
	\ifthenelse{\equal{#1}{fortran}}{}{%
	\ifthenelse{\equal{#1}{fsharp}}{}{%
	\ifthenelse{\equal{#1}{glsl}}{}{%
	\ifthenelse{\equal{#1}{gnuplot}}{}{%
	\ifthenelse{\equal{#1}{go}}{}{%
	\ifthenelse{\equal{#1}{haskell}}{}{%
	\ifthenelse{\equal{#1}{html}}{}{%
	\ifthenelse{\equal{#1}{ini}}{}{%
	\ifthenelse{\equal{#1}{java}}{}{%
	\ifthenelse{\equal{#1}{javascript}}{}{%
	\ifthenelse{\equal{#1}{json}}{}{%
	\ifthenelse{\equal{#1}{julia}}{}{%
	\ifthenelse{\equal{#1}{kotlin}}{}{%
	\ifthenelse{\equal{#1}{latex}}{}{%
	\ifthenelse{\equal{#1}{lisp}}{}{%
	\ifthenelse{\equal{#1}{llvm}}{}{%
	\ifthenelse{\equal{#1}{lua}}{}{%
	\ifthenelse{\equal{#1}{make}}{}{%
	\ifthenelse{\equal{#1}{maple}}{}{%
	\ifthenelse{\equal{#1}{mathematica}}{}{%
	\ifthenelse{\equal{#1}{matlab}}{}{%
	\ifthenelse{\equal{#1}{mercury}}{}{%
	\ifthenelse{\equal{#1}{modula2}}{}{%
	\ifthenelse{\equal{#1}{objectivec}}{}{%
	\ifthenelse{\equal{#1}{octave}}{}{%
	\ifthenelse{\equal{#1}{opencl}}{}{%
	\ifthenelse{\equal{#1}{opensees}}{}{%
	\ifthenelse{\equal{#1}{pascal}}{}{%
	\ifthenelse{\equal{#1}{perl}}{}{%
	\ifthenelse{\equal{#1}{php}}{}{%
	\ifthenelse{\equal{#1}{plaintext}}{}{%
	\ifthenelse{\equal{#1}{postscript}}{}{%
	\ifthenelse{\equal{#1}{powershell}}{}{%
	\ifthenelse{\equal{#1}{prolog}}{}{%
	\ifthenelse{\equal{#1}{promela}}{}{%
	\ifthenelse{\equal{#1}{pseudocode}}{}{%
	\ifthenelse{\equal{#1}{pseudocodecolor}}{}{%
	\ifthenelse{\equal{#1}{python}}{}{%
	\ifthenelse{\equal{#1}{qsharp}}{}{%
	\ifthenelse{\equal{#1}{r}}{}{%
	\ifthenelse{\equal{#1}{racket}}{}{%
	\ifthenelse{\equal{#1}{reil}}{}{%
	\ifthenelse{\equal{#1}{ruby}}{}{%
	\ifthenelse{\equal{#1}{rust}}{}{%
	\ifthenelse{\equal{#1}{scala}}{}{%
	\ifthenelse{\equal{#1}{scheme}}{}{%
	\ifthenelse{\equal{#1}{scilab}}{}{%
	\ifthenelse{\equal{#1}{simula}}{}{%
	\ifthenelse{\equal{#1}{sparql}}{}{%
	\ifthenelse{\equal{#1}{sql}}{}{%
	\ifthenelse{\equal{#1}{swift}}{}{%
	\ifthenelse{\equal{#1}{tcl}}{}{%
	\ifthenelse{\equal{#1}{vbscript}}{}{%
	\ifthenelse{\equal{#1}{verilog}}{}{%
	\ifthenelse{\equal{#1}{vhdl}}{}{%
	\ifthenelse{\equal{#1}{xml}}{}{%
		\errmessage{LaTeX Warning: Estilo de codigo desconocido. Valores esperados: abap,ada,assemblerx64,assemblerx86,awk,bash,basic,c,caml,cmake,cobol,cpp,csharp,css,csv,cuda,dart,docker,elisp,elixir,erlang,fortran,fsharp,glsl,gnuplot,go,haskell,html,ini,java,javascript,json,julia,kotlin,latex,lisp,llvm,lua,make,maple,mathematica,matlab,mercury,modula2,objectivec,octave,opencl,opensees,pascal,perl,php,plaintext,postscript,powershell,prolog,promela,pseudocode,pseudocodecolor,python,qsharp,r,racket,reil,ruby,rust,scala,scheme,scilab,simula,sparql,sql,swift,tcl,vbscript,verilog,vhdl,xml}%
		\stop%
	}}}}}}}}}}}}}}}}}}}}}}}}}}}}}}}}}}}}}}}}}}}}}}}}}}}}}}}}}}}}}}}}}}}}}}}}}}}}}%
}

\newcommand{\inlinesourcecodeboxed}[3][]{%
	\emptyvarerr{\inlinesourcecodeboxed}{#2}{Estilo de codigo no definido}%
	\emptyvarerr{\inlinesourcecodeboxed}{#3}{Codigo no definido}%
	\lstset{%
		basicstyle={\sourcecodeilfonts\sourcecodeilfontf\color{\maintextcolor}}%
	}%
	\checkvalidsourcecodestyle{#2}%
	\ifthenelse{\equal{#1}{}}{%
		\Colorbox{\sourcecodebgcolor}{\lstinline[style=#2]!#3!}%
	}{%
	\ifthenelse{\equal{#1}{NOCOLOR}}{%
		\lstinline[style=#2]!#3!%
	}{%
		\Colorbox{#1}{\lstinline[style=#2]!#3!}%
	}}%
	\lstset{%
		basicstyle={\sourcecodefonts\sourcecodefontf\color{\maintextcolor}}%
	}%
}

\makeatletter
\def\greek#1{\expandafter\@greek\csname c@#1\endcsname}
\def\Greek#1{\expandafter\@Greek\csname c@#1\endcsname}
\def\@greek#1{%
	\ifcase#1%
		\or $\alpha$%
		\or $\beta$%
		\or $\gamma$%
		\or $\delta$%
		\or $\epsilon$%
		\or $\zeta$%
		\or $\eta$%
		\or $\theta$%
		\or $\iota$%
		\or $\kappa$%
		\or $\lambda$%
		\or $\mu$%
		\or $\nu$%
		\or $\xi$%
		\or $o$%
		\or $\pi$%
		\or $\rho$%
		\or $\sigma$%
		\or $\tau$%
		\or $\upsilon$%
		\or $\phi$%
		\or $\chi$%
		\or $\psi$%
		\or $\omega$%
	\fi%
}
\def\@Greek#1{%
	\ifcase#1%
		\or $\mathrm{A}$%
		\or $\mathrm{B}$%
		\or $\Gamma$%
		\or $\Delta$%
		\or $\mathrm{E}$%
		\or $\mathrm{Z}$%
		\or $\mathrm{H}$%
		\or $\Theta$%
		\or $\mathrm{I}$%
		\or $\mathrm{K}$%
		\or $\Lambda$%
		\or $\mathrm{M}$%
		\or $\mathrm{N}$%
		\or $\Xi$%
		\or $\mathrm{O}$%
		\or $\Pi$%
		\or $\mathrm{P}$%
		\or $\Sigma$%
		\or $\mathrm{T}$%
		\or $\mathrm{Y}$%
		\or $\Phi$%
		\or $\mathrm{X}$%
		\or $\Psi$%
		\or $\Omega$%
	\fi%
}
\makeatother
\AddEnumerateCounter{\greek}{\@greek}{24}
\AddEnumerateCounter{\Greek}{\@Greek}{12}

\checkvardefined{\documenttitle}

\makeatletter
	\g@addto@macro\documenttitle\xspace
\makeatother

\ifthenelse{\isundefined{\documentsubtitle}}{
	\errmessage{LaTeX Warning: Se borro la variable \noexpand\documentsubtitle, creando una vacia}
	\def\documentsubtitle {}}{
}

\ifthenelse{\equal{\documentsubtitle}{}}{
	\def\documenttitlehf {\documenttitle}
}{
	\def\documenttitlehf {\documentsubtitle}
}

\ifthenelse{\equal{\cfgpdfsecnumbookmarks}{true}}{
	\bookmarksetup{numbered}}{
}

\ifthenelse{\equal{\cfgshowbookmarkmenu}{true}}{
	\def\cfgpdfpagemode {UseOutlines}
	}{
	\def\cfgpdfpagemode {UseNone}
}
\ifthenelse{\equal{\usepdfmetadata}{true}}{
	\def\pdfmetainfotitle {\documenttitle}
}{
	\def\pdfmetainfotitle {}
}
\hypersetup{
	keeppdfinfo,
	bookmarksopen={\cfgpdfbookmarkopen},
	bookmarksopenlevel={\cfgbookmarksopenlevel},
	bookmarkstype={toc},
	pdfcenterwindow={\cfgpdfcenterwindow},
	pdfcopyright={\cfgpdfcopyright},
	pdfcreator={LaTeX},
	pdfdisplaydoctitle={\cfgpdfdisplaydoctitle},
	pdfencoding={unicode},
	pdffitwindow={\cfgpdffitwindow},
	pdfinfo={
		Document.Title={\pdfmetainfotitle},
		Template.Author.Alias={ppizarror},
		Template.Author.Email={pablo@ppizarror.com},
		Template.Author.Web={https://ppizarror.com},
		Template.Author={Pablo Pizarro R.},
		Template.Date={20/11/2023},
		Template.Encoding={UTF-8},
		Template.Latex.Compiler={pdflatex},
		Template.License.Type={MIT},
		Template.License.Web={https://opensource.org/licenses/MIT},
		Template.Name={Template-Articulo},
		Template.Type={Normal},
		Template.Version.Dev={1.3.2-ARTC},
		Template.Version.Hash={67FCD402563599FD99E1EBCFCFD6E2D7},
		Template.Version.Release={1.3.2},
		Template.Web.Dev={https://github.com/Template-Latex/Template-Articulo},
		Template.Web.Manual={https://latex.ppizarror.com/articulo},
	},
	pdfkeywords={\cfgpdfkeywords},
	pdfmenubar={\cfgpdfmenubar},
	pdfpagelayout={\cfgpdflayout},
	pdfpagemode={\cfgpdfpagemode},
	pdfproducer={Template-Articulo v1.3.2 | (Pablo Pizarro R.) ppizarror.com},
	pdfremotestartview={Fit},
	pdfstartpage={1},
	pdfstartview={\cfgpdfpageview},
	pdftitle={\pdfmetainfotitle},
	pdftoolbar={\cfgpdftoolbar}
}

\graphicspath{{./\defaultimagefolder}}

\makeatletter
\ifthenelse{\equal{\fpremovetopbottomcenter}{true}}{
	\setlength{\@fptop}{0pt}
	\setlength{\@fpbot}{0pt}
}{}
\makeatother

\ifthenelse{\equal{\showlinenumbers}{true}}{
	\setlength{\linenumbersep}{\marginlinenumbers pt}
	
	}{
}

\color{\maintextcolor} 
\arrayrulecolor{\tablelinecolor} 
\sethlcolor{\highlightcolor} 
\ifthenelse{\equal{\showborderonlinks}{true}}{
	\hypersetup{
		citebordercolor=\numcitecolor,
		linkbordercolor=\linkcolor,
		urlbordercolor=\urlcolor
	}
}{
	\hypersetup{
		hidelinks,
		colorlinks=true,
		citecolor=\numcitecolor,
		filecolor=\urlcolor,
		linkcolor=\linkcolor,
		urlcolor=\urlcolor
	}
}
\ifthenelse{\equal{\pagescolor}{white}}{}{
	\pagecolor{\pagescolor}
}

\setcaptionmargincm{\captionlrmargin}
\ifthenelse{\equal{\captiontextbold}{true}}{
	\renewcommand{\captiontextbold}{bf}}{
	\renewcommand{\captiontextbold}{}
}
\ifthenelse{\equal{\captiontextsubnumbold}{true}}{
	\renewcommand{\captiontextsubnumbold}{bf}}{
	\renewcommand{\captiontextsubnumbold}{}
}

\corecheckfontsize{\captionfontsize}
\corecheckfontsize{\subcaptionfsize}
\makeatletter
\renewcommand\p@subfigure{\thefigure\captionsubchar}
\renewcommand\p@subtable{\thetable\captionsubchar}
\makeatother

\ifthenelse{\equal{\figurecaptiontop}{true}}{
	\floatsetup[figure]{position=above}}{
}

\ifthenelse{\equal{\tablecaptiontop}{true}}{
	\floatsetup[table]{position=top}
	}{
	\floatsetup[table]{position=bottom}
}

\ifthenelse{\equal{\captionalignment}{justified}}{
	\captionsetup{
		format=plain,
		justification=justified
	}
}{
\ifthenelse{\equal{\captionalignment}{centered}}{
	\captionsetup{
		justification=centering
	}
}{
\ifthenelse{\equal{\captionalignment}{left}}{
	\captionsetup{
		justification=raggedright,
		singlelinecheck=false
	}
}{
\ifthenelse{\equal{\captionalignment}{right}}{
	\captionsetup{
		justification=raggedleft,
		singlelinecheck=false
	}
}{
	\throwbadconfig{Posicion de leyendas desconocida}{\captionalignment}{justified,centered,left,right}}}}
}

\ifthenelse{\equal{\stylecitereferences}{natbib}}{
	\def\twocolumnreferencesmargin {-0.35cm}
	\bibliographystyle{\natbibrefstyle}
	\setlength{\bibsep}{\natbibrefsep pt}
	\newcommand{\shortcite}[1]{\citep{#1}}
	\newcommand{\fullcite}[1]{\citet{#1}}
	\setcitestyle{open={\natbibrefcitecharopen},close={\natbibrefcitecharclose}}
	\ifthenelse{\equal{\natbibrefcitesepcomma}{true}}{
		\setcitestyle{comma}
	}{
		\setcitestyle{semicolon}
	}
	\ifthenelse{\equal{\natbibrefcitetype}{numbers}}{
		\setcitestyle{numbers}
	}{
	\ifthenelse{\equal{\natbibrefcitetype}{authoryear}}{
		\setcitestyle{authoryear}
	}{
	\ifthenelse{\equal{\natbibrefcitetype}{super}}{
		\setcitestyle{super}
	}{
		\throwbadconfig{Tipo cita natbib desconocido}{\natbibrefcitetype}{numbers,authoryear,super}}}
	}
}{
\ifthenelse{\equal{\stylecitereferences}{apacite}}{
	\def\twocolumnreferencesmargin {-0.39cm}
	\bibliographystyle{\apacitestyle}
	\setlength{\bibitemsep}{\apaciterefsep pt}
	\newcommand{\citep}[1]{\fullcite{#1}}
	\newcommand{\citet}[1]{\shortcite{#1}}
}{
\ifthenelse{\equal{\stylecitereferences}{bibtex}}{
	\def\twocolumnreferencesmargin {-0.35cm}
	\bibliographystyle{\bibtexstyle}
	\newlength{\bibitemsep}
	\setlength{\bibitemsep}{.2\baselineskip plus .05\baselineskip minus .05\baselineskip}
	\newlength{\bibparskip}\setlength{\bibparskip}{0pt}
	\let\oldthebibliography\thebibliography
	\renewcommand\thebibliography[1]{
		\oldthebibliography{#1}
		\setlength{\parskip}{\bibitemsep}
		\setlength{\itemsep}{\bibparskip}
	}
	\setlength{\bibitemsep}{\bibtexrefsep pt}
}{
\ifthenelse{\equal{\stylecitereferences}{custom}}{
	\coretemplatemessage{Usando estilo citas referencias custom, importar librerias y configuraciones posterior al llamado de template.tex en archivo principal}
}{
	\throwbadconfig{Estilo citas desconocido}{\stylecitereferences}{bibtex,apacite,natbib,custom}}}}
}

\makeatletter
\ifthenelse{\equal{\stylecitereferences}{apacite}}{
	\ifthenelse{\equal{\apaciterefnumber}{true}}{
		\newcounter{apaciteNumberCounter}
		\renewcommand{\theapaciteNumberCounter}{
			\apaciterefcitecharopen\arabic{apaciteNumberCounter}\apaciterefcitecharclose
		}
		\patchcmd{\@lbibitem}{\item[}{\item[\stepcounter{apaciteNumberCounter}{\hss\llap{\theapaciteNumberCounter}\quad}}{}{}
		\setlength{\bibleftmargin}{2.54em}
		\setlength{\bibindent}{-0.54em}
	}{}
}{}
\makeatother

\ifthenelse{\equal{\stylecitereferences}{apacite}}{
	\ifthenelse{\equal{\apaciteshowurl}{false}}{
		
		\AtBeginEnvironment{APACrefURL}{\renewcommand{\url}[1]{}}
		\renewcommand{\doiprefix}{doi:~\kern-1pt}
	}{}
}{}

\makeatletter
\ifthenelse{\equal{\twocolumnreferences}{true}}{
	\renewenvironment{thebibliography}[1]
	{\begin{multicols}{2}[\section*{\refname}\vspace{\twocolumnreferencesmargin}]
		\@mkboth{\MakeUppercase\refname}{\MakeUppercase\refname}
		\list{\@biblabel{\@arabic\c@enumiv}}
		{\settowidth\labelwidth{\@biblabel{#1}}
			\leftmargin\labelwidth
			\advance\leftmargin\labelsep
			\@openbib@code
			\usecounter{enumiv}
			\let\p@enumiv\@empty
			\renewcommand\theenumiv{\@arabic\c@enumiv}}
		\sloppy
		\clubpenalty 4000
		\@clubpenalty \clubpenalty
		\widowpenalty 4000
		\sfcode`\.\@m}
		{\def\@noitemerr
		{\@latex@warning{Ambiente `thebibliography' no definido}}
		\endlist\end{multicols}}}{}
\makeatother

\patchcmd{\appendices}{\quad}{\charappendixsection\spacingaftersection}{}{}

\begingroup
	\makeatletter
	\let\newcounter\@gobble\let\setcounter\@gobbletwo
	\globaldefs\@ne\let\c@loldepth\@ne
	\newlistof{listings}{lol}{\lstlistlistingname}
	\newlistentry{lstlisting}{lol}{0}
	\makeatother
\endgroup

\newcommand{\listindexequationsname}{\namelteqn}
\newlistof{myindexequations}{equ}{\listindexequationsname}
\newcommand{\myindexequations}[1]{
	\addcontentsline{equ}{myindexequations}{\protect\numberline{\theequation}#1}
}
\DeclareTotalCounter{templateIndexEquations}

\makeatletter
	\def\ifGm@preamble#1{\@firstofone}
	\appto\restoregeometry{
		\pdfpagewidth=\paperwidth
		\pdfpageheight=\paperheight}
	\apptocmd\newgeometry{
		\pdfpagewidth=\paperwidth
		\pdfpageheight=\paperheight}{}{}
\makeatother

\hbadness=\maxdimen
\vbadness=\maxdimen

\makeatletter
\def\Hv@scale {0.95}
\makeatother

\makeatletter 
\preto\tabular{\global\rownum=\z@}
\preto\tabularx{\global\rownum=\z@}
\makeatother

\ttfamily \hyphenchar\the\font=`\-

\ifthenelse{\equal{\compilertype}{pdf2latex}}{
	\pdfcompresslevel=\pdfcompilecompression
	
	\pdfdecimaldigits=2
	
	\pdfinclusionerrorlevel=0
	
	\pdfminorversion=\pdfcompileversion
	
	\pdfobjcompresslevel=\pdfcompileobjcompression
}{
\ifthenelse{\equal{\compilertype}{xelatex}}{
}{
\ifthenelse{\equal{\compilertype}{lualatex}}{
}{
	\throwbadconfig{Compilador desconocido}{\compilertype}{pdf2latex,xelatex,lualatex}}}
}

\newcounter{subsubsubsection}[subsubsection]

\makeatletter
	\def\toclevel@subsubsubsection {4}
	\def\toclevel@paragraph {5}
	\def\toclevel@subparagraph {6}
	\ifthenelse{\equal{\charaftersectionnum}{}}{
		\def\l@subsubsubsection {\@dottedtocline{4}{6.97em}{4em}}
		\def\l@paragraph {\@dottedtocline{5}{10.97em}{5em}}
		\def\l@subparagraph {\@dottedtocline{6}{14em}{6em}}
	}{
		\def\l@subsubsubsection {\@dottedtocline{4}{7.83em}{4.15em}}
		\def\l@paragraph {\@dottedtocline{5}{11.98em}{4.92em}}
		\def\l@subparagraph {\@dottedtocline{6}{14.65em}{5.69em}}
	}
\makeatother

\makeatletter
\newcommand\sectionpunct[2]{%
	\expandafter\def\csname @seccntfmt@#1\endcsname##1{%
		\csname the##1\endcsname#2%
	}%
}
\def\@seccntformat#1{\@ifundefined{#1@cntformat}%
	{\csname the#1\endcsname} 
	{\csname #1@cntformat\endcsname} 
}
\newcommand\section@cntformat{\GLOBALtitlepresectionstr\thesection\charaftersectionnum\spacingaftersection}
\newcommand\subsection@cntformat{\GLOBALtitlepresubsectionstr\thesubsection\charaftersectionnum\spacingaftersection}
\newcommand\subsubsection@cntformat{\GLOBALtitlepresubsubsectionstr\thesubsubsection\charaftersectionnum\spacingaftersection}
\makeatother

\titlespacing*{\section}{\sectionspacingleft pt}{\sectionspacingtop pt plus 0pt minus 4pt}{\sectionspacingbottom pt plus 0pt minus 2pt}
\titlespacing*{\subsection}{\ssectionspacingleft pt}{\ssectionspacingtop pt plus 0pt minus 2pt}{\ssectionspacingbottom pt plus 0pt minus 2pt}
\titlespacing*{\subsubsection}{\sssectionspacingleft pt}{\sssectionspacingtop pt plus 0pt minus 2pt}{\sssectionspacingbottom pt plus 0pt minus 2pt}
\titlespacing*{\subsubsubsection}{\ssssectionspacingleft pt}{\ssssectionspacingtop pt plus 0pt minus 2pt}{\ssssectionspacingbottom pt plus 0pt minus 2pt}
\makeatletter
\renewcommand\paragraph{\@startsection{paragraph}{5}{\paragspacingleft pt}
	{\paragspacingtop pt \@plus 0pt \@minus 2pt}
	{\paragspacingbottom pt \@plus 0pt \@minus 2pt}
	{\color{\paragcolor}\normalfont\paragfontsize\paragfontstyle}}
\renewcommand\subparagraph{\@startsection{subparagraph}{6}{\paragsubspacingleft pt}
	{\paragsubspacingtop pt \@plus 0pt \@minus 2pt}
	{\paragsubspacingbottom pt \@plus 0pt \@minus 2pt}
	{\color{\paragsubcolor}\normalfont\paragsubfontsize\paragsubfontstyle}}
\makeatother

\ifthenelse{\equal{\footnoterestart}{none}}{
}{
\ifthenelse{\equal{\footnoterestart}{sec}}{
	\counterwithin*{footnote}{section}
}{
\ifthenelse{\equal{\footnoterestart}{ssec}}{
	\counterwithin*{footnote}{subsection}
}{
\ifthenelse{\equal{\footnoterestart}{sssec}}{
	\counterwithin*{footnote}{subsubsection}
}{
\ifthenelse{\equal{\footnoterestart}{ssssec}}{
	\counterwithin*{footnote}{subsubsubsection}
}{
\ifthenelse{\equal{\footnoterestart}{page}}{
	\counterwithin*{footnote}{page}
}{
\ifthenelse{\equal{\footnoterestart}{chap}}{
	\counterwithin*{footnote}{chapter}
}{
	\throwbadconfig{Formato reinicio numero footnote desconocido}{\footnoterestart}{none,chap,page,sec,ssec,sssec,ssssec}}}}}}}
}

\ifthenelse{\equal{\footnoterulefigure}{false}}{
	\floatsetup[figure]{footnoterule=none}}{
}
\ifthenelse{\equal{\footnoteruletable}{false}}{
	\floatsetup[table]{footnoterule=none}}{
}

\ifthenelse{\equal{\equationrestart}{none}}{
}{
\ifthenelse{\equal{\equationrestart}{chap}}{
}{
\ifthenelse{\equal{\equationrestart}{sec}}{
}{
\ifthenelse{\equal{\equationrestart}{ssec}}{
}{
\ifthenelse{\equal{\equationrestart}{sssec}}{
}{
\ifthenelse{\equal{\equationrestart}{ssssec}}{
}{
	\throwbadconfig{Formato reinicio numero ecuacion desconocido}{\equationrestart}{none,chap,sec,ssec,sssec,ssssec}}}}}}
}

\newtheoremstyle{templatetheorem}{\baselineskip}{3pt}{\itshape}{}{\bfseries}{}{.5em}{}
\newtheoremstyle{templateobs}{\baselineskip}{3pt}{}{}{\bfseries}{}{.5em}{}
\theoremstyle{templatetheorem}

\ifthenelse{\equal{\showsectioncaptionmat}{none}}{
	\newtheorem{defn}{\namemathdefn}
	\newtheorem{teo}{\namemaththeorem}
	\newtheorem{cor}{\namemathcol}
	\newtheorem{lema}{\namemathlem}
	\newtheorem{prop}{\namemathprp}
}{
\ifthenelse{\equal{\showsectioncaptionmat}{chap}}{
	\newtheorem{defn}{\namemathdefn}[chapter]
	\newtheorem{teo}{\namemaththeorem}[chapter]
	\newtheorem{cor}{\namemathcol}[chapter]
	\newtheorem{lema}{\namemathlem}[chapter]
	\newtheorem{prop}{\namemathprp}[chapter]
}{
\ifthenelse{\equal{\showsectioncaptionmat}{sec}}{
	\newtheorem{defn}{\namemathdefn}[section]
	\newtheorem{teo}{\namemaththeorem}[section]
	\newtheorem{cor}{\namemathcol}[section]
	\newtheorem{lema}{\namemathlem}[section]
	\newtheorem{prop}{\namemathprp}[section]
}{
\ifthenelse{\equal{\showsectioncaptionmat}{ssec}}{
	\newtheorem{defn}{\namemathdefn}[subsection]
	\newtheorem{teo}{\namemaththeorem}[subsection]
	\newtheorem{cor}{\namemathcol}[subsection]
	\newtheorem{lema}{\namemathlem}[subsection]
	\newtheorem{prop}{\namemathprp}[subsection]
}{
\ifthenelse{\equal{\showsectioncaptionmat}{sssec}}{
	\newtheorem{defn}{\namemathdefn}[subsubsection]
	\newtheorem{teo}{\namemaththeorem}[subsubsection]
	\newtheorem{cor}{\namemathcol}[subsubsection]
	\newtheorem{lema}{\namemathlem}[subsubsection]
	\newtheorem{prop}{\namemathprp}[subsubsection]
}{
\ifthenelse{\equal{\showsectioncaptionmat}{ssssec}}{

}{
	\throwbadconfig{Valor configuracion incorrecto}{\showsectioncaptionmat}{none,chap,sec,ssec,sssec,ssssec}}}}}}
}
\theoremstyle{templateobs}

\let\cleardoublepage\clearpage

\ifthenelse{\equal{\twopagesclearformat}{blank}}{
	\let\emptypagespredocformat\insertblankpage
}{
\ifthenelse{\equal{\twopagesclearformat}{empty}}{
	\let\emptypagespredocformat\insertemptypage
}{
	\throwbadconfig{Valor configuracion incorrecto}{\twopagesclearformat}{blank,empty}}
}

\unaccentedoperators

\AtEndDocument{
	\addtocounter{equation}{\value{templateEquations}}
	\addtocounter{figure}{\value{templateFigures}}
	\addtocounter{lstlisting}{\value{templateListings}}
	\addtocounter{table}{\value{templateTables}}
}

\newcolumntype{C}[1]{>{\centering\let\newline\\\arraybackslash\hspace{0pt}}m{#1}}
\newcolumntype{\CColor}[2]{>{\columncolor{#1}\centering\let\newline\\\arraybackslash\hspace{0pt}}m{#2}}

\newcolumntype{P}[1]{>{\centering\let\newline\\\arraybackslash\hspace{0pt}}p{#1}}
\newcolumntype{\PColor}[2]{>{\columncolor{#1}\centering\let\newline\\\arraybackslash\hspace{0pt}}p{#2}}

\newcolumntype{B}[1]{>{\centering\let\newline\\\arraybackslash\hspace{0pt}}b{#1}}
\newcolumntype{\BColor}[2]{>{\columncolor{#1}\centering\let\newline\\\arraybackslash\hspace{0pt}}b{#2}}

\newcolumntype{L}[1]{>{\raggedright\let\newline\\\arraybackslash\hspace{0pt}}m{#1}}
\newcolumntype{\LColor}[2]{>{\columncolor{#1}\raggedright\let\newline\\\arraybackslash\hspace{0pt}}m{#2}}
\newcolumntype{T}[1]{>{\raggedright\let\newline\\\arraybackslash\hspace{0pt}}p{#1}}
\newcolumntype{\TColor}[2]{>{\columncolor{#1}\raggedright\let\newline\\\arraybackslash\hspace{0pt}}p{#2}}
\newcolumntype{F}[1]{>{\raggedright\let\newline\\\arraybackslash\hspace{0pt}}b{#1}}
\newcolumntype{\FColor}[2]{>{\columncolor{#1}\raggedright\let\newline\\\arraybackslash\hspace{0pt}}b{#2}}

\newcolumntype{R}[1]{>{\raggedleft\let\newline\\\arraybackslash\hspace{0pt}}m{#1}}
\newcolumntype{\RColor}[2]{>{\columncolor{#1}\raggedleft\let\newline\\\arraybackslash\hspace{0pt}}m{#2}}
\newcolumntype{H}[1]{>{\raggedleft\let\newline\\\arraybackslash\hspace{0pt}}p{#1}}
\newcolumntype{\HColor}[2]{>{\columncolor{#1}\raggedleft\let\newline\\\arraybackslash\hspace{0pt}}p{#2}}
\newcolumntype{G}[1]{>{\raggedleft\let\newline\\\arraybackslash\hspace{0pt}}b{#1}}
\newcolumntype{\GColor}[2]{>{\columncolor{#1}\raggedleft\let\newline\\\arraybackslash\hspace{0pt}}b{#2}}

\BeforeBeginEnvironment{tablenotes}{%
	\tablenotesfontsize\selectfont%
}
\AfterEndEnvironment{tablenotes}{%
	\normalsize\selectfont%
}

\let\SOURCEcaptionlrmargin\captionlrmargin
\newcounter{multicoldepth}
\BeforeBeginEnvironment{multicols}{%
	\def\captionlrmargin {\captionlrmarginmc}%
	\global\def\GLOBALenvmulticol {true}%
	\setcaptionmargincm{\captionlrmargin}%
	\addtocounter{multicoldepth}{1}%
}
\AfterEndEnvironment{multicols}{%
	\def\captionlrmargin {\SOURCEcaptionlrmargin}%
	\setcaptionmargincm{\captionlrmargin}%
	\addtocounter{multicoldepth}{-1}
	\ifnumequal{\number\value{multicoldepth}}{0}{%
		\global\def\GLOBALenvmulticol {false}
	}{}
}

\checkmodulenotloaded{tcolorbox}

\soulregister\cite7
\soulregister\eqref7
\soulregister\eqref7
\soulregister\footnote7
\soulregister\href7
\soulregister\pageref7
\soulregister\quotes7
\soulregister\ref7
\soulregister\scite7

\AtBeginDocument{%
	\normalfont%
	\setlength{\parindent}{\documentparindent pt}%
	\setlength{\parskip}{\documentparskip pt}%
}

\newcommand{\templatePagecfg}{%
	
	\clearpage
	\ifthenelse{\equal{\predocpageromannumber}{true}}{
		\ifthenelse{\equal{\predocpageromanupper}{true}}{%
			\pagenumbering{Roman}
		}{%
			\pagenumbering{roman}
		}}{%
		\pagenumbering{arabic}
	}
	\setcounter{page}{1}
	\setcounter{footnote}{0}
	
	\setpagemargincm{\pagemarginleft}{\pagemargintop}{\pagemarginright}{\pagemarginbottom}
	\resettablecellpadding
	
	\ifthenelse{\equal{\pointdecimal}{true}}{%
		\decimalpoint}{%
	}
	
	\renewcommand{\abstractname}{\nameabstract} 
	\renewcommand{\appendixname}{\nameltappendixsection} 
	\renewcommand{\appendixpagename}{\nameappendixsection} 
	\renewcommand{\appendixtocname}{\nameappendixsection} 
	\renewcommand{\contentsname}{\nameltcont} 
	\renewcommand{\figurename}{\nameltwfigure} 
	\renewcommand{\listfigurename}{\nameltfigure} 
	\renewcommand{\listtablename}{\namelttable} 
	\renewcommand{\lstlistingname}{\nameltwsrc} 
	\renewcommand{\lstlistlistingname}{\nameltsrc} 
	\renewcommand{\refname}{\namereferences} 
	\renewcommand{\bibname}{\namereferences} 
	\renewcommand{\tablename}{\nameltwtable} 
	
	\sectionfont{%
		\color{\sectioncolor} \sectionfontsize \sectionfontstyle \selectfont%
	}
	\subsectionfont{%
		\color{\ssectioncolor} \ssectionfontsize \ssectionfontstyle \selectfont%
	}
	\subsubsectionfont{%
		\color{\sssectioncolor} \sssectionfontsize \sssectionfontstyle \selectfont%
	}
	\titleformat{\subsubsubsection}{%
		\color{\ssssectioncolor} \ssssectionfontsz \ssssectionfontstyle%
	}{%
		\GLOBALtitlepresubsubsubsectionstr\thesubsubsubsection\charaftersectionnum\spacingaftersection%
		\corepatchaftersubsubsubsection%
	}{0em}{%
	}
	\def\bibfont {\fontsizerefbibl} 
	
	\ifthenelse{\equal{\stylecitereferences}{apacite}}{%
		\renewcommand{\BOthers}[1]{\apacitebothers\hbox{}}%
	}{}
	
	\ifthenelse{\isundefined{\authorshf}}{%
		\def\authorshf {}}{%
	}
	\fancyheadoffset{0pt} 
	\def\hfheaderimageparamsA {height=\baselineskip} 
	\ifthenelse{\equal{\hfstyle}{style1}}{%
		\pagestyle{fancy}
		\newcommand{\COREstyledefinition}{%
			\fancyhf{}
			\fancyfoot[C]{\thepage}
			\renewcommand{\headrulewidth}{0pt}
			\renewcommand{\footrulewidth}{0pt}
		}
		\setlength{\headheight}{49pt}
		\COREstyledefinition
	}{%
	\ifthenelse{\equal{\hfstyle}{style2}}{%
		\pagestyle{fancy}
		\newcommand{\COREstyledefinition}{%
			\fancyhf{}
			\fancyfoot[R]{\thepage}
			\renewcommand{\headrulewidth}{0pt}
			\renewcommand{\footrulewidth}{0pt}
		}
		\setlength{\headheight}{49pt}
		\COREstyledefinition
	}{%
	\ifthenelse{\equal{\hfstyle}{style3}}{%
		\pagestyle{fancy}
		\newcommand{\COREstyledefinition}{%
			\fancyhf{}
			\fancyhead[LE,RO]{\authorshf: \documenttitlehf}
			\fancyhead[RE,LO]{\thepage}
			\renewcommand{\headrulewidth}{0pt}
			\renewcommand{\footrulewidth}{0pt}
		}
		\fancypagestyle{portraitstyle}{%
			\fancyhf{}
			\fancyhead[L]{\journalname}
			\renewcommand{\headrulewidth}{0pt}
			\renewcommand{\footrulewidth}{0pt}
		}
		\thispagestyle{portraitstyle}
		\COREstyledefinition
	}{%
	\ifthenelse{\equal{\hfstyle}{style4}}{%
		\pagestyle{fancy}
		\newcommand{\COREstyledefinition}{%
			\fancyhf{}
			\fancyhead[LE,RO]{\small \textit{\authorshf}}
			\fancyhead[RE,LO]{\small \textit{\journalname}}
			\fancyfoot[C]{\small \thepage}
			\renewcommand{\headrulewidth}{0pt}
			\renewcommand{\footrulewidth}{0pt}
		}
		\fancypagestyle{portraitstyle}{%
			\fancyhf{}
			\fancyhead[C]{\small \journalname}
			\renewcommand{\headrulewidth}{0pt}
			\renewcommand{\footrulewidth}{0pt}
		}
		\thispagestyle{portraitstyle}
		\COREstyledefinition
	}{%
	\ifthenelse{\equal{\hfstyle}{style5}}{%
		\pagestyle{fancy}
		\newcommand{\COREstyledefinition}{%
			\fancyhf{}
			\fancyhead[LE,RO]{\thepage}
			\fancyhead[RE]{\authorshf}
			\fancyhead[LO]{\documenttitlehf}
			\renewcommand{\headrulewidth}{0.5pt}
			\renewcommand{\footrulewidth}{0pt}
		}
		\fancypagestyle{portraitstyle}{%
			\fancyhf{}
			\fancyhead[L]{\journalname}
			\renewcommand{\headrulewidth}{0.5pt}
			\renewcommand{\footrulewidth}{0pt}
		}
		\thispagestyle{portraitstyle}
		\COREstyledefinition
	}{%
	\ifthenelse{\equal{\hfstyle}{style6}}{%
		\pagestyle{fancy}
		\newcommand{\COREstyledefinition}{%
			\fancyhf{}
			\fancyhead[R]{\journalname}
			\fancyfoot[C]{\thepage}
			\renewcommand{\headrulewidth}{0pt}
			\renewcommand{\footrulewidth}{0pt}
		}
		\COREstyledefinition
	}{%
	\ifthenelse{\equal{\hfstyle}{style7}}{%
		\pagestyle{fancy}
		\newcommand{\COREstyledefinition}{%
			\fancyhf{}
			\fancyhead[LE,RO]{\textbf{\thepage}}
			\fancyhead[RE]{\textbf{\authorshf}}
			\fancyhead[LO]{\textbf{\documenttitlehf}}
			\renewcommand{\headrulewidth}{0pt}
			\renewcommand{\footrulewidth}{0pt}
		}
		\fancypagestyle{portraitstyle}{%
			\fancyhf{}
			\renewcommand{\headrulewidth}{0pt}
			\renewcommand{\footrulewidth}{0pt}
		}
		\thispagestyle{portraitstyle}
		\COREstyledefinition
	}{%
	\ifthenelse{\equal{\hfstyle}{style8}}{%
		\pagestyle{fancy}
		\newcommand{\COREstyledefinition}{%
			\fancyhf{}
			\fancyfoot[C]{\thepage}
			\renewcommand{\headrulewidth}{0pt}
			\renewcommand{\footrulewidth}{0pt}
		}
		\fancypagestyle{portraitstyle}{%
			\fancyhf{}
			\fancyhead[C]{\journalname}
			\fancyfoot[C]{\thepage}
			\renewcommand{\headrulewidth}{0pt}
			\renewcommand{\footrulewidth}{0pt}
		}
		\thispagestyle{portraitstyle}
		\COREstyledefinition
	}{%
	\ifthenelse{\equal{\hfstyle}{style9}}{%
		\pagestyle{fancy}
		\newcommand{\COREstyledefinition}{%
			\fancyhf{}
			\fancyhead[LE]{\thepage\quad\quad\authorshf}
			\fancyhead[RO]{\documenttitlehf\quad\quad\thepage}
			\renewcommand{\headrulewidth}{0pt}
			\renewcommand{\footrulewidth}{0pt}
		}
		\fancypagestyle{portraitstyle}{%
			\fancyhf{}
			\fancyfoot[L]{\journalname}
			\renewcommand{\headrulewidth}{0pt}
			\renewcommand{\footrulewidth}{0pt}
		}
		\thispagestyle{portraitstyle}
		\COREstyledefinition
	}{%
	\ifthenelse{\equal{\hfstyle}{style10}}{%
		\pagestyle{fancy}
		\newcommand{\COREstyledefinition}{%
			\fancyhf{}
			\fancyhead[C]{\journalname}
			\fancyfoot[R]{\thepage}
			\renewcommand{\headrulewidth}{0pt}
			\renewcommand{\footrulewidth}{0pt}
		}
		\COREstyledefinition
	}{%
	\ifthenelse{\equal{\hfstyle}{style11}}{%
		\pagestyle{fancy}
		\newcommand{\COREstyledefinition}{%
			\fancyhf{}
			\fancyhead[LE,RO]{\small \journalname}
			\fancyhead[LO,RE]{\small \authorshf}
			\renewcommand{\headrulewidth}{0pt}
			\renewcommand{\footrulewidth}{0pt}
		}
		\fancypagestyle{portraitstyle}{%
			\fancyhf{}
			\renewcommand{\headrulewidth}{0pt}
			\renewcommand{\footrulewidth}{0pt}
		}
		\thispagestyle{portraitstyle}
		\COREstyledefinition
	}{%
	\ifthenelse{\equal{\hfstyle}{style12}}{%
		\pagestyle{fancy}
		\newcommand{\COREstyledefinition}{%
			\fancyhf{}
			\fancyhead[LE,RO]{\small \thepage}
			\fancyhead[C]{\small \textit{\authorshf / \journalname}}
			\renewcommand{\headrulewidth}{0pt}
			\renewcommand{\footrulewidth}{0pt}
		}
		\fancypagestyle{portraitstyle}{%
			\fancyhf{}
			\fancyhead[C]{\small \journalname}
			\renewcommand{\headrulewidth}{0pt}
			\renewcommand{\footrulewidth}{0pt}
		}
		\thispagestyle{portraitstyle}
		\COREstyledefinition
	}{%
	\ifthenelse{\equal{\hfstyle}{style13}}{%
		\pagestyle{fancy}
		\newcommand{\COREstyledefinition}{%
			\fancyhf{}
			\fancyhead[R]{\small \textit{\documenttitlehf}}
			\fancyfoot[L]{\small \journalname}
			\fancyfoot[R]{\small \thepage\ / \pageref{TotPages}}
			\renewcommand{\headrulewidth}{0.5pt}
			\renewcommand{\footrulewidth}{0.5pt}
		}
		\fancypagestyle{portraitstyle}{%
			\fancyhf{}
			\fancyfoot[L]{\small \journalname}
			\renewcommand{\headrulewidth}{0pt}
			\renewcommand{\footrulewidth}{0.5pt}
		}
		\thispagestyle{portraitstyle}
		\COREstyledefinition
	}{%
	\ifthenelse{\equal{\hfstyle}{style14}}{%
		\pagestyle{fancy}
		\newcommand{\COREstyledefinition}{%
			\fancyhf{}
			\fancyhead[C]{\small \journalname}
			\fancyfoot[R]{\small \thepage\ / \pageref{TotPages}}
			\renewcommand{\headrulewidth}{0pt}
			\renewcommand{\footrulewidth}{0pt}
		}
		\fancypagestyle{portraitstyle}{%
			\fancyhf{}
			\fancyhead[C]{\small \journalname}
			\renewcommand{\headrulewidth}{0pt}
			\renewcommand{\footrulewidth}{0pt}
		}
		\thispagestyle{portraitstyle}
		\COREstyledefinition
	}{%
	\ifthenelse{\equal{\hfstyle}{style15}}{%
		\pagestyle{fancy}
		\newcommand{\COREstyledefinition}{%
			\fancyhf{}
			\fancyhead[L]{\small \textit{\journalname}}
			\fancyhead[R]{\small \thepage\ \namepageof \pageref{TotPages}}
			\renewcommand{\headrulewidth}{0.5pt}
			\renewcommand{\footrulewidth}{0pt}
		}
		\fancypagestyle{portraitstyle}{%
			\fancyhf{}
			\fancyfoot[C]{\small \journalname}
			\renewcommand{\headrulewidth}{0pt}
			\renewcommand{\footrulewidth}{0.5pt}
		}
		\thispagestyle{portraitstyle}
		\COREstyledefinition
	}{%
	\ifthenelse{\equal{\hfstyle}{style16}}{%
		\pagestyle{fancy}
		\newcommand{\COREstyledefinition}{%
			\fancyhf{}
			\fancyhead[L]{\small \textit{\journalname}}
			\fancyhead[R]{\small \thepage\ \namepageof \pageref{TotPages}}
			\renewcommand{\headrulewidth}{0pt}
			\renewcommand{\footrulewidth}{0pt}
		}
		\fancypagestyle{portraitstyle}{%
			\fancyhf{}
			\fancyfoot[C]{\small \journalname}
			\renewcommand{\headrulewidth}{0pt}
			\renewcommand{\footrulewidth}{0pt}
		}
		\thispagestyle{portraitstyle}
		\COREstyledefinition
	}{%
	\ifthenelse{\equal{\hfstyle}{style17}}{%
		\pagestyle{fancy}
		\newcommand{\COREstyledefinition}{%
			\fancyhf{}
			\renewcommand{\headrulewidth}{0pt}
			\renewcommand{\footrulewidth}{0pt}
		}
		\renewcommand{\sectionmark}[1]{\markboth{##1}{}}
		\COREstyledefinition
	}{%
		\throwbadconfigondoc{Estilo de header-footer incorrecto}{\hfstyle}{style1 .. style17}}}}}}}}}}}}}}}}}
	}
	\fancypagestyle{plain}{%
		\fancyheadoffset{0pt}
		\COREstyledefinition
	}
	\floatpagestyle{plain}
	\rotfloatpagestyle{plain}

	\ifthenelse{\equal{\showlinenumbers}{true}}{%
		\linenumbers}{%
	}
	
}

\newcommand{\templateFinalcfg}{%
	
	\markboth{}{}
	\clearpage
	
	\ifthenelse{\equal{\showsectioncaptioncode}{none}}{%
		\def\sectionobjectnumcode {}
	}{%
	\ifthenelse{\equal{\showsectioncaptioncode}{sec}}{%
		\def\sectionobjectnumcode {\thesection\sectioncaptiondelimiter}
	}{%
	\ifthenelse{\equal{\showsectioncaptioncode}{ssec}}{%
		\def\sectionobjectnumcode {\thesubsection\sectioncaptiondelimiter}
	}{%
	\ifthenelse{\equal{\showsectioncaptioncode}{sssec}}{%
		\def\sectionobjectnumcode {\thesubsubsection\sectioncaptiondelimiter}
	}{%
	\ifthenelse{\equal{\showsectioncaptioncode}{ssssec}}{%
		\def\sectionobjectnumcode {\thesubsubsubsection\sectioncaptiondelimiter}
	}{%
	\ifthenelse{\equal{\showsectioncaptioncode}{chap}}{%
		\def\sectionobjectnumcode {\thechapter\sectioncaptiondelimiter}
	}{%
		\throwbadconfig{Valor configuracion incorrecto}{\showsectioncaptioncode}{none,chap,sec,ssec,sssec,ssssec}}}}}}
	}
	
	\ifthenelse{\equal{\showsectioncaptioneqn}{none}}{%
		\def\sectionobjectnumeqn {}
	}{%
	\ifthenelse{\equal{\showsectioncaptioneqn}{sec}}{%
		\def\sectionobjectnumeqn {\thesection\sectioncaptiondelimiter}
	}{%
	\ifthenelse{\equal{\showsectioncaptioneqn}{ssec}}{%
		\def\sectionobjectnumeqn {\thesubsection\sectioncaptiondelimiter}
	}{%
	\ifthenelse{\equal{\showsectioncaptioneqn}{sssec}}{%
		\def\sectionobjectnumeqn {\thesubsubsection\sectioncaptiondelimiter}
	}{%
	\ifthenelse{\equal{\showsectioncaptioneqn}{ssssec}}{%
		\def\sectionobjectnumeqn {\thesubsubsubsection\sectioncaptiondelimiter}
	}{%
	\ifthenelse{\equal{\showsectioncaptioneqn}{chap}}{%
		\def\sectionobjectnumeqn {\thechapter\sectioncaptiondelimiter}
	}{%
		\throwbadconfig{Valor configuracion incorrecto}{\showsectioncaptioneqn}{none,chap,sec,ssec,sssec,ssssec}}}}}}
	}
	
	\ifthenelse{\equal{\showsectioncaptionfig}{none}}{%
		\def\sectionobjectnumfig {}
	}{%
	\ifthenelse{\equal{\showsectioncaptionfig}{sec}}{%
		\def\sectionobjectnumfig {\thesection\sectioncaptiondelimiter}
	}{%
	\ifthenelse{\equal{\showsectioncaptionfig}{ssec}}{%
		\def\sectionobjectnumfig {\thesubsection\sectioncaptiondelimiter}
	}{%
	\ifthenelse{\equal{\showsectioncaptionfig}{sssec}}{%
		\def\sectionobjectnumfig {\thesubsubsection\sectioncaptiondelimiter}
	}{%
	\ifthenelse{\equal{\showsectioncaptionfig}{ssssec}}{%
		\def\sectionobjectnumfig {\thesubsubsubsection\sectioncaptiondelimiter}
	}{%
	\ifthenelse{\equal{\showsectioncaptionfig}{chap}}{%
		\def\sectionobjectnumfig {\thechapter\sectioncaptiondelimiter}
	}{%
		\throwbadconfig{Valor configuracion incorrecto}{\showsectioncaptionfig}{none,chap,sec,ssec,sssec,ssssec}}}}}}
	}
	
	\ifthenelse{\equal{\showsectioncaptiontab}{none}}{%
		\def\sectionobjectnumtab {}
	}{%
	\ifthenelse{\equal{\showsectioncaptiontab}{sec}}{%
		\def\sectionobjectnumtab {\thesection\sectioncaptiondelimiter}
	}{%
	\ifthenelse{\equal{\showsectioncaptiontab}{ssec}}{%
		\def\sectionobjectnumtab {\thesubsection\sectioncaptiondelimiter}
	}{%
	\ifthenelse{\equal{\showsectioncaptiontab}{sssec}}{%
		\def\sectionobjectnumtab {\thesubsubsection\sectioncaptiondelimiter}
	}{%
	\ifthenelse{\equal{\showsectioncaptiontab}{ssssec}}{%
		\def\sectionobjectnumtab {\thesubsubsubsection\sectioncaptiondelimiter}
	}{%
	\ifthenelse{\equal{\showsectioncaptiontab}{chap}}{%
		\def\sectionobjectnumtab {\thechapter\sectioncaptiondelimiter}
	}{%
		\throwbadconfig{Valor configuracion incorrecto}{\showsectioncaptiontab}{none,chap,sec,ssec,sssec,ssssec}}}}}}
	}
	
	\ifthenelse{\equal{\captionnumcode}{arabic}}{%
		\renewcommand{\thelstlisting}{\sectionobjectnumcode\arabic{lstlisting}}
	}{%
	\ifthenelse{\equal{\captionnumcode}{alph}}{%
		\renewcommand{\thelstlisting}{\sectionobjectnumcode\alph{lstlisting}}
	}{%
	\ifthenelse{\equal{\captionnumcode}{Alph}}{%
		\renewcommand{\thelstlisting}{\sectionobjectnumcode\Alph{lstlisting}}
	}{%
	\ifthenelse{\equal{\captionnumcode}{roman}}{%
		\renewcommand{\thelstlisting}{\sectionobjectnumcode\roman{lstlisting}}
	}{%
	\ifthenelse{\equal{\captionnumcode}{Roman}}{%
		\renewcommand{\thelstlisting}{\sectionobjectnumcode\Roman{lstlisting}}
	}{%
		\throwbadconfig{Tipo numero codigo fuente desconocido}{\captionnumcode}{arabic,alph,Alph,roman,Roman}}}}}
	}
	
	\ifthenelse{\equal{\captionnumequation}{arabic}}{%
		\renewcommand{\theequation}{\sectionobjectnumeqn\arabic{equation}}
	}{%
	\ifthenelse{\equal{\captionnumequation}{alph}}{%
		\renewcommand{\theequation}{\sectionobjectnumeqn\alph{equation}}
	}{%
	\ifthenelse{\equal{\captionnumequation}{Alph}}{%
		\renewcommand{\theequation}{\sectionobjectnumeqn\Alph{equation}}
	}{%
	\ifthenelse{\equal{\captionnumequation}{roman}}{%
		\renewcommand{\theequation}{\sectionobjectnumeqn\roman{equation}}
	}{%
	\ifthenelse{\equal{\captionnumequation}{Roman}}{%
		\renewcommand{\theequation}{\sectionobjectnumeqn\Roman{equation}}
	}{%
		\throwbadconfig{Tipo numero ecuacion desconocido}{\captionnumequation}{arabic,alph,Alph,roman,Roman}}}}}
	}
	
	\ifthenelse{\equal{\captionnumfigure}{arabic}}{%
		\renewcommand{\thefigure}{\sectionobjectnumfig\arabic{figure}}
	}{%
	\ifthenelse{\equal{\captionnumfigure}{alph}}{%
		\renewcommand{\thefigure}{\sectionobjectnumfig\alph{figure}}
	}{%
	\ifthenelse{\equal{\captionnumfigure}{Alph}}{%
		\renewcommand{\thefigure}{\sectionobjectnumfig\Alph{figure}}
	}{%
	\ifthenelse{\equal{\captionnumfigure}{roman}}{%
		\renewcommand{\thefigure}{\sectionobjectnumfig\roman{figure}}
	}{%
	\ifthenelse{\equal{\captionnumfigure}{Roman}}{%
		\renewcommand{\thefigure}{\sectionobjectnumfig\Roman{figure}}
	}{%
		\throwbadconfig{Tipo numero figura desconocido}{\captionnumfigure}{arabic,alph,Alph,roman,Roman}}}}}
	}
	
	\ifthenelse{\equal{\captionnumsubfigure}{arabic}}{%
		\renewcommand{\thesubfigure}{\arabic{subfigure}}
	}{%
	\ifthenelse{\equal{\captionnumsubfigure}{alph}}{%
		\renewcommand{\thesubfigure}{\alph{subfigure}}
	}{%
	\ifthenelse{\equal{\captionnumsubfigure}{Alph}}{%
		\renewcommand{\thesubfigure}{\Alph{subfigure}}
	}{%
	\ifthenelse{\equal{\captionnumsubfigure}{roman}}{%
		\renewcommand{\thesubfigure}{\roman{subfigure}}
	}{%
	\ifthenelse{\equal{\captionnumsubfigure}{Roman}}{%
		\renewcommand{\thesubfigure}{\Roman{subfigure}}
	}{%
		\throwbadconfig{Tipo numero subfigura desconocido}{\captionnumsubfigure}{arabic,alph,Alph,roman,Roman}}}}}
	}
	
	\ifthenelse{\equal{\captionnumtable}{arabic}}{%
		\renewcommand{\thetable}{\sectionobjectnumtab\arabic{table}}
	}{%
	\ifthenelse{\equal{\captionnumtable}{alph}}{%
		\renewcommand{\thetable}{\sectionobjectnumtab\alph{table}}
	}{%
	\ifthenelse{\equal{\captionnumtable}{Alph}}{%
		\renewcommand{\thetable}{\sectionobjectnumtab\Alph{table}}
	}{%
	\ifthenelse{\equal{\captionnumtable}{roman}}{%
		\renewcommand{\thetable}{\sectionobjectnumtab\roman{table}}
	}{%
	\ifthenelse{\equal{\captionnumtable}{Roman}}{%
		\renewcommand{\thetable}{\sectionobjectnumtab\Roman{table}}
	}{%
		\throwbadconfig{Tipo numero tabla desconocido}{\captionnumtable}{arabic,alph,Alph,roman,Roman}}}}}
	}
	
	\ifthenelse{\equal{\captionnumsubtable}{arabic}}{%
		\renewcommand{\thesubtable}{\arabic{subtable}}
	}{%
	\ifthenelse{\equal{\captionnumsubtable}{alph}}{%
		\renewcommand{\thesubtable}{\alph{subtable}}
	}{%
	\ifthenelse{\equal{\captionnumsubtable}{Alph}}{%
		\renewcommand{\thesubtable}{\Alph{subtable}}
	}{%
	\ifthenelse{\equal{\captionnumsubtable}{roman}}{%
		\renewcommand{\thesubtable}{\roman{subtable}}
	}{%
	\ifthenelse{\equal{\captionnumsubtable}{Roman}}{%
		\renewcommand{\thesubtable}{\Roman{subtable}}
	}{%
		\throwbadconfig{Tipo numero subtabla desconocido}{\captionnumsubtable}{arabic,alph,Alph,roman,Roman}}}}}
	}
	
	\let\cleardoublepage\corecleardoublepage
	
	\ifthenelse{\equal{\predocpageromannumber}{true}}{%
		\renewcommand{\thepage}{\arabic{page}}}{%
	}
	
	\ifthenelse{\equal{\predocresetpagenumber}{true}}{%
		\setcounter{page}{1}}{%
	}
	
	\setcounter{section}{0}
	\setcounter{footnote}{0}
	
	\ifthenelse{\equal{\showlinenumbers}{true}}{%
		\linenumbers}{%
	}
	
	\titleclass{\subsubsubsection}{straight}[\subsection]
	
	\global\def\GLOBALtitlerequirechapter {false}
	\global\def\GLOBALtitleinitchapter {false}
	\global\def\GLOBALtitleinitsection {false}
	\global\def\GLOBALtitleinitsubsection {false}
	\global\def\GLOBALtitleinitsubsubsection {false}
	\global\def\GLOBALtitleinitsubsubsubsection {false}
	
}

\usepackage{authblk}
\title{Temporal Stamp Classifier: 

Classifying Short Sequences of Astronomical Alerts}
\author[1,2]{Daniel Neira O.}
\author[1,2]{Pablo A. Estévez}
\author[2,3,4,5]{Francisco Förster}
\affil[1]{Department of Electrical Engineering, Universidad de Chile, Santiago, Chile} 
\affil[2]{Millennium Institute of Astrophysics, Santiago, Chile} 
\affil[3]{Data and Artificial Intelligence Initiative (IDIA), Universidad de Chile, Santiago, Chile}
\affil[4]{Center for Mathematical Modeling (CMM), Universidad de Chile, Santiago, Chile}
\affil[5]{Department of Astronomy, Universidad de Chile, Santiago, Chile}
\date{\vspace{-5ex}}

\begin{document}

\templatePagecfg

\templateFinalcfg



\changepagesizeformat{A4}

\pagenumbering{gobble}

\maketitle


\begin{multicols}{2}
    \begin{abstract}
    \end{abstract}
    \textbf{In this work, we propose a deep learning-based classification model of astronomical objects using alerts reported by the Zwicky Transient Facility (ZTF) survey. The model takes as inputs sequences of stamp images and metadata contained in each alert, as well as features from the AllWISE catalog. The proposed model, called temporal stamp classifier, is able to discriminate between three classes of astronomical objects:  Active Galactic Nuclei (AGN), SuperNovae (SNe) and Variable Stars (VS), with an accuracy of approximately $98\%$ in the test set, when using 2 to 5 detections. The results show that the model performance improves with the addition of more detections. Simple recurrence models obtain competitive results with those of more complex models such as LSTM. We also propose changes to the original stamp classifier model, which only uses the first detection. The performance of the latter model improves with changes in the architecture and the addition of random rotations, achieving a $1.46\%$ increase in test accuracy.   }
    \section{Introduction}

Astronomy is facing the challenge of increasingly large streams of data produced by large survey telescopes. The collection of this high volume of data is managed by astronomical surveys such as the Zwicky Transient Facility Survey (ZTF)  [1] and, in the near future, by the Legacy Survey of Space and Time (LSST) [2]. The latter will generate over 30TB of data per night with information on astronomical objects that change in time and/or position, e.g., supernovae (SNe). This increase in data production has opened the way for a lot of efforts to obtain fast and accurate methods capable of managing large amounts of data in real-time. This task is being managed by intermediary agents called astronomical alert brokers, who have to receive alert data streams, process them, and then classify and report astronomical objects to facilitate their follow-up and study by the astronomical community.  \blfootnote{The authors acknowledge support from the National Agency of Research and Development (ANID) Millennium Science Initiative through grant ICN12\_009, awarded to the Millennium Institute of Astrophysics. DN and PE acknowledge support from ANID grant FONDECYT Regular 1220829. FF acknowledge support from ANID grant BASAL Center of Mathematical Modeling PAI AFB-170001, FONDECYT Regular 1200710 and infrastructure funds QUIMAL190012.}

The Automatic Learning for the Rapid Classification of Events (ALeRCE) broker [3] is currently processing ZTF data in preparation for the LSST era. ZTF alerts come in a triad of images: the most recent observation of the object (science), an averaging of multiple images of the region at the beginning of the survey (template), and the difference between the two previous images showing the change in flux of the event (difference). These images are cropped around an object of interest and are called stamps, e.g., in ZTF, the stamps have $63\times63$ pixels. Alerts also contain additional information, called metadata, such as position in the sky, magnitude of the flux, time of observation, observation error, information from other catalogs, etc. ALeRCE has developed a pipeline with two models to classify astronomical events: The real-time stamp classifier \cite{stamp} and the light curve classifier \cite{light_curve}. The former focuses on discriminating 5 classes of events using the first detection only. In what follows, we use alerts and detections as synonyms. The light-curve classifier processes time series having 6 or more samples.  This means that there is a gap between the first detection and six or more detections. In this work we develop a new deep learning-based model to fill this gap by processing sequences from two to five alerts. Notice that we are dealing with irregularly sampled time-series, which are a standard practice in astronomy.

We propose a model based on a convolutional neural network (CNN) followed by a recurrent neural network (RNN). This model uses all the information available from the first detection up to 2,3,4 and 5 alerts of an astronomical object, which includes the stamp images of the objects plus metadata. The stamp images used for the model are the science and template images (we do not use the difference images, which are more useful when there is a single detection, especially to detect bogus events). The proposed model focuses on the classification of three classes of astronomical objects: active galactic nuclei (AGN), supernovae (SNe) and variable stars (VS). 

An additional contribution of this work is the improvement of the performance of the original Stamp Classifier model. In this work we introduce random rotations around fixed angles, in order to exploit the rotational invariance of astronomical images. This modified stamp classifier is used as a basis for developing the temporal stamp classifier. The main contributions of this work are the following:

\begin{itemize}
    \item Propose a deep learning-based model for processing sequences of 2 to 5 alerts, using image stamps and metadata to classify 3 classes of astronomical objects: AGN, SNe, VS.
    \item Compare simple recurrent models with complex ones such as GRU and LSTM, in the context of short sequences of astronomical alerts.
    \item Improve the performance of the original Stamp Classifier model for classifying the first alert of astronomical events, by using random rotations.
\end{itemize}
    \section{Related work}

\subsection{Zwicky Transient Facility Survey}

The Zwicky Transient Facility (ZTF) is a time-domain survey currently in operation, that aims at extending our knowledge of temporal and dynamic sky. The real-time pipeline of ZTF generates a stream of image stamps of $63\times63$ pixels centered around the detected event. Figure \ref{fig:stamp_example} shows examples of science, template, and difference images for five classes of astronomical objects.
The image on the left is the science image and corresponds to the most recent measurement of the source. The image at the center is the reference image, which is fixed for a given region and bandpass. It is usually based on images taken at the beginning of the survey and it is built by averaging multiple images to improve its signal-to-noise ratio. The image at the right is the difference image, which shows the change in flux between the reference and science images, removing other sources with constant brightness \cite{ztf_difference}.

\subsection{Stamp Classifier}

The original Stamp Classifier \cite{stamp} is a real-time classifier based on a convolutional neural network. It was trained using the first alert of astronomical events reported by the ZTF survey. The science, template, and difference images are used as inputs, along with metadata of the alert, including features and crossmatches with other catalogs like PanSTARRS1, DAOPhot, and PSF-catalog. Each image from the ZTF survey is normalized between 0 and 1, followed by a replacement of NaN values with 0. Then all images are cropped at the center, obtaining images of $21 \times 21$ pixels. These images are rotated in angles of $0°$, $90°$, $180°$, and $270°$ and are injected separately to the convolutional layers, followed by a cyclic pooling process that takes average at the output to exploit the rotational invariance of astronomical images. This model classifies the alerts into five classes: active galactic nuclei (AGN), supernovae (SNe), variable stars (VS), asteroids, and bogus.

\subsection{Light Curve Classifier}

The light curve classifier \cite{light_curve} is a two-level scheme classifier based on a Balanced Random Forest algorithm. It uses astronomical alerts generated by the ZTF survey to extract features computed directly from the alerts, colors obtained from the AllWISE catalog, and ZTF photometry. The top-level classifies each example into one of three classes: periodic, stochastic, or transient. The bottom level further resolves the three hierarchical classes into 15 subclasses. To compute features from the alerts, it is necessary to have at least 6 detections of an event in the g-band or in the r-band.

\section{Enhancing the Original Stamp Classifier}

In this work, we introduce improvements to the original Stamp Classifier model \cite{stamp} by including random rotations and a different image size. This modification led us to an improvement in the performance of the model for every class.

Our aim is to improve the first detection model by modifying the way it process the images and, in consequence, obtain the best configuration to process astronomical images that is going to be used in the classification of sequence of detections. 

\subsection{Data}

The data used in this section was reported by the ZTF survey and curated by the ALeRCE broker and correspond to the same dataset used in \cite{stamp}. It contains samples from objects detected at least one time. The data considers 5 classes of objects AGN, SN, VS, asteroids and bogus with a number of examples of $14966$ ($29\%$), $1620$ ($3\%$), $14996$ ($29\%$), $9899$ ($19\%$) and $10763$ ($20\%$), respectively, with a total of $52,244$ samples. 

\begin{figure}
    \centering
    \includegraphics[scale = 0.6]{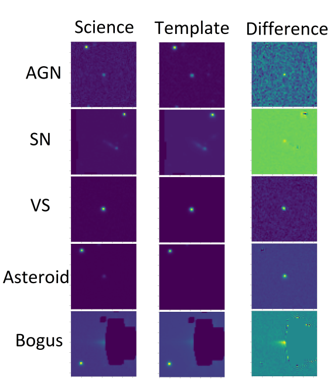}
    \caption{Examples of ZTF alerts of five classes of astronomical objects: AGN, SN, VS, Asteroid and Bogus.}
    \label{fig:stamp_example}
\end{figure}

Each sample contains the first alert of an astronomical object. It is composed of three images called stamps and metadata related to the source, the observation conditions of the exposure, and other useful information. The stamps are cropped at 63 pixels on a side from the original image and centered at the position of the object. Figure \ref{fig:stamp_example} shows an example of the three images contained in an alert. The metadata used in this section is the same as in \cite{stamp}.

The dataset described aboved was pre-processed as follows. First, alerts with an image with a shape different from $63\times 63$ were discarded, because it is considered an error in capture. Images were cropped at 45 pixels per side centered at the point-wise objects of interest, and normalized between 0 and 1. For metadata used as features, we limit their values as shown in Table \ref{limmetadatos}, using the training set, and then we normalize each feature by subtracting its mean value in the training set and dividing the

\noindent result by its standard deviation. We split our data into train, validation and test sets, taking randomly 100 samples per class for the validation set and 100 samples per class for the test set.

\subsection{Enhanced stamp classifier model}

The random rotations procedure is described in what follows. The algorithm has three parameters. The first parameter is the number of rotations $k$, which are equally spaced, following the function $r = [360/k]°$ which rotates the images by $r$ degrees. The second parameter is the rotation range $\theta$, which defines how much an image can rotate around a fixed point. For instance, if $k = 5$ and $\theta = 10°$, then the first rotation will take any value in $72° \pm 10°$, as shown in Figure 3. This operation is carried out at each iteration of the model. Using this notation, the original stamp classifier is described by $k = 4$ and $\theta = 0°$.

The convolutional neural network model receives a triad of images of $45 \times 45$, which includes k-rotated versions of each alert image, resulting in an augmented batch $B(x) = [x, rx, ..., r^{k-1}x, r^k x]$; and then, each rotated image is cropped at the center in $a \times a$, where $a$ is a search parameter. All the convolutional layers, except for the first and the last one, use zero padding in order to keep the same dimension at the output of each layer, and are followed by ReLU activation functions. The output of the last Max Pooling layer is flattened to form the vectors that enter the Fully Connected layer, as shown at the middle of Figure 3. This is the last layer that works independently with the different rotations. At the output of this Max Pooling layer, the vectors of each rotation are stacked, obtaining a $k \times 64$ tensor, and the values in the stacked dimension are averaged, finally obtaining a 64-dimensional vector, this process is called Cyclic Pooling. The metadata is concatenated at the output of the cyclic pooling, obtaining a vector of 87 dimensions, which is inputted to the Fully Connected layers, to obtain the estimated probabilities for the 5 classes. The enhanced stamp classifier is shown in Figure 3.

\subsection{Experiments}

\begin{figure}
    \centering
    \includegraphics[width = 0.7 \textwidth]{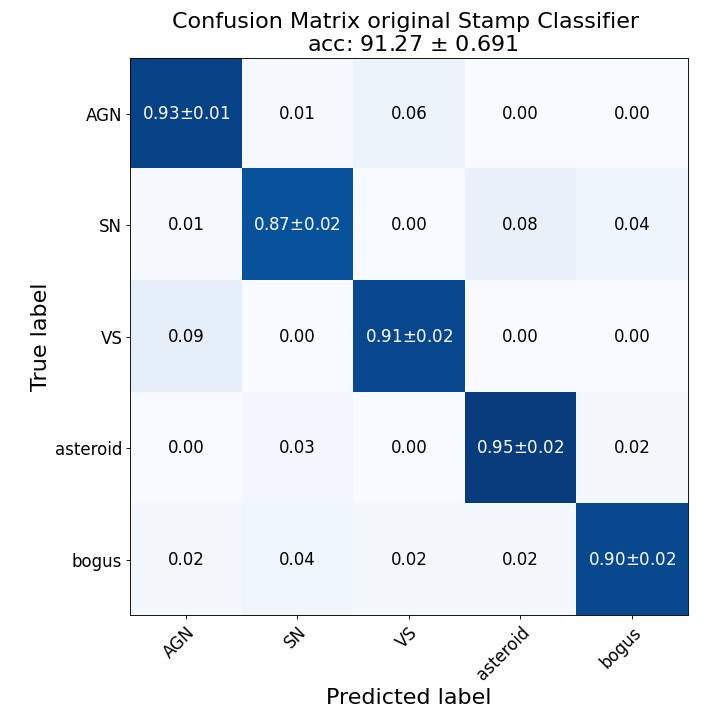}
    \caption{Confusion matrix of the original Stamp Classifier \cite{stamp} using our own implementation.}
    \label{fig:stamp_original}
\end{figure}

\begin{figure}
    \centering
    \includegraphics[width = 0.7 \textwidth]{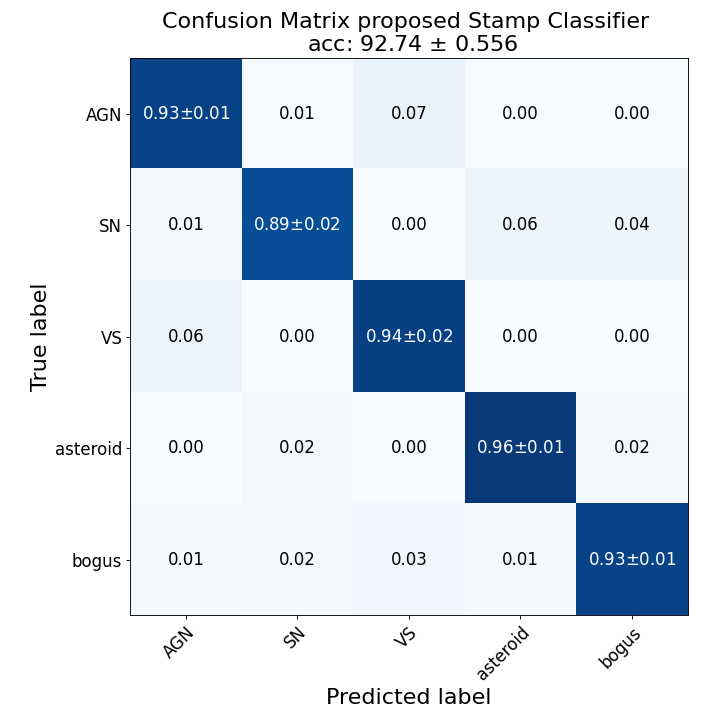}
    \caption{Confusion matrix of the enhanced Stamp Classifier that includes an image size of $33 \times 33$ and 6 rotations with $0.08[rad]$ of rotation range.}
    \label{fig:stamp_original_propuesto}
\end{figure}

We tested all the hyperparameter combinations shown in Table \ref{param-stampcla} with 5 initializations for every combination including a total rotation range $\theta = 360°$. The latter means random rotations between $0°$ and $360°$, i.e., total random rotations without restrictions. We obtained the average accuracy in the validation set, and selected the top 5 results shown in Table \ref{table:stamp_models}.  At the bottom of this table, we present the results of the baseline model $M_b$ and the model with fully random rotations $M_r$ for comparison purposes. Next, using the best hyperparameters, each model was trained 20 times to get a more robust result. Figure \ref{fig:stamp_original} shows the confusion matrix obtained with the original Stamp Classifier, and Figure \ref{fig:stamp_original_propuesto} shows the confusion matrix obtained with the enhanced Stamp Classifier, which includes random rotations and a different image cropping.

Table \ref{table:stamp_models} shows that the best image sizes are between 27 and 39 pixels per side, achieving the best results with an image of $33\times33$. Likewise, the best number of rotations and rotation range, are 4 or more rotations and a rotation range close to $0.1$ [rad]. This is corroborated by the p-value obtained comparing the proposed models with the baseline model, obtaining values smaller than $0.05$ for the top 5 models in the table. These p-values were obtained with 5 repetitions of the same experiments and using the permutation test \cite{hipotesis}. In the case of fully random rotations ($\theta = 360°$), i.e., the model $M_r$, the results were inconsistent between different tests. The results could go from a low accuracy ($91.3\%$) to a high accuracy ($93\%$) depending on the iteration that was made. For instance, the best model with $360°$ of rotation range (using 5 rotations and a image size of $33 \times 33$) give us a good result the first time that we tested it, but when the process was repeated, the results varied. When the rotation range is $360°$, we obtain a totally different variation every time that the model is called, one call could give rotations equispaced and in another one rotations closer to each other.

Figures \ref{fig:stamp_original} and \ref{fig:stamp_original_propuesto} show robust statistical results when using 20 repetitions for each experiment. The difference between both experiments is statistically significant with a p-value of $0.01$. This led us to conclude that the inclusion of random rotations with a larger image crop size is a relevant improvement to the Stamp Classifier.

\begin{table}[H]
\resizebox{0.8\columnwidth}{!}{%
\begin{tabular}{c|c}
Hyperparameter                   & Search values        \\ \hline 
k rotations                     & 4; 5; 6; 7; 8          \\ \hline
Rotation range $\theta$                   & $0^o$; $7^o$; $14^o$, $22^o$; $29^o$; $36^o$; $360°$  \\ \hline
$a$                    & 21, 27, 33, 39               \\ \hline
\end{tabular}
}
\caption{Hyperparameter search for enhancing the original Stamp Classifier through random rotations.}
\label{param-stampcla}
\end{table}

\begin{table}[H]
\resizebox{\columnwidth}{!}{%
\begin{tabular}{|l|l|l|l|}
\hline
Model & Hyperparameter & Test accuracy & p-value \\ \hline
$M_1$ & $a: 33$ ; $k:6$ ; $\theta:$ $0.08$ & $93.36 \pm 0.4$  & $0.007$ \\ \hline
$M_2$ & $a: 27$ ; $k:6$ ; $\theta:$ $0.1$ & $92.96 \pm 0.5$  & $0.008$  \\ \hline
$M_3$ & $a: 27$ ; $k:5$ ; $\theta:$ $0.1$ & $92.84 \pm 0.4$   & $0.008$ \\ \hline
$M_4$ & $a: 39$ ; $k:8$ ; $\theta:$ $0.08$ & $92.84 \pm 0.3$   & $0.007$ \\ \hline
$M_5$ &  $a: 27$ ; $k:7$ ; $\theta:$ $0.08$ & $92.72 \pm 0.3$  & $0.007$  \\ \hline \hline
$M_r$ &  $a: 33$ ; $k:5$ ; $\theta:$ $1$ & $92.22 \pm 0.8$   & $0.3$  \\ \hline
$M_b$ &  $a: 21$ ; $k:4$ ; $\theta:$ 0 & $91.44 \pm 0.4$ & Reference  \\ \hline
\end{tabular}%
}
\caption{Top-5 models with the highest validation accuracies obtained from the hyperparameter search, ranked from $M_1$ to $M_5$. $M_r$ stands for the best model for a fully random rotation. $M_b$ stands for the baseline Stamp Classifier.}
\label{table:stamp_models}
\end{table}

\begin{imagesmc}{top}{Enhanced Stamp Classifier model with $k = 5$ rotations and a rotation range $\theta$.}
		\addimageanum{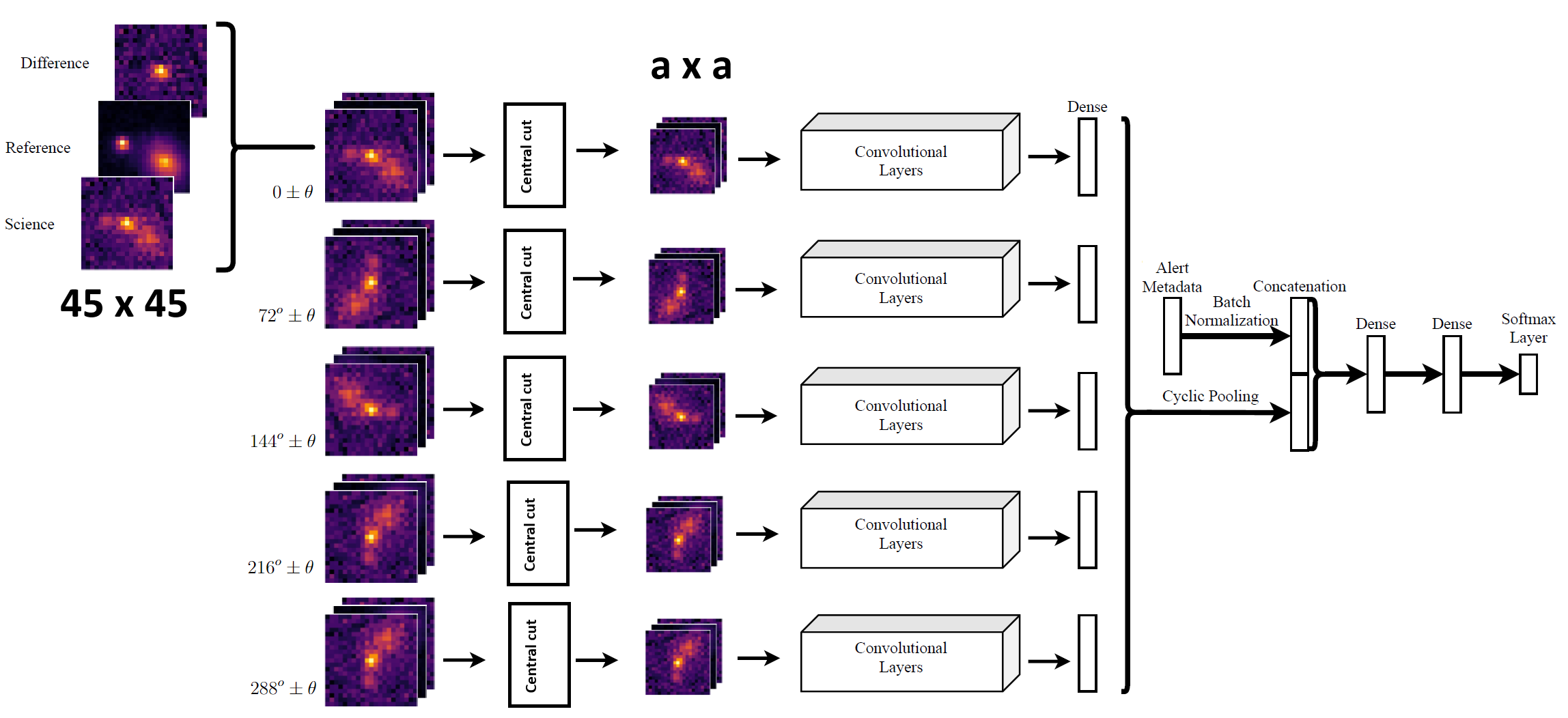}{width = 0.9 \textwidth}
\end{imagesmc}
\label{stamp_propuesto}

\section{Temporal Stamp Classifier: Extending the original model.}

In this work, we use sequence of alerts reported by ZTF and classify them into three categories: AGN, SN and VS classes. We assume that alerts with more than one detection correspond only to real objects, so the bogus class can be discarded. In section B we discuss about how to detect bogus with more than one detection. In addition, as we are looking repeatedly at specific locations in the sky, asteroids can also be discarded because they are moving objects. 

Our aim is to use sequences of 2 to 5 alerts and improve the results obtained when using just the first detection and approach the results of the light curve classifier \cite{light_curve}, which uses 6 or more detections. We use the information given by the images and the features, with the inclusion of crossmatch with the AllWISE catalog\cite{allwise}, as inputs to a neural network model that includes convolutional neural networks and recurrent neural networks. We compare the results obtained with different recurrent neural networks from the vanilla model to LSTM. 

\subsection{Data}

The data used in this section was reported by the ZTF survey. It contains samples from objects detected once or more times. The data considers 4 classes of objects AGN, SN, VS, and bogus with a number of examples of $48262$ ($28.7\%$), $2524$ ($1.5\%$), $101038$ ($60.1\%$), and $16332$ ($9.7\%$), respectively, with a total of $168,156$ examples.

Each sample contains 1 to 10 alerts of the objects. Each alert is composed of three images called stamps and metadata related to the source, the observation conditions of the exposure, and other useful information. The stamps are cropped at 63 pixels on a side from the original image and centered at the position of the object. The metadata used in this work is the same as in \cite{stamp}, except for the inclusion of AllWISE colors \cite{allwise} information through crossmatch by positions in the sky.

The dataset described above was pre-processed. First, all bogus examples were discarded. Alerts with images with a shape different from $63\times 63$ were discarded, because it is considered an error in capture. Then all consecutive alerts with less than 6 hours of difference between each other were discarded. In addition, all difference images were discarded, and the remaining images were cropped at $33$ pixels centered at the point of interest, and normalized between 0 and 1. For metadata used as features, we limit their values as shown in Table \ref{limmetadatos}, using the training set, and then we normalize each feature by subtracting its mean value in the training set and dividing the result by its standard deviation. Each number of detections has its own limit value for each metadata. Additionally, we separated metadata according to its time variability into static metadata and dynamic metadata. For instance, the PSF-Magnitude is dynamic because it's related with the magnitude of the flux measured in the given detection, while the Star Galaxy Score (sgscore) stands for star near the source that depends on the position of the object and this value shouldn't change with the detections. The static/dynamic category of every metadata is shown in Table \ref{limmetadatos}.

\begin{table}[H]
\centering
\resizebox{\columnwidth}{!}{
\begin{tabular}{c|c|c}
Metadata       & [min value, max value] & Category \\ \hline \hline
sgscore 1      & [-1, max]     & Static        \\ \hline
distpsnr1      & [-1, max]     & Static         \\ \hline
sgscore2       & [-1, max]             & Static \\ \hline
distpsnr2      & [-1, max]      & Static        \\ \hline
sgscore3       & [-1, max]      & Static        \\ \hline
distpsnr3      & [-1, max]      & Static        \\ \hline
ifwhm          & [min, 10]      & Dynamic        \\ \hline
ndethist       & [min, 20]      & Dynamic        \\ \hline
ncovhist       & [min, 3000]    & Dynamic        \\ \hline
chinr          & [-1, 15]       & Static        \\ \hline
sharpnr        & [-1, 1.5]      & Static        \\ \hline
no-detections & [min, 2000]     & Dynamic       \\ \hline
isdiffpos & [min, max]     & Dynamic       \\ \hline
magpsf & [min, max]     & Dynamic       \\ \hline
sigmapsf & [min, max]     & Dynamic       \\ \hline
RA, DEC & [min, max]     & Static       \\ \hline
diffmaglim & [min, max]     & Dynamic       \\ \hline
classtar & [min, max]     & Dynamic       \\ \hline
Galactic Coordinates & [min, max]     & Static       \\ \hline
Ecliptic Coordinates & [min, max]     & Static       \\ \hline
\end{tabular}
}
\caption{Range of values for each  metadata. Max and min stand for the maximum and minimum values obtained in the training set. The definition of the metadata can be found in \cite{stamp}.}
\label{limmetadatos}
\end{table}

\subsection{Bogus with more than one detection}

The dataset described above has $16,332$ bogus samples, and only $253$ of these samples has more than one detection. This is because most bogus, e.g. cosmic rays, don't appear in the same position twice. In order to detect bogus with more than one detection, we propose to use the Enhanced Stamp Classifier model with the first and the second detection and, if the model predicts a sample as bogus in the first and second detection, it can be considered as bogus and discarded. We tested $253$ bogus with the enhanced stamp classifier model, obtaining an accuracy of $86\%$ for bogus samples in the first detection and $84\%$ in the second detection. This result indicates that a total of $32$ out of $16,332$ bogus samples could go undetected. Representing less than $0.2\%$ of bogus. This small number of cases does not allow to build a Deep Learning based classifier. This problem should be addressed directly using other approaches.

\subsection{Temporal Stamp Classifier model}

As mentioned above, the goal is to develop a deep learning-based model that processes a short sequence of alerts, i.e., where the number of detections is $n \in \{2,3,4,5\}$. The enhanced stamp classifier uses convolutional neural networks (CNNs) to process a triad of images and their rotations. In the proposed model, CNNs are the first step, as shown in Figure 5.

The random rotations procedure is the same as the enhanced stamp classifier, with the addition of a third parameter that consists of the number of detections $n$.

The convolutional neural network model receives a set of images for each detection, which includes k-rotated versions of each alert image, resulting in an augmented batch $B(x) = [x, rx, ..., r^{k-1}x, r^k x]$; and generates an output for each detection. All the convolutional layers, except for the first and the last one, use zero padding in order to keep the same dimension at the output of each layer, and are followed by ReLU activation functions. The output of the last Max Pooling layer is flattened to form the vectors that enter the Fully Connected layers, as shown at the middle of figure 5. This is the last layer that works independently with the different rotations. At the output of this Max Pooling layer, the vectors of each rotation are stacked, obtaining a $k \times 64$ tensor, and the values in the stacked dimension are averaged, finally obtaining a 64-dimensional vector for each detection, this process is called Cyclic Pooling. The output of the n detections, a vector of size $n \times 64$, is concatenated with a vector of time differences among detections, plus the metadata, resulting in a vector of size  $ n \times 72$. This concatenated vector is the input of a recurrent neural network (RNN), obtaining a final output vector of 72 dimensions. After completing this recurrent processing, the static metadata is concatenated, obtaining a vector of 87 dimensions, which is inputted to the Fully Connected layers, to obtain the estimated probabilities for the 3 classes.
As shown in Figure 5 and Table \ref{tab:temporal_stamp}, the recurrent layer is not defined beforehand, allowing us to test several different recurrent models from vanilla RNN to LSTM.

\begin{imagesmc}{top}{Temporal Stamp Classifier model with $k = 5$ and a range of rotation $\theta$. Convolutional layers refers to the convolutional filters of table \ref{tab:temporal_stamp} from the first Convolution to Flatten layers. For each sample, we use \textit{science} and \textit{template} images as inputs and metadata that are separated into dynamic and static features.}
		\addimageanum{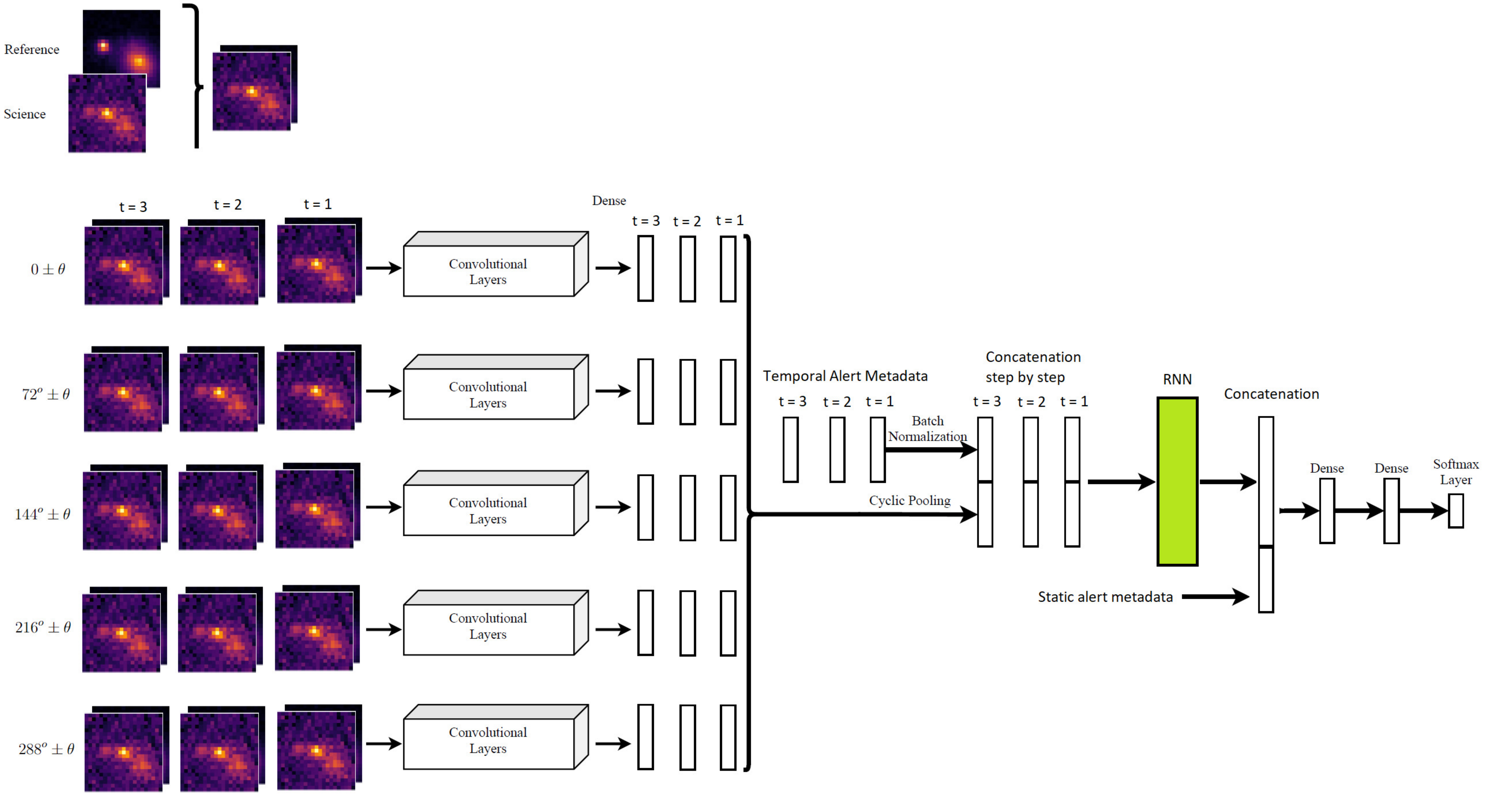}{width = 0.89\textwidth}
  \label{temporal_stamp}
\end{imagesmc}

\subsection{Experiments}

In this section, we first describe the experiments to select the best hyperparameters of our proposed model. Every experiment run was trained up to $30,000$ iterations, the loss was evaluated in the validation set every 10 iterations, and the best model so far was saved.

\begin{table}[H]
\centering
\resizebox{\columnwidth}{!}{%
\begin{tabular}{c|c|c}
Layer                                                                                     & Parameters & Output size \\ \hline \hline
Input                                                                                  & -                     & $t \times 33 \times 33 \times 3$  \\ \hline
\begin{tabular}[c]{@{}c@{}}Rotation \\ augmentation\end{tabular}          & -                     & $t \times 33 \times 33 \times 3$  \\ \hline
Convolution                                                                              & $4 \times 4$, 32             & $t \times 24 \times 24 \times 32$ \\ \hline
Convolution                                                                              & $5 \times 5$, 32             & $t \times 24 \times 24 \times 32$ \\ \hline
Max pooling                                                                              & $2 \times 2$, stride 2       & $t \times 12 \times 12 \times 32$ \\ \hline
Convolution                                                                              & $5 \times 5$, 64             & $t \times 12 \times 12 \times 64$ \\ \hline
Convolution                                                                              & $5 \times 5$, 64             & $t \times 12 \times 12 \times 64$ \\ \hline
Convolution                                                                              & $5 \times 5$, 64             & $t \times 12 \times 12 \times 64$ \\ \hline
Max pooling                                                                              & $2 \times 2$, stride 2       & $t \times 6 \times 6 \times 64$   \\ \hline
Flatten                                                                                 & -                     & $t \times 2304$         \\ \hline
Fully connected                                                                          & $2304 \times 64$                  & $t \times 64$           \\ \hline
\begin{tabular}[c]{@{}c@{}}Rotation \\ concatenation\end{tabular}                                                              & -                     & $t \times 6 \times 64$       \\ \hline
Cyclic pooling                                                                           & -                     & $t \times 64$           \\ \hline
\begin{tabular}[c]{@{}c@{}}Concat with dynamic\\  metadata after BN\end{tabular} & -                     & $t \times 64 + 8$       \\ \hline
RNN                                                                                      & Variable                    & $72$               \\ \hline
\begin{tabular}[c]{@{}c@{}}Concat with static \\ metadata after BN\end{tabular} & -                     & $87$               \\ \hline
\begin{tabular}[c]{@{}c@{}}Fully Connected \\ with dropout\end{tabular}                                                              & $90 \times 64$               & $64$               \\ \hline
Fully connected                                                                          & $64 \times 64$               & $64$               \\ \hline
\begin{tabular}[c]{@{}c@{}}Fully Connected \\ with softmax\end{tabular}                                                            & $64 \times 3$                & $3$ (n° classes)    \\ \hline
\end{tabular}
}
\caption{Temporal Stamp Classifier architecture.}
\label{tab:temporal_stamp}
\end{table}

\noindent After the first $20000$ iterations, if a lower loss is obtained on the validation set, another $10000$ iterations are performed. The validation and testing sets are randomly sampled just once before starting any experiment, and these sets are kept the same for all experiments. $200$ samples per class are taken from the whole dataset, totaling $600$ samples for each of the two subsets validation, and testing; the rest of the samples were used for training. Every experiment was trained 10 times, and the average and standard deviation are reported. The Adam optimizer was used, with its default hyperparameters, a learning rate of 1e-4 and batch size of $128$. The training was performed in a $GTX1080Ti$ GPU.

\subsubsection{Recurrent model}

In order to determine the best temporal model for each number of alerts, we searched the hyperparameter values shown in table \ref{param-recurr} for six models: LSTM \cite{lstm}, GRU \cite{GRU}, SimpleRNN, Gamma Memory \cite{gamma}, Convolution in one dimension (Conv1D) \cite{conv} and TDNN \cite{tdnn}.

\noindent To process the images, we used the configuration of the enhanced stamp classifier in Section III that consists of $k = 6$ rotations with a range of rotation of $\theta = 0.08[rad]$ and an image size $a = 33$. The output of the convolutional layers was stacked with the time differences between alerts, forming the input to the recurrent layers. No metadata was used in this experiment. Next, the output of the recurrent layer was injected into an MLP with softmax activation to determine the class of each sample.

Table \ref{table:recurrent} shows the best configuration obtained for each model for each number of alerts. We present the accuracy achieved in the test set and the p-value obtained between the proposed model and the LSTM model for each number of alerts, as well as the inference time (IT) for the model. LSTM was chosen as the baseline model for comparison purposes because it is the most complex model used in our experiments. In table 6, $A$ stands for the number of alerts, $L$ the number of layers, $U$ the number of units, $k$ the number of Leaky Filters, $V$ stands for time windows, and $Ke$ is the kernel size for the 1D convolution.

The results show that there are no significant differences among the tested recurrent models in terms of accuracy, resulting in similar values for every number of alerts used in this work. This is ratified by the p-values of the permutation hypothesis test \cite{hipotesis} obtained between the different models and LSTM. The p-values are bigger than 0.05 for every pairwise of models.

On the other hand, the best configuration for most recurrent model tends to be the simplest one. A single layer in the case of layer-dependent models (LSTM, GRU and SimpleRNN), and a small kernel size for Conv1D. In contrast, Gamma Memory requires a substantial number of Leaky Filters. In terms of inference time (IT), the Gamma Memory and Conv1D tend to be 4 times faster in processing a sample for 2, 3 and 4 alerts, while the rest of the models have similar computational times for every configuration used. Figure 6 shows that the difference between inference times of these models is important, having Gamma Memory and Conv1D separated by a considerable distance from the others models, which are grouped together. This result is similar for all number of alerts as shown in Table 6. This difference in time inference is relevant for the use of the model in a production environment.

As the number of alerts increased, the accuracy obtained by the models increased too, but not substantially. This may be due to the fact that the information given by stamp images is not enough to solve the problem and additional metadata is needed. These results confirm that simple recurrent models and even Conv1D are competitive with complex models such as LSTM and GRU for short sequences of data and are more effective since their inference time is shorter. In order to choose the best model, as the results are similar and has no statistically significant difference, we choose the model with the lower inference time, in order to use it later in a production environment.

\begin{table}[H]
\resizebox{0.8\columnwidth}{!}{%
\begin{tabular}{c|cc}
Model          & \multicolumn{1}{c|}{Cells}       & Layers  \\ \hline \hline
LSTM          & \multicolumn{1}{c|}{32, 64, 128}  & 1, 2, 3   \\ \hline
GRU           & \multicolumn{1}{l|}{32, 64, 128}  & 1, 2, 3   \\ \hline
SimpleRNN    & \multicolumn{1}{c|}{32, 64, 128}  & 1, 2, 3   \\ \hline
Model          & \multicolumn{2}{c}{Filters k}                 \\ \hline \hline
Gamma Memory & \multicolumn{2}{c}{1, 2, 3, 4, 5, 6, 7, 8} \\ \hline
Model          & \multicolumn{2}{c}{Filter size}                 \\ \hline \hline
TDNN  & \multicolumn{2}{c}{2} \\ \hline
Conv1D  & \multicolumn{2}{c}{2,...,A} \\ \hline
\end{tabular}
}
\caption{Hyperparameter search to determine the best model for processing short sequence of alerts. \textit{A} represent the number of alerts and $k$ stands for the number of Leaky Filters for the Gamma Memory model.}
\label{param-recurr}
\end{table}

\begin{table}[H]
\resizebox{\columnwidth}{!}{%
\begin{tabular}{c|c|c|c|c|c|c|c|}
\cline{2-8}
 & Model & GRU & LSTM & G. M. & TDNN & RNN & Conv1D \\ \hline
\multicolumn{1}{|c|}{\multirow{4}{*}{2 A}} & Acc. & $91.3 \pm 0.4$ & $91.7 \pm 0.3$ & $91.3 \pm 0.2$ & $91 \pm 0.1$ & $91.2 \pm 0.7$ & $91.3 \pm 0.6$ \\ \cline{2-8} 
\multicolumn{1}{|c|}{} & p-v. & $0.35$ & Ref. & $0.1$ & $0.1$ & $0.68$ & $0.3$ \\ \cline{2-8} 
\multicolumn{1}{|c|}{} & Arc. & 1L 256U & 1L 256U & $k = 7$ & $V = 2$ & 1L 128U & $Ke=2$ \\ \cline{2-8} 
\multicolumn{1}{|c|}{} & IT & 21.04 & 20.57 & 5.47 & 22.68 & 24.63 & 5.75 \\ \hline
\multicolumn{1}{|c|}{\multirow{4}{*}{3 A}} & Acc. & $91.8 \pm 0.4$ & $92 \pm 0.3$ & $91.4 \pm 0.5$ & $91.6 \pm 0.1$ & $91.8 \pm 0.4$ & $91.6 \pm 0.6$ \\ \cline{2-8} 
\multicolumn{1}{|c|}{} & p-v. & $0.28$ & Ref. & $0.06$ & $0.07$ & $0.08$ & $0.1$ \\ \cline{2-8} 
\multicolumn{1}{|c|}{} & Arc. & 1L 256U & 1L 64U & $K = 7$ & $V = 2$ & 1L 128U & $Ke = 3$ \\ \cline{2-8} 
\multicolumn{1}{|c|}{} & IT & 30.42 & 29.63 & 7.84 & 32.88 & 30.42 & 8.33 \\ \hline
\multicolumn{1}{|c|}{\multirow{4}{*}{4 A}} & Acc. & $92.5 \pm 0.4$ & $92.2 \pm 0.6$ & $92.2 \pm 0.5$ & $92.4 \pm 0.5$ & $92.4 \pm 0.4$ & $92.5 \pm 0.6$ \\ \cline{2-8} 
\multicolumn{1}{|c|}{} & p-v. & $0.31$ & Ref. & $0.81$ & $0.68$ & $0.69$ & $0.48$ \\ \cline{2-8} 
\multicolumn{1}{|c|}{} & Arc. & 1L 256U & 3L 256U & $K=6$ & $V = 2$ & 1L 64U & $Ke = 2$ \\ \cline{2-8} 
\multicolumn{1}{|c|}{} & IT & 39.31 & 46.06 & 10.86 & 40.8 & 49.02 & 11.13 \\ \hline
\multicolumn{1}{|c|}{\multirow{4}{*}{5 A}} & Acc. & $93.2 \pm 0.5$ & $92.9 \pm 0.3$ & $92.6 \pm 0.4$ & $92.9 \pm 0.3$ & $92.6 \pm 0.6$ & $93.2 \pm 0.6$ \\ \cline{2-8} 
\multicolumn{1}{|c|}{} & p-v. & $0.23$ & Ref. & $0.24$ & $0.98$ & $0.2$ & $0.29$ \\ \cline{2-8} 
\multicolumn{1}{|c|}{} & Arc. & 1L 64U & 1L 128U & $k = 8$ & $V = 2$ & 1L 64U & $Ke = 4$ \\ \cline{2-8} 
\multicolumn{1}{|c|}{} & IT & 55 & 50.54 & 55.06 & 57.16 & 57.44 & 71.48 \\ \hline
\end{tabular}%
}
\caption{Best results for each of the five recurrent models, plus Conv1D, tested per number of alerts ($\#A$). Here only images were used. For each model, we show the best configuration of hyperparameters (Arc), the average test accuracy and standard deviation of 10 iterations (Acc), the inference time (IT), and the p-value of each model against the LSTM model corresponding to the number of alerts.}
\label{table:recurrent}
\end{table}

\begin{figure}[H]
\includegraphics[scale = 0.45]{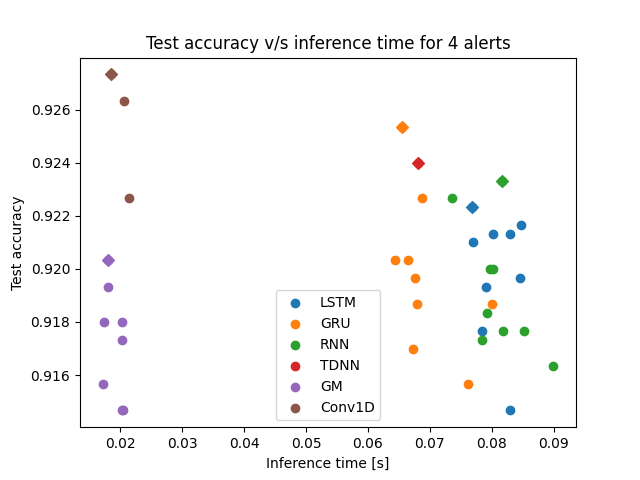}
\caption{Test accuracy v/s inference time for 4 alerts. Diamonds represent the best configuration for a given model in terms of test accuracy.}
\label{fig:accuracydetect}
\end{figure}

\subsubsection{Cross-matching detections with AllWISE catalog}

In \cite{light_curve}, it was shown that the information from the AllWISE catalog improves the classification of stochastic objects, being the most valuable information according to random forest importance values.

In order to improve the discrimination between AGN and VS, we included information on the match radius of 2 arcseconds, obtaining W1, W2, W3, and W4 photometry, and calculating the difference between these magnitudes W1-W2, W2-W3, and W3-W4, which are used as new metadata. We trained a model using only metadata as input, where the features are separated into static and dynamic metadata, according to its temporal behavior. Dynamic metadata are injected in an LSTM layer, that processes the recurrent dependences present on the metadata. The output of the LSTM layer is concatenated with static metadata and then injected into MLP layers to obtain the final classification score.

We trained models with AllWISE information and without it. The model use the static and dynamic metadata, and we consider AllWISE as a static metadata due to it's dependence with the position of the object. Table \ref{table:allwise} shows the average accuracy and standard deviation of 5 random initializations, obtained for each number of alerts with and without AllWISE information. In addition, the p-value between both models is given.

The results of table \ref{table:allwise} show that the crossmatch between the original dataset and the AllWISE catalog improves the performance of the proposed model for every number of alerts, achieving a $3\%$ increase in accuracy for the case of 2 alerts and $1.3\%$ for 5 alerts. This result is corroborated by the p-values obtained for 2, 3, and 4 alerts, being all smaller than $0.05$, meaning that the distributions of both models are different and have statistical significance. For 5 alerts, the p-value is 0.08, meaning that the improvement by the AllWISE catalog is not statistically significant in this case. This improvement in the accuracy of the model comes mostly from a better discrimination between AGN and VS. In \cite{light_curve}, it is stated that AllWISE colors are essential to discriminate among AGN subclasses and to identify these kinds of objects from others. In the case of this work, the AllWISE information can be used with 2 or more alerts for real objects with a fixed position in the sky.

\subsubsection{Complete model}

The complete temporal stamp classifier model for the classification of sequences of astronomical alerts is a concatenation of the best models found in the previous experiments. Each number of detections has it's own model trained separately. The complete model uses the best parameters found for the convolutional filters, followed by concatenation of the dynamic metadata, thus being processed by the best model for every number of detection found in Table \ref{table:recurrent} in terms of inference time and summarized in Table \ref{tab:complete}. The output of the recurrent layer is then concatenated with the static metadata including AllWISE catalog information obtained through crossmatch, then inputted to MLP layers with ReLU as their function activation except for the last layer, which have a softmax activation. The architecture of the proposed temporal stamp classifier is shown in Figure 5.

\begin{table}[H]
\resizebox{\columnwidth}{!}{%
\begin{tabular}{|l|l|l|l|l|}
\hline
\multicolumn{1}{|c|}{Model / Alerts} & \multicolumn{1}{c|}{2 alerts} & \multicolumn{1}{c|}{3 alerts} & \multicolumn{1}{c|}{4 alerts} & 5 alerts \\ \hline
Without AllWISE & $95.07 \pm 0.57$ & $96.1 \pm 0.32$ & $96.9 \pm 0.25$ & $97.37 \pm 0.55$ \\ \hline
With AllWISE & $98 \pm 0.56$  & $98.27 \pm 0.27$ &  $98.37 \pm 0.29$ & $98.07 \pm 0.29$ \\ \hline
p-value & $0.01$ & $0.01$ & $0.008$ & $0.08$ \\ \hline
\end{tabular}%
}
\caption{Results of recurrent models that use only metadata as input with and without AllWISE information for every number of alerts. The average accuracy and standard deviation of 5 iterations for each model, as well as the p-value between the model with and without AllWISE information, are shown.}
\label{table:allwise}
\end{table}

Table \ref{tab:complete} shows the test accuracy obtained with the best model for each number of alerts when using recurrent models or Conv1D in terms of inference time. We also present the confusion matrices for each number of alerts in Figure \ref{fig:recurrente}.

The models shown in table \ref{tab:complete} corresponds to the best configuration obtained in the previous experiments, and it's dominated by Conv1D model for the case of 2, 3 and 4 detections, and by LSTM in 5 detections. The configuration for every model tends to be the simplest one and the accuracy obtained in test partition tends to be similar between the number of alerts except for 4 alerts. These results can be interpreted as the model reaching a limit in terms of its performance in the task at hand, which has as a consequence that there is no improvement in accuracy as we add more alerts to the model. An interesting option is to increase the number of classes, expanding in a controlled way some of the classes (SN for example) into their subclassifications like the light curve classifier. The decrease in accuracy for the 4 alerts model could be just a statistical coincidence because the model with less and more detections have better performances and the decrease it is just of $0.5\%$. These results are competitive with the ones obtained in \cite{light_curve}, reaching a comparable value of accuracy with a different test set. The proposed temporal stamp classifier achieves more successful prediction than the original stamp classifier for the first detection for the classification of the three classes considered, filling the gap between the light curve model and the first detection model. See Figure 8. 

\subsubsection{Comparison with baseline models}

In order to corroborate the effectiveness of the proposed model, we compare its performance with those of the ALeRCE models \cite{stamp} and \cite{light_curve}. We took all the test samples and made calls to the ALeRCE API \cite{api}, obtaining the classification made by the stamp classifier and the light curve classifier, and comparing the results with the true class. In this way, we obtained the confusion matrices shown in Figures \ref{fig:comparison_stamp} and \ref{fig:comparison_lc}.

\begin{figure}[H]
\centering
\begin{subfigure}[b]{0.4\textwidth}
\includegraphics[scale = 0.17]{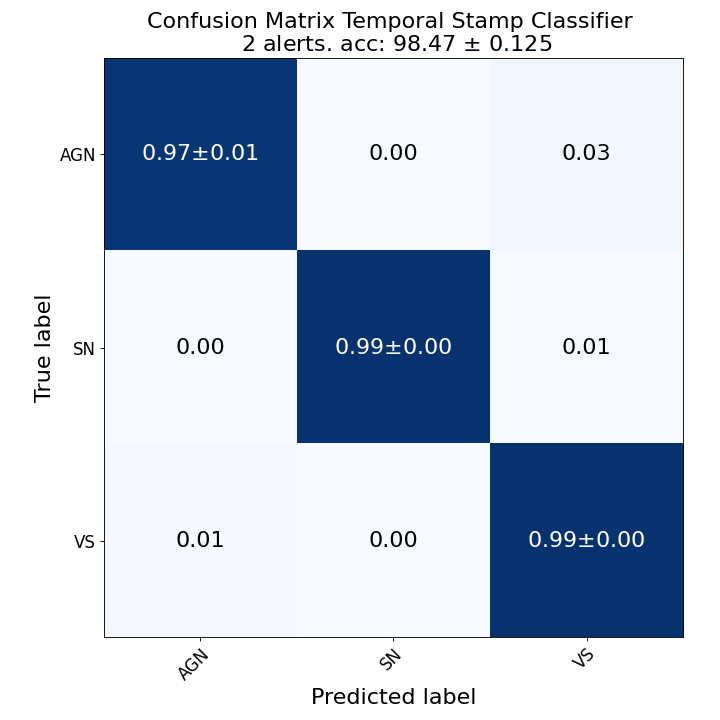}
\caption{2 alerts}
\end{subfigure}
\begin{subfigure}[b]{0.4\textwidth}
    \includegraphics[scale = 0.17]{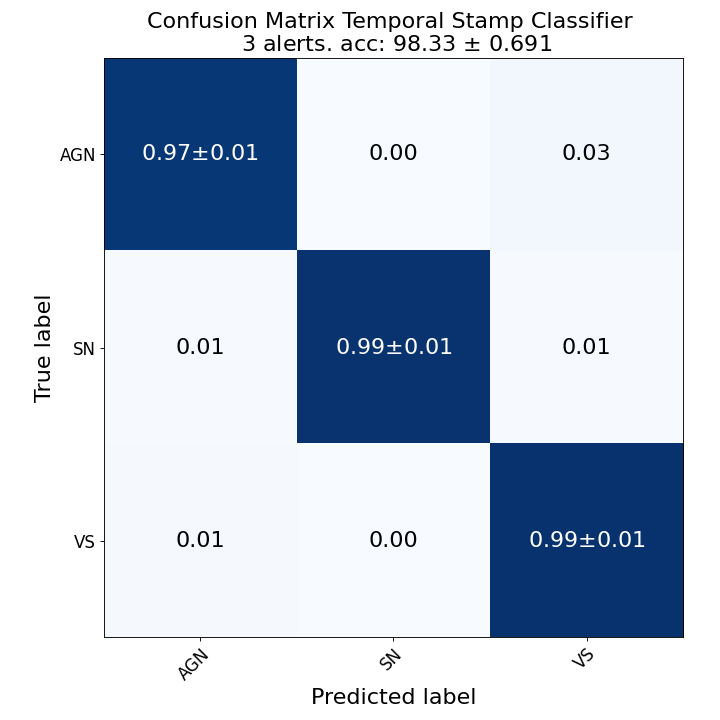}
\caption{3 alerts}
\end{subfigure}
\begin{subfigure}[b]{0.4\textwidth}
\includegraphics[scale = 0.17]{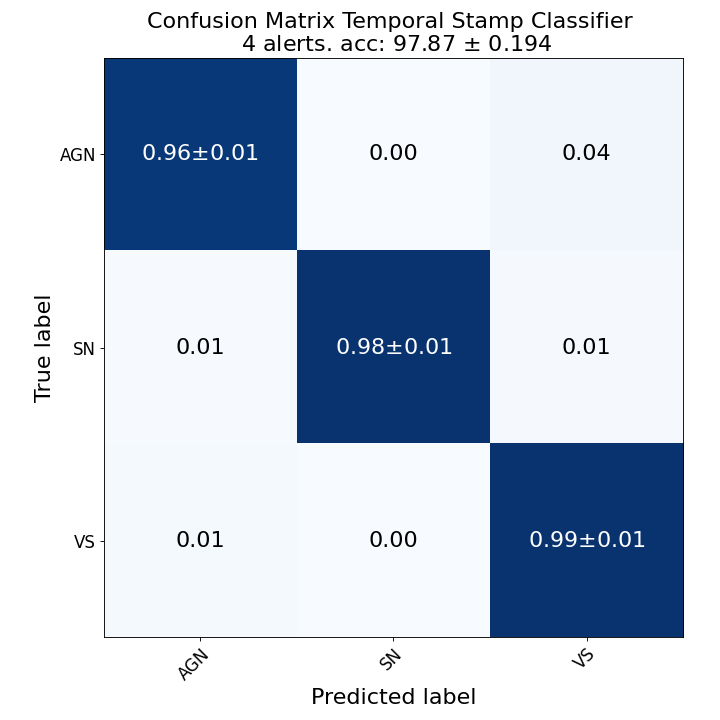}
\caption{4 alerts}
\end{subfigure}
\begin{subfigure}[b]{0.4\textwidth}
    \includegraphics[scale = 0.17]{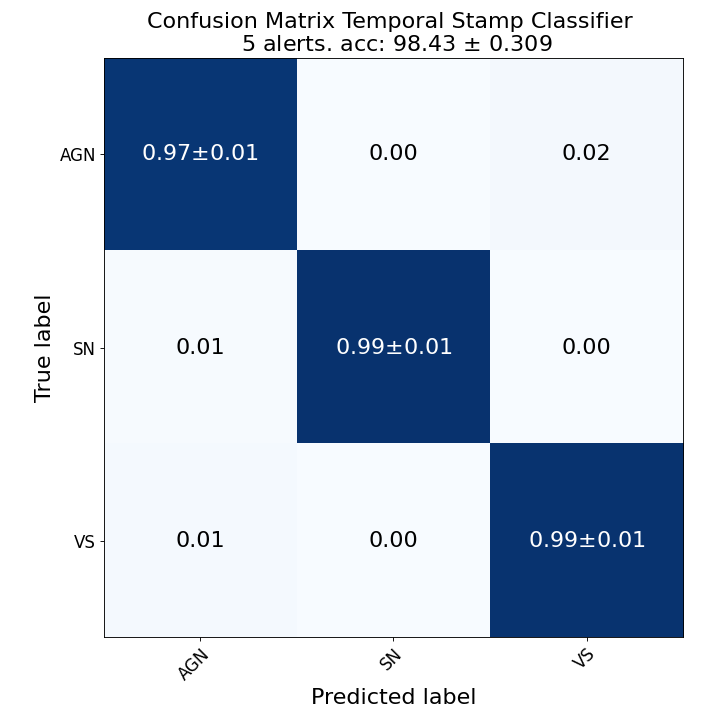}
\caption{5 alerts}
\end{subfigure}
\caption{Average confusion matrix for the test set using 5 different realizations of the temporal stamp classifier for each number of alerts using the model configurations of Table \ref{table:recurrent}. }
\label{fig:recurrente}
\end{figure}

\begin{figure}[H]
\includegraphics[width =0.8 \textwidth]{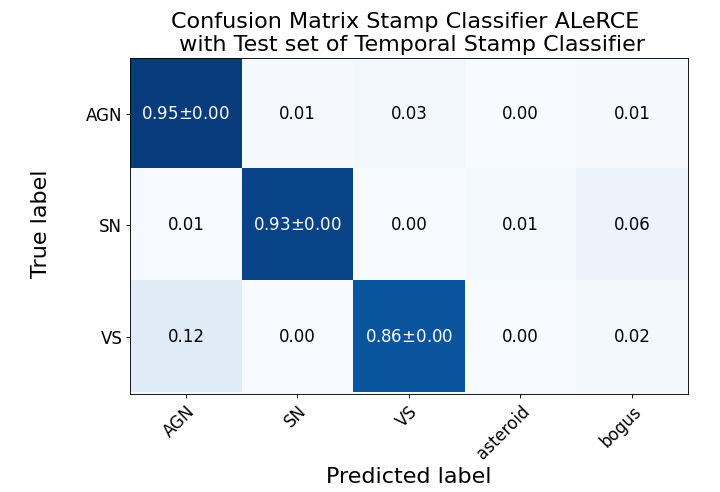}
\caption{Confusion matrix of original Stamp Classifier \cite{stamp} in the test set.}
\label{fig:comparison_stamp}
\end{figure}

\begin{figure}[H]
\centering
    \includegraphics[width = 0.7\textwidth]{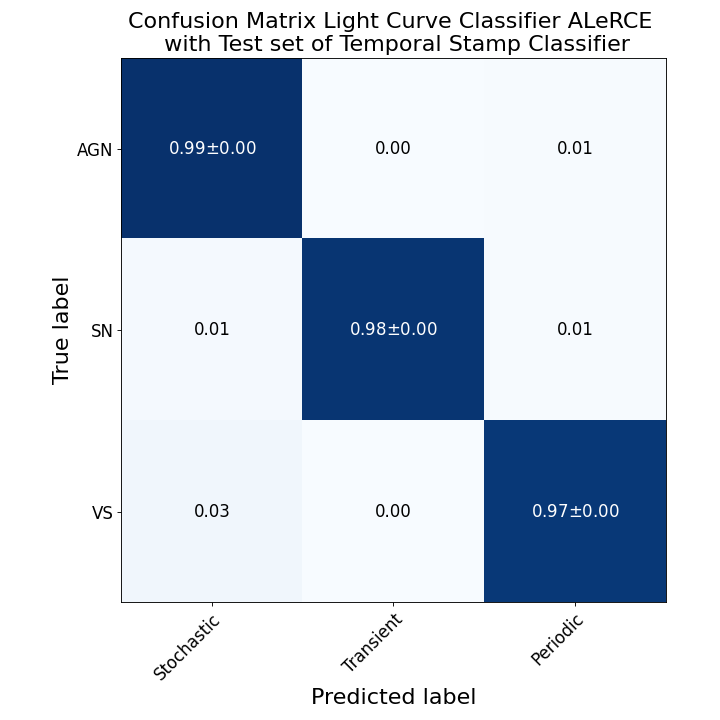}
\caption{Confusion matrix of Light Curve Classifier \cite{light_curve} in the test set.}
\label{fig:comparison_lc}
\end{figure}

The results of the original Stamp Classifier \cite{stamp} show that it can predict relatively well the SN and AGN classes, but the class VS is confused with AGN in $12\%$ of the cases. This difficulty is mainly solved in our model by including the crossmatch with the AllWISE catalog. Notice that the original stamp classifier deals with 5 classes instead of 3, because it includes asteroids and bogus that maybe present in a single detection. In the case of the Light Curve Classifier \cite{light_curve}, Figure 9 shows that the predictions with our test samples are pretty similar to our own results, shown in Figure \ref{fig:recurrente} for different numbers of alerts. These results show that the proposed temporal stamp classifier is able to fill the gap between the two ALeRCE models currently in production. This allows improving the confidence in the classification when the object is detected more than one time, and merge the information given by the images with the metadata.

\begin{table}[H]
\resizebox{0.8\columnwidth}{!}{%
\begin{tabular}{c|c|c|c|}
\cline{2-4}
                                    & Model &  Architecture & Test accuracy \\ \hline
\multicolumn{1}{|c|}{2 alerts} & Conv1D   & $Ke = 2$ & $98.47 \pm 0.1$    \\ \hline
\multicolumn{1}{|c|}{3 alerts} & Conv1D  & $Ke = 3$  & $98.33 \pm 0.7$           \\ \hline
\multicolumn{1}{|c|}{4 alerts} & Conv1D    & $Ke = 2$ & $97.87 \pm 0.2$   \\ \hline
\multicolumn{1}{|c|}{5 alerts} & LSTM   & 1L 128U  & $98.43 \pm 0.3$      \\ \hline
\end{tabular}
}
\caption{Best model configuration by inference time obtained by number of alerts, according to Table \ref{table:recurrent}. Test accuracy stands for the accuracy of the complete model that uses images and metadata. Ke stands for Kernel size, L the number of layers and U stands for Units. }
\label{tab:complete}
\end{table}

\section{Conclusions}

The proposed temporal stamp classifier is a significant improvement for the early classification of astronomical events, allowing us to go beyond the prediction of the first alert of an astronomical event using sequences of images from the alerts, the time information, metadata, and crossmatch with the AllWISE catalog. In addition, we proposed an enhanced stamp classifier for a single detection, by adding random rotations and changing the size of the stamps (cropped images). This led to a statistically significant increase in the performance of the stamp classifier.

As future work, we propose to split the current three classes into known subgroups in order to take advantage of the extra information given by a larger number of alerts. For instance, we can split the SN class into SN type Ia and all other types of SN. Another idea comes from \cite{oscar}, where the time is modulated by a function, which allows improving the results in the classification of SNe, in contrast with using the time difference between detections. The proposed temporal stamp classifier might be used to find outliers like \cite{anomaly,outliers}, e.g., using the model uncertainty operator proposed in \cite{principe} to find examples that are in the tail of the distributions. Finally, deep attention models (transformers) could be used instead of recurrent models or to process images directly \cite{vision,attention}.

    \bibliography{main}
\end{multicols}

\end{document}